\documentclass[journal]{vgtc}

\usepackage[export]{adjustbox}
\usepackage{amsmath}
\usepackage{array}
\usepackage{blindtext}
\usepackage{booktabs}
\usepackage{calc}
\usepackage{ccicons}
\usepackage{comment}
\usepackage{enumitem}
\usepackage{epstopdf}
\usepackage{eurosym}
\usepackage{float}
\usepackage[T1]{fontenc}
\usepackage[nomessages]{fp}
\usepackage[utf8]{inputenc}
\usepackage{makecell}
\usepackage{marginnote}
\usepackage{mathptmx}
\usepackage{microtype}
\usepackage{multicol}
\usepackage{multirow}
\usepackage{mwe}
\usepackage{nicematrix}
\usepackage[percent]{overpic}
\usepackage{setspace}
\usepackage{soul}
\usepackage{tabu}
\usepackage{tcolorbox}
\usepackage{textcomp}
\usepackage{tikz}
\usepackage{times}
\usepackage{todonotes}
\usepackage{wrapfig}
\usepackage{xspace}

\newcommand{\changes}[1]{#1}
\newcommand{\eg}{e.\,g.}
\newcommand{\etal}{et al.\xspace}

\newcommand{\ie}{i.\,e.}
\newcommand{\mytitle}{
\rvision{What Makes a Visualization Image Complex? }}

\newcommand{\osflink} {\href{https://osf.io/bdet6/} {\texttt{osf\discretionary{}{.}{.}io\discretionary{/}{}{/}bdet6}}}
\newcommand{\osfpreregistration} {\href{https://osf.io/5xe8a} {\texttt{osf\discretionary{}{.}{.}io\discretionary{/}{}{/}5xe8a}}}

\newcommand{\sm}{\textcolor{gray}{Apdx.}\xspace}

\newcommand{\tochange}[1]{#1}
\newcommand{\totaltrials}{27,571\xspace}
\newcommand{\totaltrialsperrun}{1,969\xspace}
\newcommand{\vccolablink} {\href{https://drive.google.com/file/d/1XDqI51N7CrGhw5aSn6F6qXzvwt7XsFrJ/view}{\texttt{go.osu.edu/vcmodel}}}
\newcommand{\vcdatalink} {\href{https://docs.google.com/spreadsheets/d/1ZxN8kLq9Hhf1nTjuc5epv6bLhicfPt-UofcBbEchLS4/edit?usp=sharing} {\texttt{go.osu.edu/viscomplexitydata}}\xspace}
\newcommand{\vcdataset}{VisComplexity2K\xspace}
\newcommand{\vcgoogledrive}{\href{https://go.osu.edu/vcgoogledrive}{\texttt{go.osu.edu/vcgoogledrive}}}

\newcommand{\zfq}[1]{#1}

\definecolor{Black}{RGB}{0,0,0}
\definecolor{jccolor}{RGB}{159,89,53}
\definecolor{mscolor}{rgb}{0,0,0.7}
\definecolor{rlcolor}{rgb}{0.7,0,0}

\onlineid{1919}

\vgtccategory{Research}

\vgtcpapertype{theory/model}

\newcommand\rvision[1]{\protect\textcolor{black}{#1}}
\title{\mytitle}

\author{\authororcid{Mengdi Chu} {0000-0003-0533-7801},
\authororcid{Zefeng Qiu}{0009-0002-8451-8001},
\authororcid{Meng Ling}{0000-0001-6597-5448},
\authororcid{Shuning Jiang}{0000-0002-6706-2818},
\authororcid{Robert~S.~Laramee}{0000-0002-3874-6145}, \texorpdfstring{\\}{ }
\authororcid{Michael~Sedlmair}{0000-0001-7048-9292}, and
\authororcid{Jian~Chen }{0000-0002-3874-6145} }

\authorfooter{

\item Mengdi Chu, Zefeng Qiu, Meng Ling, Shuning Jiang, and Jian~Chen are with Ohio State University. E-mail: \{chu.752, qiu.573, ling.253, jiang.2126, chen.8028\}@osu.edu.
\item Robert~S.~Laramee is with University of Nottingham. E-mail: R.S.Laramee@swansea.ac.uk.
\item Michael~Sedlmair is with University of Stuttgart. E-mail: michael.sedlmair@visus.uni-stuttgart.de.
\\ }

\abstract{We investigate the perceived visual complexity (VC) in data visualizations using objective image-based metrics. We collected VC scores through a large-scale crowdsourcing experiment involving 349 participants and 1,800 visualization images. We then examined how these scores align with 12 image-based metrics spanning pixel-based and statistic-information-theoretic (clutter), color, shape, and our two new object-based metrics (meaningful-color-count (MeC) and text-to-ink ratio (TiR)). Our results show that both low-level edges and high-level elements affect perceived VC in visualization images; the number of corners and distinct colors are robust metrics across visualizations. Second, feature congestion, a statistical information-theoretic metric capturing color and texture patterns, is the strongest predictor of perceived complexity in visualizations rich in the same continuous color/texture stimuli; edge density effectively explains VC in node-link diagrams. Additionally, we observe a bell-curve effect for texts: increasing TiR initially reduces complexity, reaching an optimal point, beyond which further text increases VC. Our quantification model is also interpretable---enabling metric-based explanations---grounded in the \vcdataset dataset, bridging computational metrics with human perceptual responses. 
\osfpreregistration ~has the preregistration and \osflink ~has the \vcdataset dataset, source code, and all \sm and figures. 
}

\keywords{Perceived visual complexity, image-based metrics, scene-like, text-ink-ratio, meaningful-color-count}

\teaser{
\centering
\includegraphics[width=\textwidth]{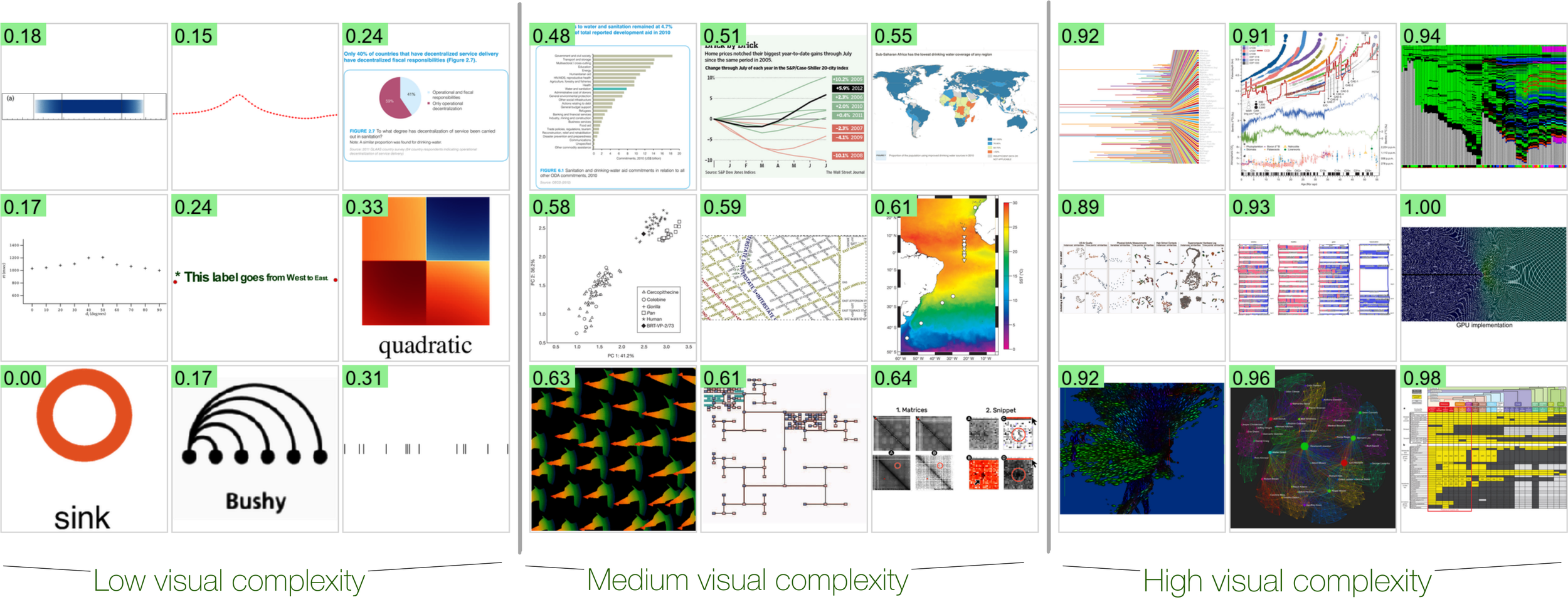}
\centering
\caption{Examples of low, medium, and high visual complexity (VC) images from the \vcdataset dataset. VC scores labeled in the upper-left corner are gathered from human observer ratings. }
\label{fig:teaserVis2025} }

\graphicspath{{figs/}{figures/}{pictures/}{images/}{./}} \PassOptionsToPackage{warn}{textcomp} 
\usetikzlibrary{patterns,shadings}

\PassOptionsToPackage{warn}{textcomp} 
\usetikzlibrary{patterns,shadings}

\newlength{\pictureheight}
\newlength{\picturewidth}

\newlength\myheight
\newlength\mydepth
\newlength\maxlen
\def\header{(262) graphical element}
\newcommand\databar[4][4CD9F6]{ \FPeval\result{round(#2/#3:4)} \rlap{\textcolor[HTML]{#1}{\hspace*{\dimexpr-\tabcolsep+.5\arrayrulewidth+5pt} \rule[-.3\ht\strutbox]{\result\maxlen}{1.3\ht\strutbox}}}
\makebox[\dimexpr\maxlen-2\tabcolsep+\arrayrulewidth][l]{#4} }

\begin{document}

\firstsection{Introduction}

\maketitle Understanding the visual complexity (VC) of an image is crucial in a variety of applications. It can affect the ability of an observer in understanding an environment~\cite{heaps1999similarity,snodgrass1980standardized} or navigate through it~\cite{hartmann2008towards}. VC affects visibility~\cite{feixas1999information}, aesthetics~\cite{he2022beauvis,reppa2008visual}, clutter~\cite{rosenholtz2005feature, rosenholtz2007measuring, ajani2021declutter}, and thus understandability~\cite{duffy2013measuring} and memory~\cite{tuch2009visual}. VC also impacts visual search~\cite{rosenholtz2007measuring}, design~\cite{windhager2024complexity}, and metaphorical thinking~\cite{borgo2012empirical}.

Evaluating perceived VC can also contribute to foundational research in visualization. VC has been defined as ``\textit{the amount of detail or intricacy}''~\cite{snodgrass1980standardized, purchase2012exploration}. Most evaluations use simple images, that are less complex than those found in practical, real-world uses~\cite{yoghourdjian2018exploring}. In contrast, practitioners foraging for information need to handle realistic visual designs to achieve more generalizable outcomes~\cite{windhager2024complexity}. This gap introduces at least two evaluation challenges. First, the vast majority of visualization images contain an array of features, patterns, and contexts. We have a limited understanding of their feature space. Even though high-level features (\eg, number of elements~\cite{mack2004perceptual}, text~\cite{hearst2023show}, and the scene gist~\cite{olivia2004identifying}) and a computational equivalent (\eg, feature congestion~\cite{rosenholtz2005feature}) contribute to VC, studied previously primarily for natural scenes~\cite{sonkusare2019naturalistic}, it is unclear if any of the existing metrics and theories are generalizable across domains. Second, we also lack an understanding of observers' experiences when examining these images.

Our goal is to identify some of the image features that influence VC. In this work, we explore how VC can be viewed through both perceived and objective lenses. To this end, we draw inspiration from the extensive work in vision science that collects large-scale absolute VC scores~\cite{saraee2020visual}. Furthermore, we draw on a precedent for such metric-based explanations~\cite{purchase2012exploration, rogowitz1998perceptual, ramanarayanan2008dimensionality},
\rvision{to study what metrics can represent VC.} In the longer term, measuring and explaining a human
\rvision{judgment} with an extensive set of metrics can help reverse engineer high-level multifaceted understanding of the perceptual mechanisms~\cite{kramer2023features}.

\begin{figure*}[!t]
    \centering
    \includegraphics[width=0.95\textwidth]{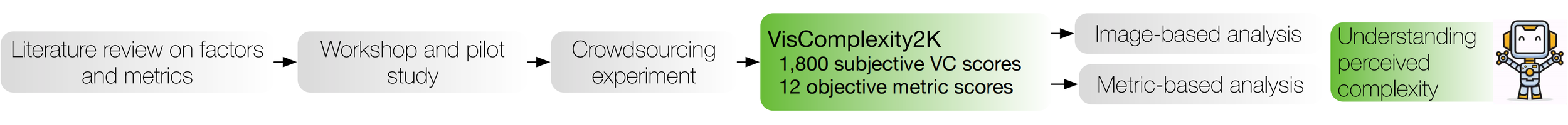}
    \caption{\textbf{Overview of our process to study perceived visual complexity.} We used objective quality metric to measure the subjective high-order perception. Activities (in gray) and outcomes (in green). }
    \label{fig:overview}
\end{figure*}

\textbf{Method}~ (\autoref{fig:overview}). We begin with an extensive literature review on subjective VC factors in~\autoref{tab:svc} to guide our selection of 12 objective metrics in \autoref{tab:ovc}. Some metrics, such as information-theoretic metrics, capture mathematical uncertainty, while others are sensitive to clutter (feature congestion)~\cite{rosenholtz2007measuring}, shape and structure, or color. We further introduce two new metrics to represent visualization elements: (1) MeC (the number of \rvision{meaningful} colors), to represent {the number of component parts} in an image---by applying color-as-object-boundaries~\cite{greene2019information}; (2) TiR (text-ink-ratio)---\textit{the proportion of annotation in a visualization image}, to investigate how context influences VC, given its prominent role in visualization~\cite{rahman2024qualitative} and its power to assist viewers to comprehend concepts that are otherwise difficult to interpret. We collect perceived VC scores of 1,800 visualization images and then use image-based feature metrics to capture VC judgment.
\rvision{While a precedent for metric-based explanations of specific visualizations~(\eg, networks~\cite{purchase2012exploration}) and stimulus-specific clutter (\eg, color and texture~\cite{rosenholtz2005feature}) exists within our community, measuring and explaining human perceived VC using an extensive set of metrics offers a pathway towards finding a low-dimensional proxy to engineer high-dimensional VC.}

\textbf{Results.}
We collected perceived VC scores from 1,800 images~(\autoref{fig:teaserVis2025}). Our analysis reveals that a range of metrics contribute to explaining perceived VC. Among them, edge density (ED) emerges as a dominant factor across visualization images. Object-level metrics such as color count (MeC) and text-ink-ratio (TiR) capture high-level aspects of visual complexity but have little correlation with pixel-based metrics. A notable finding is that clutter-based feature congestion (FC) ~\cite{rosenholtz2007measuring, rosenholtz2005feature}, an information-theoretic measure adapted to account for statistical scene structures of color and texture, was particularly effective in distinguishing VC in visualizations rich in the same color and texture variations. Finally, we observe that the amount of annotation text, as measured by TiR, has a non-linear U-shaped relation with perceived VC: moderate levels of text reduce VC while excessive annotations increase it,
extending the current literature. 

In summary, our main \textbf{contributions} are: (1) A crowdsourcing study collecting perceived VC scores for 1,800 visualization images from 349 participants. (2) A metrics study that analyzes how 10 existing and two new metrics help us interpret perceived VC. (3) We provide the human and metric scores from both studies, along with the 1,800 visualization images, in a dataset called \vcdataset, openly available at \osflink.

\section{\rvision{Background and Related Work}}
\label{sec:relatedwork}

This section reviews the definitions of VC and the associated metrics. As outlined in~\autoref{tab:svc}, our literature review identifies five key categories, which serve as the foundation for organizing the metrics summarized in~\autoref{tab:ovc}.

\subsection{Definitions of Visual Complexity}
\label{sec:lrdefinition}

\noindent Defining visual complexity is challenging due to its subjective and multifaceted nature. VC arguably plays a key role in comprehension. For example, vision science has broadly explored human understanding of VC, and defined it as ``\textit{the degree of difficulty in providing a verbal description of an image}''~\cite{heylighen1999growth, olivia2004identifying}, related to \textit{recall of the (natural) scene} or to make sense of the structure of it~\cite{olivia2004identifying}, or the naming~\cite{farhadi2009describing}. A key factor contributing to VC is the representation of structure and attributes, such as clutter. Oliva \etal~\cite{olivia2004identifying} studied VC as both a high-level ``\textit{scene gist}'' and a set of low-level features, finding that humans adapt their perception based on the task. VC can be ``\textit{principally represented by perceptions of quantifying objects, clutter, openness, symmetry, organization and variety of colors}.'' These structural elements influence how people later remember or forget about an environment~\cite{kyle2025scene}.

In visualization, a core challenge our community faces is understanding how human observers interpret visualizations in terms of readability, legibility, and understanding~\cite{cabouat2024previs, chen2012effects}. \textit{Perceived} experience often refers to the subjective impression or feeling evoked by a visualization, regardless of tasks or familiarity~\cite{he2022beauvis}. This understanding is influenced by intrinsic factors such as the nature of the data~\cite{duffy2013measuring}, the task at hand, and high-order perceptual and cognitive processes~\cite{yoghourdjian2018exploring}, all of which may be influenced by the same set of design variables that govern VCs~\cite{donderi2006visual}. As a result, VC has been described in a variety of ways across the literature: (1) as ``\textit{a form of visual uncertainty}''~\cite{duffy2013measuring, chen2010information}, where data variations enable its evaluation through information-theoretical measures~\cite{chen2010information, janicke2007multifield}; (2) as ``\textit{a measure of the degree of difficulty for humans to interpret a visual presentation correctly}''; (3) as feature congestion (FC), framed as the antonym of \textit{`simplicity'}~\cite{mack2004perceptual}, or ``\textit{allowing little room to put in more information}''~\cite{rosenholtz2005feature}; (4) as ``\textit{the amount of detail or intricacy}''~\cite{snodgrass1980standardized, purchase2012exploration}; and (5) as a spatial metaphor, where Borgo et al.~\cite{borgo2012empirical} described complexity as ``\textit{a direct function of the degree of perceivable structure, variety of parts and separation of parts vs. their conceptualization as a whole}''.

\begin{table}[!t]
    \centering
    \caption{\textbf{Five factors that affect perceived visual complexity (VC).}
    VC was studied along
    five dimensions below, defined in different scopes.
    This table is a compact summary of the full-paper list in \sm~\autoref{tab:paperlist}.}
    \scriptsize
    \begin{tabular}{p{1.05cm}p{4.5cm}>{\raggedright\arraybackslash}p{2cm}}
        \toprule
        \textbf{Factors}  & \textbf{Scope} & \textbf{Source}\\
        \midrule
        Info.-theoretic     &
        Information amount
        as the size of digital file, inverse of~\cite{shannon2001mathematical}, and as saliency.
        &
        \cite{miniukovich2018visual, olivia2004identifying, saraee2020visual, chen2010information, janicke2010salience, janicke2007multifield}
        \\   \midrule
        Clutter (Info.-theoretic)         & Summary statistics over the entire
        image, \eg,
        there exists
        ``\textit{little room for a new feature to draw attention}''~\cite{rosenholtz2007measuring}.
        &
        \cite{saraee2020visual, rosenholtz2007measuring,olivia2004identifying}
        \\
        \midrule
        Color
        & The variety of colors (hue, saturation, and brightness),
        the level of continuity, intensity, and range of colors. &\cite{ purchase2012exploration, rosenholtz2007measuring, fernandez2019visual}
        \\
        \midrule
        Shape and Structure         &
        The variety and intricacy of edges and turns within an image, encompassing the contours and boundaries and
        Structural clarity (\eg, regularity, heterogeneity, region segmentation by color or textures).
        &\cite{attneave1957physical, purchase2012exploration, le2012representing, sun2021curious, corchs2016predicting, oliva2001modeling, ramanarayanan2008dimensionality}
        \\     \midrule
        Object
        &
        The number (set-size) of unique objects
        (\eg, graphical elements, color and other ensemble attributes~\cite{rahman2024qualitative}, and text).
        &  \cite{olivia2004identifying, chai2010scene, snodgrass1980standardized, ramanarayanan2008dimensionality}
        \\
        \bottomrule
    \end{tabular}
    \label{tab:svc}
\end{table}

\begin{table*}[!t]
    \centering
    \caption{Twelve objective image metrics associated with factors that may influence perceived VC. The first 10 metrics are from the literature where their influence on VC was previously reported. We define two new object-level metrics to quantify the number of elements in a visualization image. }
    \small
    \begin{flushleft} \begin{tabular}
            {p{0.04\textwidth}
            p{0.002\textwidth}
            >{\raggedright\arraybackslash}p{1.5cm}
            p{5.3cm}
            p{7.2cm}
            p{0.3cm}}
            \toprule
            &  & \textbf{Metrics} & \textbf{Description} & \textbf{Equation} &   \\
            \midrule
            Info.-theoretic
            (O.Info)
            &   (1) & Shannon Image entropy (O.IE) &
            The amount of information or randomness in the image,
            calculated by applying the entropy formula to the grayscale pixel intensity values of the image.
            &
            $ - \sum_{i=1}^N p(i)\log_2p(i)$,
            { where \( p(i) \), the probability of occurrence of the $i^{th}$ grayscale pixel intensity in the image, and
            \( N \), the total number of different pixel intensities.
            }
            &        \cite{shannon1948mathematical,tsai2008information}
            \\%\hline
            & (2) &
            Kolmogorov complexity (O.KC) & Image size after removing redundant information.
            &$S_{\text{origin}}- S_{\text{redundancy}}$, where $S_{\text{origin}}$, the original size of the image, and $S_{\text{redundancy}}$, the size of redundant compressible
            information.
            &~\cite{collet2018zstandard,perkio2009modelling}
            \\%\hline
            & (3) &
            Subband entropy (O.SE)&
            The entropy of the image content within subbands, calculated using the probability distribution of the wavelet coefficients within each subband.
            &
            $- \sum_{i=1}^{N} p(i)\log_2p(i)$, where \( p(i) \), the probability distribution of intensity values or coefficients within the $i^{th}$ subband, and the sum runs over all  $N$  subbands.
            &~\cite{rosenholtz2007measuring}
            \\
            & (4) & Information gain (O.IG) &
            The amount of additional information needed based on the mean information gain to determine the color of an adjacent cell given the color of a particular cell.
            & $-\sum_{i,j} p_{ij} \log_2(p_{i \rightarrow j})$, where $p_{ij}$, the joint probability that a given cell has the $i^{th}$ color and the adjacent cell has the $j^{th}$ color, and $p_{i \rightarrow j} = p_{ij}/p_i$, the conditional probability that the adjacent cell has the $j^{th}$ color given that a cell has the $i^{th}$ color.
            &\cite{andrienko2000complexity}
            \\%[3pt]\hline
            \midrule
            Clutter
            (O.CL) & (5) &     Feature congestion (O.FC) &

            The clutter estimate in the image computed by assessing feature viability based on the entropy of color, luminance contrast, and orientation.
            & $-\sum_{i=1}^{N} p_{i} \log_{2} p_{i}  (p_i = \frac{E_i}{{\sum\nolimits_{i=1}^{N}} E_i})$,  where \( p_i \), the proportion of energy \( E_i \) of the $i^{th}$ feature relative to the total energy of all  $N$ features in the image. & ~\cite{rosenholtz2007measuring}
            \\%\hline
            & (6) & Heterogeneity (O.H) &
            A gray-level concurrent
            matrix
            to
            represent
            the texture and contrast of an image.
            Higher heterogeneity values indicate more contrast and variation in
            gray levels, suggesting a more complex or textured image.
            &$ \sum^{L-1}{i=1}\sum^{L-1}{j=1}\frac{P(i,j)}{1+(i-j)^2}$,
            \( P(i, j) \) represents the probability that a pixel with intensity \( i \) occurs adjacent to a pixel with intensity \( j \), and \( L \) is the total number of gray levels in the image.
            & ~\cite{de2013multi}
            \\
            \midrule
            Color (O.CD)
            & (7) &     Colorfulness (O.CF) & Overall saturation and the variety and extent of colors.
            &
            $\sqrt{\sigma_{rg}^2 + \sigma_{yb}^2} + \kappa \cdot \sqrt{\mu_{rg}^2 + \mu_{yb}^2}$, where \(\sigma_{rg}\) and \(\sigma_{yb}\), the standard deviations of the chromaticity of red-green and yellow-blue, \(\mu_{rg}\) and \(\mu_{yb}\), the mean of the chromaticity of red-green and yellow-blue, and \(\kappa\), the weight coefficient.
            & ~\cite{hasler2003measuring}
            \\%\hline
            & (8) &        Color RGB entropy (O.ERGB)&
            The variability in the distribution of colors calculated
            by applying Shannon's entropy to the color histogram of an image.
            &
            $-\sum_{i=1}^{N} p_{\text{color}}(i) \log_2(p_{\text{color}})(i)$, where \( p_{\text{color}}(i) \), the probability of occurrence of the $i^{th}$ color in the histogram, and \( N \), the total number of different colors histogram.
            &    \cite{shannon1948mathematical,tsai2008information}
            \\%[3pt]\hline
            \midrule
            Shape (O.ShD) & (9) & Edge density (O.ED) &
            The proportion of pixels identified as edges within an image, quantified by the Canny edge detector.
            &
            $\frac{P_{\text{edge}}}{P_{\text{total}}}$, where ${P_{\text{edge}}}$, the total number of edge pixels,
            and ${P_{\text{total}}}$, the total number of pixels        in the image.
            & ~\cite{canny1986computational}

            \\%\hline
            & (10) &     Feature point (O.FP) &
            Points where edges intersect, identified by measuring changes in image brightness.
            &
            $\frac{P_{\text{junction}}}{{P_{\text{total}}}}$, where ${P_junction}$, the total number of pixels of junction points, and ${P_{\text{total}}}$, the total number of pixels        in the image.&
            ~\cite{harris1988combined
            }
            \\[2pt]\hline
            {Object (O.OD)} & (11) & Text-ink ratio (O.TiR) &
            The proportion of text pixels in an image.
            & $\frac{P_{\text{text}}}{P_{\text{total}}}$, where ${P_{\text{text}}}$, the total number of text pixels in the image, and ${P_{\text{total}}}$, the total number of pixels         in the image.
            \\      &   (12) & Number-of-color (O.MeC)
            & The number of meaningful colors count.
            &
            \rvision{The curated human namable colors of Heer and Stone~\cite{heer2012color} in an image.}
            \\%\hline
            \bottomrule
        \end{tabular}
    \end{flushleft}
    \label{tab:ovc}
\end{table*}

Our work builds on these definitions and refocuses them within the context of visualization images, adopting the definition from Purchase et al.~\cite{purchase2012exploration}, which describes VC as ``the amount of detail or intricacy''.

\subsection{Five Factors that Govern Visual Complexity}

We draw inspiration from the extensive research on characterizing complexity to select a set of metrics that represent these factors.

\subsubsection{Data-driven Information-Theoretic Measure}
\label{sec:lrit} The information-theoretic measure, often computed using Shannon's entropy~\cite{shannon2001mathematical}, is a purely mathematical representation of pixel data in images. In data visualization, Yang-Pel\"aez and Flowers (Y-F)~\cite{yang2000information} extended Shannon's classic entropy by partitioning `information' into three types: \textit{syntactic} (marks, lines, regions, and structures thereof), \textit{semantics} (meaning of those marks or other syntactic information), and \textit{pragmatic} (the value or usefulness of that information). The number of \textit{bits} is subsequently derived from \textit{marks} by searching neighboring pixels. Every bit/pixel carries an equal weight representing the meaning of the message. Later, Chen and J\"anicke~\cite{chen2010information} (C-J) conceptualized the Y-F method to take data variations in every stage of the visualization workflow, by \textit{weighting} the neighboring pixels to compute the information capacity from data in an image. Both Y-F and C-J's methods require knowing the information flow in the image space. To automate the process to compute \textit{data-bits}, J\"anicke \etal~\cite{janicke2007multifield,lin2022saliency} used color entropy to highlight and sort flow features. Here, they treated sensory perception independent of data saliency to analyze flow features by information-theoretic solutions. Furthermore, J\"anicke and Scheuermann~\cite{janicke2009visual} computed statistical complexity in data~\cite{janicke2007multifield} to compare, analyze, and highlight features that assist human observers~\cite{miksch2020knowledge}.

What inspired us in these works is the transition from the low-level design element to key probabilistic data-specific knowledge in images, modeled under the same information-theoretic framework of mathematical uncertainty. However, we were unable to directly extend their work, as there is currently no established method for defining bits in the context of general visualizations beyond flow fields. A broader idea drawn from this line of research is that more complex images tend to exhibit lower redundancy and thus higher entropy. For this reason, the compression rate of an image---aligned with Kolmogorov complexity~\cite{rigau2007conceptualizing}---serves as reasonable approximation of pixel-level uncertainty (Item (1) O.KC,~\autoref{tab:ovc}). Building on this idea, we introduced several additional metrics (Items (2)-(4),~\autoref{tab:ovc}) that adopt this entropy-based approach. Furthermore, we incorporated pixel-level RGB entropy into our metric set (Item (8) O.ERGB,~\autoref{tab:ovc}).

\subsubsection{Clutter-based Feature Congestion}
\label{sec:lrfc}

The design of FC originated from an interesting question: whether the robust measure of the ``number of unique objects''~\cite{chai2010scene, mack2004perceptual} can be computed mathematically~\cite{rosenholtz2005feature} to eliminate the need for labor-intensive manual coding. Here, congestion is characterized by the lack of available space for new features~\cite{rosenholtz2007measuring}, and ``\textit{the more cluttered a display or scene is, the more difficult it would be to add a new item that would reliably draw attention}''~\cite{rosenholtz2005feature}. Subsequently, this metric assembles the spatial frequency of Gabor texture patches and colors to simulate receptive fields, extending the subband entropy onto the \textit{object-level} structure of color, orientation, and saliency~\cite{rosenholtz2005feature}. This is grounded in the understanding that an image is considered `simple'~\cite{olivia2004identifying} when it is symmetric and structured. Thus, complexity is inversely correlated with perceptual grouping and regularity~\cite{heylighen1999growth}. Empirical evaluation results also supported the FC metric (Item (5) O.FC,~\autoref{tab:ovc}), subband entropy (Item (3) O.SE,~\autoref{tab:ovc}) and edge density (Item (9) O.ED,~\autoref{tab:ovc}) on measuring VC ~\cite{rosenholtz2005feature, rosenholtz2024visual}. Borgo \etal~\cite{borgo2012empirical} further emphasized the idea of statistical attributes to characterize complexity. This line of work informs us that information amount influences VC. Both Rosenholtz in vision science and J\"anicke and Chen in visualization use color as the primary channel for information-theoretic metrics~\cite{rosenholtz2024visual,janicke2007multifield}.
\rvision{Given that color and textures are uniquely represented in visualization images, we modeled between continuity and complexity conditioned for continuous and discrete color and texture stimuli, or RR-ColorTexture of Continuous Heatmap and Discrete Heatmaps in images.}

\subsubsection {Color-based Measures}
\label{sec:lrcolor}

Beyond its role as a secondary attribute attached to other primary stimuli~\cite{borkin2013makes}, color is perhaps one of the most extensively studied visual variables for direct data mapping, as evidenced by numerous reviews~\cite{ware1988color, zhou2015survey,silva2011using, ware2023rainbow} and real-world applications ~\cite{barua2024urban, chen2019measuring}. The importance of color is further underscored by the fact that the transfer function in visualization is fundamentally a coloring process~\cite{ pfister2001transfer}.

Color also plays a significant role in shaping perceptual experiences. For example, Li and Chen~\cite{li2018toward} suggest that using too many colors can reduce memorability due to increased clutter.
Moreover, color is a leading factor in how visual groupings are perceived, supporting both segmentation and the identification of structure within complex visual scenes. In vision science, both Greene \etal~\cite{greene2019information} and Rosenholtz~\cite{rosenholtz2024visual} highlight the role of color in defining object boundaries. Many studies have linked color-related features to VC. For example, Rosenholtz \etal observed that, color variability is independent of both edge density and subband entropy~\cite{rosenholtz2005feature}. Both Oliva \etal~\cite{olivia2004identifying} and Rosenholtz \etal~\cite{rosenholtz2007measuring} reported that perceived VC increases more with color variability (the number of colors and how different they are) than with the presence of high frequency details in color channels. Corchs~\cite{corchs2016predicting, hu2014interactive} introduced metrics such as color harmony and color coherence to show that both the interaction and uniformity of colors contribute to VC. Here, we balance our metric choices to capture both statistical color properties and object-level color counts. Specifically, we used
\textit{colorfulness} (Item (7) O.CF,~\autoref{tab:ovc}), defined as ``\textit{the variety and diversity of colors present in the region}''~\cite{hasler2003measuring}, and gray-level color heterogeneity (Item (6) O.H,~\autoref{tab:ovc}) to indicate overall saturation~\cite{hasler2003measuring},
\rvision{in addition to the aforementioned O.ERGB,~\autoref{tab:ovc} (see~\autoref{sec:lrit}).}

\subsubsection{Object-based Measures}
\label{sec:lrtext} Several studies have focused on object-level measures, emphasizing that VC involves higher-level cognitive processing, which requires the analysis and integration of multiple elements along with their semantic content. Semantic content in visualization can have \textit{two} types:
\textit{graphical elements} (\eg, legend, tick marks) and \textit{annotations}~\cite{rahman2024qualitative} conveying key insights and takeaways~\cite{adar2020communicative}. The effect of texts on VC can be mixed. For one, they can effectively guide viewing behavior~\cite{bylinskii2018different, borkin2013makes} to reduce VC. For the other and from a pixel-based perspective, text is treated as a visual structure---often dark-colored, densely arranged, and high in contrast, that can increase complexity (Item (11) O.TiR,~\autoref{tab:ovc}).

Color plays \textit{two} main roles in object perception: it can function as a \textit{low-level} graphical element to encode data directly (\eg, field visualization). It also supports \textit{high-order} ensemble perception by contributing to perceptual grouping, thereby facilitating visual search. For the later, Greene \etal~\cite{greene2019information}, Oliva \etal~\cite{olivia2004identifying}, and Chai \etal~\cite{chai2010scene} suggest that color helps define object boundaries, and that the number of distinct object types in a scene influences its perceived complexity. Similarly, Snodgrass \etal~\cite{snodgrass1980standardized} and Donderi \etal~\cite{donderi2006information} consider both the number and variety of objects in their assessments of visual complexity. Finally, it is worth noting that \changes{we are not the first to introduce color count (MeC) as a way to quantify high-order perception. Borkin et al. and Li and Chen used it to quantify memorability~\cite{borkin2013makes,li2018toward}. However, a key distinction is that their MeC values were based on participant-reported perceptions of color, while our metric is derived from objective image analysis based on actual color usage.} Collectively, these studies demonstrate how the presence, quantity, diversity, and relationships of visual objects contribute to the perception of VC.

Since images are not typically treated as object-based scenes, where elements are explicitly counted or statistical attributes are quantified, we explore the use of color to differentiate visual elements (Item (12) O.MeC,~\autoref{tab:ovc}). Results from our workshop further suggest that color is a leading factor influencing image clarity. Together, these insights allow us to investigate how high-level visualization elements contribute to perceived VC.

\section{A Crowd-Sourcing Study}
\label{sec:crowdsourcingsetting}

\noindent Having established an understanding of the factors influencing VC through an extensive literature review, we proceeded to directly collect crowdsourced \tochange{VC scores. We began with a workshop and pilot studies to inform the proper procedure for our data collection. }

\subsection{Image Data Collection and Characteristics} \label{sec:imgDatabase}

We prioritize diverse stimuli and therefore used both VIS30K~\cite{chen2021vis30k} and MASSVIS~\cite{borkin2013makes} as primary data sources. They both come from real-world applications thus let us examine practical VC. Most importantly, they function differently: VIS30K contains scholarly images extracted from formal IEEE publications, and often presents scientific discoveries or new designs, where annotation is rare and the corresponding description often appears in the captions or main text. In contrast, MASSVIS images are used mainly for communication, therefore key trends are sometimes annotated, which further serve our goal of understanding non-graphical elements, \eg, text in images. Also, VIS30K contains glyphs and vector field images not found in MASSVIS.

\subsection{Workshop and Pilot Study}
\label{sec:pilotSummary}

The goal of the workshop was mainly to choose a reliable method to collect a large number of VC scores. Extensive details of the activities and outcomes are provided in \sm~\autoref{sec:pilotWork}. Here we give a high-level overview and decisions influenced by these activities.

We first adapted a recent method used in BeauVis for a direct score assignment~\cite{he2022beauvis}, but found it difficult to align scores above several 10s of images (He et al.~\cite{he2022beauvis} used 15 images). Next, we conducted an in-person exploratory workshop with 49 participants, akin to the hierarchical division method of Oliva et al.~\cite{olivia2004identifying}. Participants ranked images and captured factors influencing VC. We found that it was well-suited for collecting comments and factors, however, it also did not scale (Oliva et al.~\cite{olivia2004identifying} used 100 images). We also found that not a single participant used a single criterion in their VC assignment.

The third method we tried was an active sampling solution that could provide us absolute VC scores via pairwise 2AFC ranking successfully used in the computer vision community to collect a large set of image VC scores~(\eg,~\cite{saraee2020visual, nagle2020predicting}). The idea was similar to how game competitions assigned competing players to achieve a global score for each player. Our piloting experiments with 90 and 200 images supported that the single-stage Bradley-Terry ranking~\cite{bradley1952rank} and Microsoft's TrueSkill\texttrademark~\cite{herbrich2006trueskill} were both reliable. However, scaling up this approach to 1,000s in the global space using only one-time sampling would require many more trials, which was not feasible. We finally piloted Mikhailiuk et al.~\cite{mikhailiuk2021active}'s multi-stage active sampling algorithm (\sm~\autoref{fig:validationExps}), which could achieve the same level of reliability as the single-stage active sampling with $10\%$ of total trials. In each phase, the algorithm optimized the information obtained from all pairwise comparisons to compute reliable VC scores, while limiting the total number of comparisons. Our pilot tests achieved the same reliability as the authors reported while reducing operating costs by 9 fold.

In the pilot study, we also discussed what \textit{not} to collect. For example, we excluded other high-level subjective metrics, such as aesthetics, understandability, and readability, even though they might correlate with VC~\cite{rigau2008informational, reinecke2013predicting, wallace2022towards} because these factors are also inherently subjective, potentially relying on the same set of objective metrics.
Instead, we focused on objective image metrics, aiming to develop an understanding that remains independent of individuals' experience and expertise, for quantifying the crowdsourced perceived VC scores. Images were static and not dependent on viewers' ability to navigate. 

\subsection{Participants and Multi-stage Procedure}
\label{sec:ParticipantProcedure} At this stage, we had clear ideas on the image data and the method to collect the perceived VC scores. Now we report our crowd-sourcing experiment via the online experiment platform Prolific~\cite{palan2018prolific} to capture the absolute perceived VC scores.

We instructed participants on the experiment procedure and provided them the high-level definition of VC, ``\textit{the amount of detail or intricacy in images}''~(\sm~\autoref{fig:interface}) as used in Purchase et al.~\cite{purchase2012exploration, snodgrass1980standardized}, which was always displayed on the screen. They were instructed not to infer knowledge and not to use information other than those presented in the images, following the instruction of Oliva et al~\cite{isola2011makes}. We also had two attention test trials to identify and remove inattentive participants. In each stage, about 20-25 volunteers compared 79 pairs of images to determine the VC for that stage, viewing images recommended by the active sampling algorithm. At each stage, an absolute complexity score for each image was updated from a probabilistic process to generate a global complexity score described in more detail next.

Participants' browser type and operating system were automatically detected, and the screen resolution must be at least $1028\times764$ on a desktop operating system. Before this data collection, we first collected participants' demographic information and ensured accountability by recruiting only those with a success rate above $95\%$ on Prolific. Additionally in the post-questionnaire, we gathered participants' written reasoning from two randomly selected trials.

\begin{figure}[!t]
    \centering
    \includegraphics[width=\columnwidth]{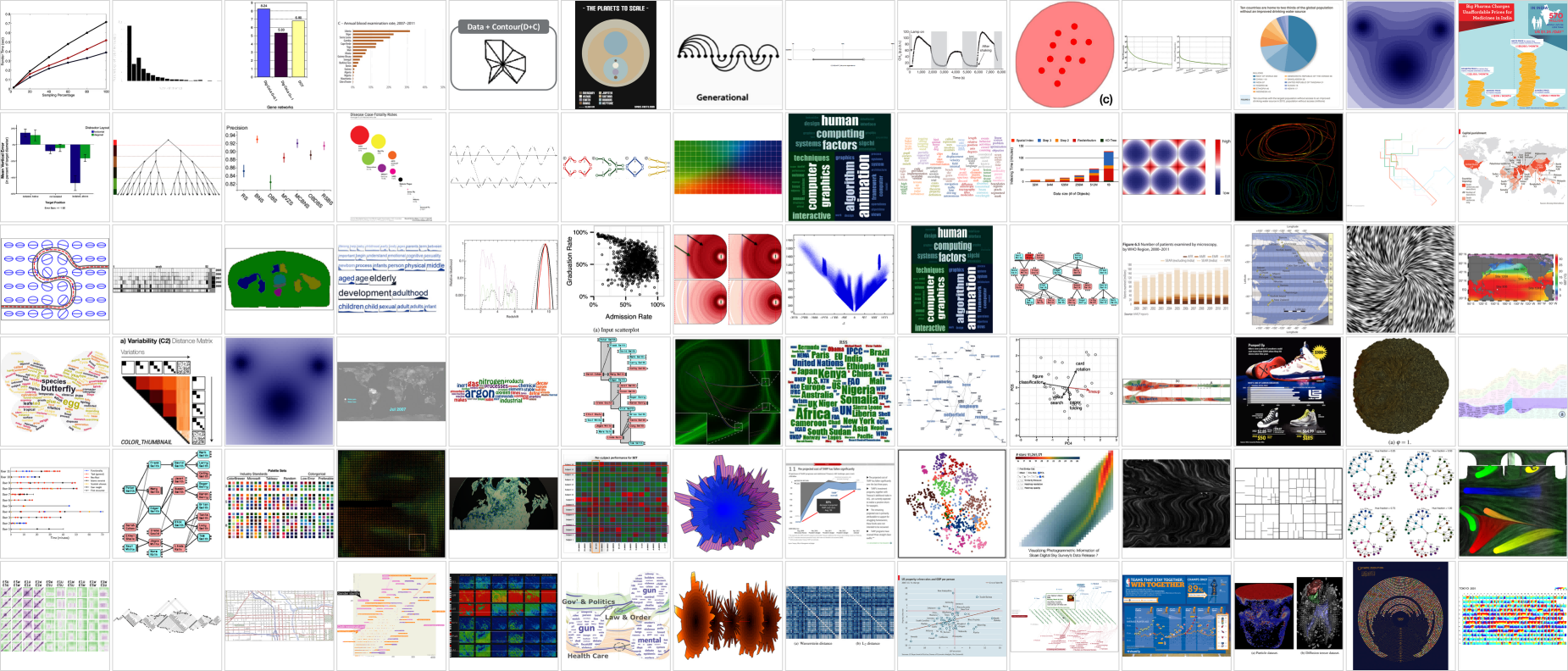}
    \caption{\textbf{Resulting visual complexity scored image examples} (least to most from top left to bottom right).
    }
    \vspace{-5pt}
    \label{fig:global200}
\end{figure}

In total, there were 349 participants with ages distributed as follows: 18-25 (111), 26-35 (143), 36-45 (51), 46-55 (27), and over 55 (17). Their education levels included graduate school (134), college (130), professional school (55), and some high school (30). The gender distribution was nearly balanced, with 168 women, 176 men, and 5 who did not disclose their gender. The participants were compensated \EUR{9.5} / hour for their participation.

\subsection{Crowdsourcing Results}
\label{sec:Crowd-sourcing-result}

The experiment was conducted over 14 stages for approximately 34 hours, resulting in a total of \totaltrials trials, with an average of \totaltrialsperrun trials per stage. Each trial lasted an average of 4.34 seconds ($95\%$ CI: [4.17, 4.50]). Each visualization image was viewed approximately 234 times ($95\%$ CI: [202, 266]).

\textbf{VC Score Calculation.} We applied the Mikhailiuk et al.'s approach~\cite{mikhailiuk2021active} to calculate the VC scores for all 1,800 images. The approach first resulted in a comparison matrix containing the outcomes of all pairwise comparisons, where each entry recorded how often an image was selected as more visually complex when compared against others, and then applied the TrueSkill{\texttrademark} algorithm~\cite{herbrich2006trueskill}, to infer relative complexity from the comparisons. TrueSkill{\texttrademark} is a Bayesian ranking algorithm that models each image’s VC as a Gaussian distribution \( \mathcal{N}(\mu, \sigma^2) \) with a mean (\( \mu \)) representing the estimated VC score and a standard deviation (\( \sigma \)) representing the uncertainty of that estimate in our context. During each comparison between two images \( i \) and \( j \), TrueSkill{\texttrademark} estimates the probability that \( i \) is more visually complex than \( j \). Specifically, if image \( i \) is judged more complex than \( j \), the mean \( \mu_i \) is increased and \( \mu_j \) is decreased; both variances \( \sigma_i^2 \) and \( \sigma_j^2 \) are reduced as the model gains confidence through increased comparisons. After many pairwise comparisons, the distributions converge.

For our analysis, the final VC score of each image is taken as the posterior mean which reflects the inferred likelihood that it would be judged more complex than a randomly chosen image from the dataset (Additional validation in \sm~\autoref{fig:validationExps}).
\autoref{fig:global200} shows a subset of 84 images, sorted in ascending order of their VC scores from left to right, then top to bottom (\sm~\autoref{fig:VCExamplesFirstSet} provides more images and scores).

\section{Metrics-Based VC Attribution Analysis}
\label{sec:12metricsStudy}

\begin{figure*}[t!]
    \centering
    \includegraphics[width=\textwidth]{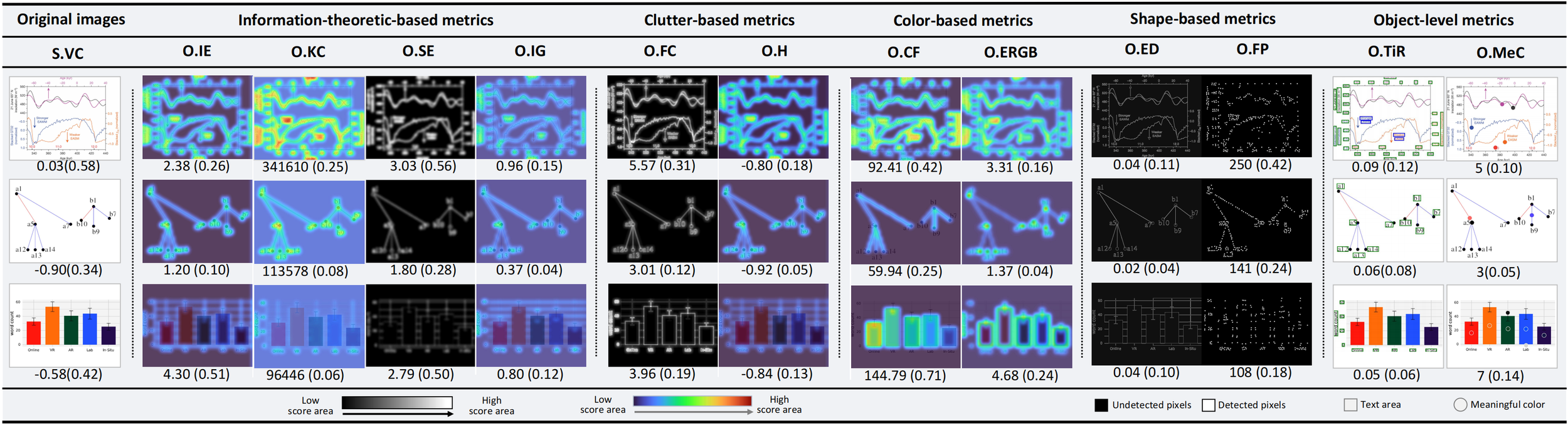}
    \caption{\textbf{Objective metric scores in response to the visualization image input.} Each row shows the original image followed by visual representations of the 12 objective metrics, with corresponding metric scores shown as raw (normalized) values below. The heatmaps highlight image regions where each metric responds most strongly—brighter and redder areas indicate higher values or more visually complex regions. }
    \vspace{-5pt}
    \label{fig:metrics}
\end{figure*}

This section presents our two new objective metrics and observations after applying these 12 metrics to our image dataset.
\changes{\autoref{fig:metrics} shows some example outputs after applying the metrics to visualization images. }

\subsection{Our Two Object-based Image Quality Metrics}
\label{sec:objMetrics}

We define two metrics related to the semantic information of a scene, building on the concept of graphical elements, where objects are recognized quickly, sometimes in a single glance ~\cite{oliva2006building}. While our community has traditionally relied on edges to define boundaries, we focus here on high-level perception of using colors. Here, we drew on insights from vision science, including the role of color in delineating element boundaries and the preattentive nature of coloring in visualizations~\cite{chen2019measuring}. Also, we saw that many images contained categorical colors in the legend. These perspectives informed our decision to use color as a means of defining boundaries between data elements, alongside text, which can be recognized and quantified with relative ease.

\begin{figure}[!t]
    \centering
    \includegraphics[width=\columnwidth]{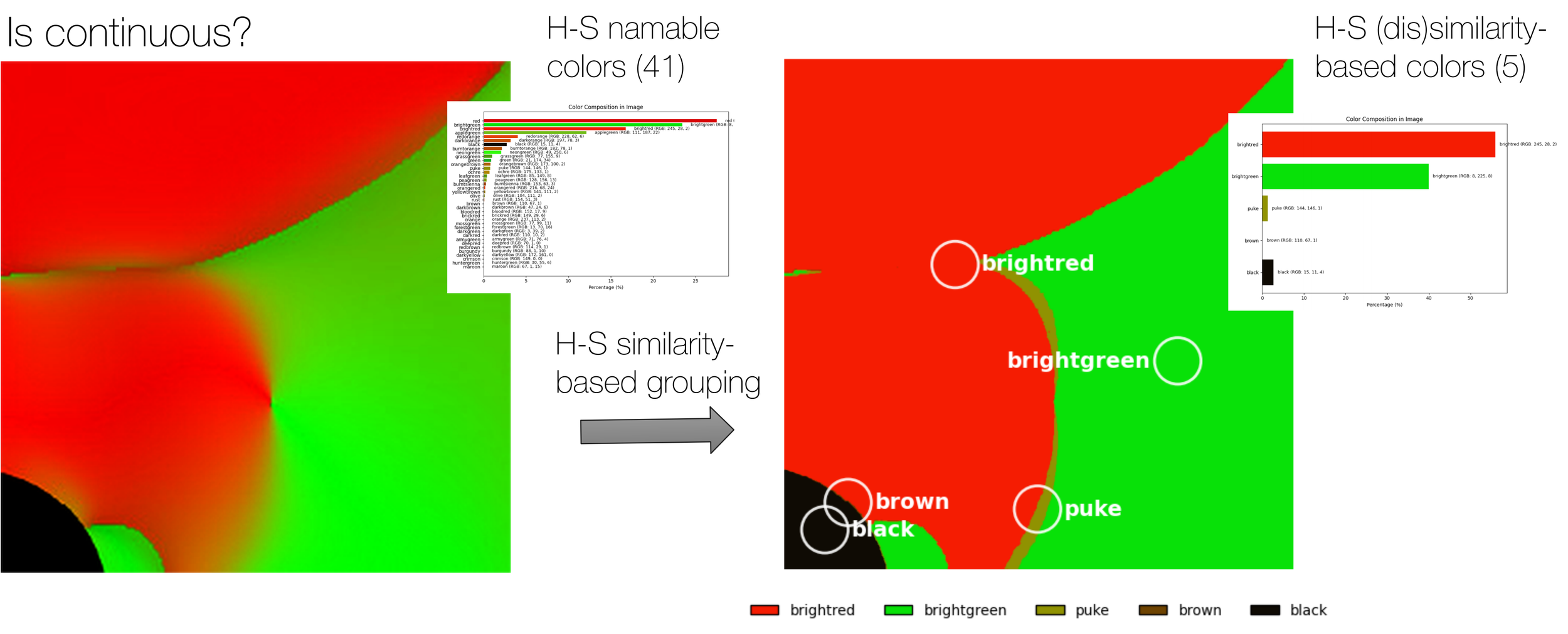}
    \caption{\textbf{An O.MeC calculation pipeline for an image with a continuous color-map.} \tochange{Left: the original image with 41 measured H-S~\cite{heer2012color} namable colors; Right: 5 colors grouped by H-S similarity to produce O.MeC=5.}}
    \vspace{-5pt}
    \label{fig.TiRMeC}
\end{figure}

\textbf{Text-ink ratio (O.TiR)} represents the proportion of text-area pixels in an image. Here, ``text'' refers to contextual information, including annotations, axis labels, or titles. Any text used to explain, clarify, or define image content is considered ``annotated''~\cite{rahman2024qualitative}. We excluded text-based visualizations where text was used to show data. There are many methods to compute O.TiR.
\rvision{In our case, we used an OCR (Optical Character Recognition) \zfq{package PP-OCR }~\cite{du2020pp} to compute text regions and then use the VIS30K bounding box labeling tool~\cite{chen2021vis30k} (\sm~\autoref{fig:OTiRInterface}) to curate annotations, legends, etc. }

\textbf{Color count (O.MeC)} defines the number of distinct colors/objects in a visualization image. Several considerations guided how we computed the MeC, due to the difficulty to name the groups as Rosenholtz et al. pointed out~\cite{rosenholtz2007measuring}. For example, would a data sample in a scatterplot be considered an item or a cluster of these an item? Here, we assume that designers would have distinguished `items' by color---if the goal is to see the cluster, they would be shown as color-based clusters. As a result, for discrete data, we simply count the number of items by looking at the legend without concerning items within that cluster, since legend captures human attention first~\cite{bylinskii2018different} and color supports fast pre-attentive processing~\cite{whitney2018ensemble, zhao2022evaluating}. For these images using discrete colors, we count the number of unique colors present in the legend, along with any additional text color.

Counting colors in continuous colormaps is also not straightforward. Our implementation prioritizes semantically meaningful colors by Heer and Stone (H-S)~\cite{heer2012color}, as these are more easily and rapidly perceived by humans. Using the H-S color dictionary, we map each unique pixel value to the closest color name in the H-S vocabulary. For example, this step shows 41 namable colors for the image in \autoref{fig.TiRMeC}. We next merge the similar colors using the H-S's similarity dictionary, where we use the nearest-centroid color of the colors in the same similarity group. For the same figure, this step trimmed the 41 to 5 colors, shown in the white circles. The name of these colors are shown next to their locations. For this figure, we used ``5'' as the final O.MeC.

One may argue that luminance areas were not consistently counted in our analysis. In practice, we addressed this on a case-by-case basis. For texture-based pattern images such as LIC, we included luminance in the color count. Since the H-S naming system is not uniformly distributed in the L*A*B space, some color names are perceptually closer than others. In cases like this, we applied a color distance threshold: colors with a $\Delta$ E $\leq 14 $ were merged (\sm\autoref{fig:threshold}). This value is slightly higher than the typical human perceptual threshold ($\Delta E= 10$~\cite{wurm1993color}), allowing us to account for pixel-level noise and small color variations. This threshold also helped reduce salt-and-pepper noise in the image. Two ($2^{nd}$ and last) co-authors curated the final answers.

\subsection{\changes{Observing 12 Objective Image Quality Metrics}}
\label{sec:12metrics}

We applied the 12 metrics to our dataset and had a few observations.

\textbf{Information-theoretic-based metrics.} We used six metrics. Among these four were the original Shannon's entropy or pixel-based framed as: (1)~the original \textit{Shannon Image Entropy (O.IE)}~\cite{shannon2001mathematical}; (2) \textit{Kolmogorov Complexity (O.KC)}, which estimates an upper bound of an image's information capacity using the \textit{zstd} compression method~\cite{collet2018zstandard}, where higher complexity indicates greater internal redundancy, resulting in larger compressed sizes; (3) \textit{Subband entropy (O.SE)} ~\cite{rosenholtz2005feature}, measuring the randomness of information within different frequency components of an image after wavelet decomposition; and (4) \textit{information gain (O.IG)}~\cite{andrienko2000complexity}, which examines neighboring pixel probabilities and is sensitive to local textures, edges, and details within the image. We observed that these four information-theoretic metrics were highly correlated emphasizing similar regions of an image~(\autoref{fig:metrics} and
\sm~\autoref{fig:metricsCorrelationfig}).

\textbf{Clutter-based Metrics} include a statistic information theoretic metric: (5) feature congestion (\textit{O.FC}) ~\cite{rosenholtz2005feature}. O.FC quantifies how busy an image appears by analyzing structural variations of color and luminance contrast using information-theoretic measure. In contrast, (6) heterogeneity (\textit{O.H})~\cite{de2013multi} considers arrangement and overall pattern distribution, and higher \textit{O.H} indicates greater contrast and variation in gray levels, which suggest a more complex or textured image.
\autoref{fig:metrics} shows that areas with a high density pattern and structure variations were emphasized by these two metrics.

\textbf{Color-based metrics.}
\textit{O.ERGB} is the last information-theoretic metric, and higher entropy indicates greater color variation. We can see that smaller regions with many color changes tend to have higher ERGB, \tochange{capturing the edges between the colored bars.}
\changes{\textit{O.ERGB} also highlighted nodes in node-link diagrams. On the other hand, colorfulness (\textit{O.CF}) tends to make the larger areas stand out more prominently.}

\textbf{Shape-based metrics.} We used (9) edge density (\textit{O.ED}), as studied in ~\cite{rosenholtz2007measuring,mack2004perceptual,rosenholtz2005feature}, which represents the ratio of edge pixels---indicating discontinuities in image intensity---to total number of pixels in an image, and (10) feature points (\textit{O.FP})~\cite{harris1988combined}, which are corner points exhibiting significant variation from surrounding pixels. Both metrics emphasize areas with densely packed lines and significant variations in color values.

\textbf{Multi-colinearity analysis.} We also performed a co-linearity analysis (see \sm~\autoref{fig:metricsCorrelationfig} for detailed analysis). We observed correlations within categories, especially among pixel-based information-theoretic metrics, and between shape-based metrics. In contrast, the object-based metrics, \textit{O.TiR} and \textit{O.MeC} showed little to no correlation with other metrics.

\subsection{Metric-informed Factors that influence Perceived VC}

We now examine how the 12 objective image quality metrics relate to perceived VC, as collected through our crowdsourcing experiment described in \autoref{sec:crowdsourcingsetting}.
\rvision{We aimed to obtain the most informative subset of these 12 input metrics to model VC and make it interpretable.}

\subsubsection{Analysis Method}

We employed Partial Least Squares Regression (PLS) to model perceived VC using the 12 objective image quality metrics as predictors. PLS was chosen for its ability to handle multicollinearity, \ie, the high correlations among predictors, shown in our data in \sm~\autoref{fig:metricsCorrelationfig}. PLS leverages the correlations to extract latent components that summarize the variance in both predictors and the response
\rvision{to identify a few variables with strong influence on the outcome (perceived VC)~\cite{mehmood2012review}.}

\rvision{Various methods can measure multiple variables for model explainability in data analysis. We did not use linear~\cite{purchase2012exploration} or least absolute shrinkage and selection operator regression (LASSO)~\cite{tibshirani1996regression}, as they either require manual intervention to address multicollinearity or shrink less important coefficients to zero, which may introduce subjectivity. We also did not use factor analysis methods such as item response theory (IRT)~\cite{baker2001basics}, because it focuses on uncovering latent constructs within the predictor ($x_i$) space. PLS constructs latent components optimized to \textit{explain} the outcome variable (perceived VC). While factor analysis or IRT are useful for understanding underlying factor structures or modeling latent traits, they do not provide direct predictive modeling frameworks aligned with regression objectives. PLS bridges this gap by estimating both the underlying structure and a modest number of explanatory variables in relation to the response.}

\rvision{We computed PLS regression coefficients to estimate the relative contributions of each metric to perceived VC. We also used bootstrap resampling to estimate the variability of each coefficient and compute $p$-values, thereby identifying which metrics are \textit{more significant} and \textit{stable} across resamples (details in~\sm~\autoref{sec:PLS}).} We used the resulting PLS regression coefficients to estimate the relative contribution of each metric to perceived VC. \rvision{Interpreting these coefficients, however, requires care.}
\rvision{Unlike in standard regression, the coefficients in PLS reflect the contribution of a variable conditional on its shared variance with other predictors, which means the following: (1) variables with larger absolute coefficient values are interpreted as having greater influence, while the sign of the coefficient indicates whether the effect is positive or negative; (2) small coefficients may not indicate that a metric is unimportant, but rather that its influence overlaps with other variables or is too unstable to predict VC. }

\subsubsection {Results: Overall Factorizing Perceived VC}
\label{sec:FactorizingPerceivedVC}

\begin{figure}[!t]
    \centering
    \fbox{\includegraphics[width=0.85\columnwidth]{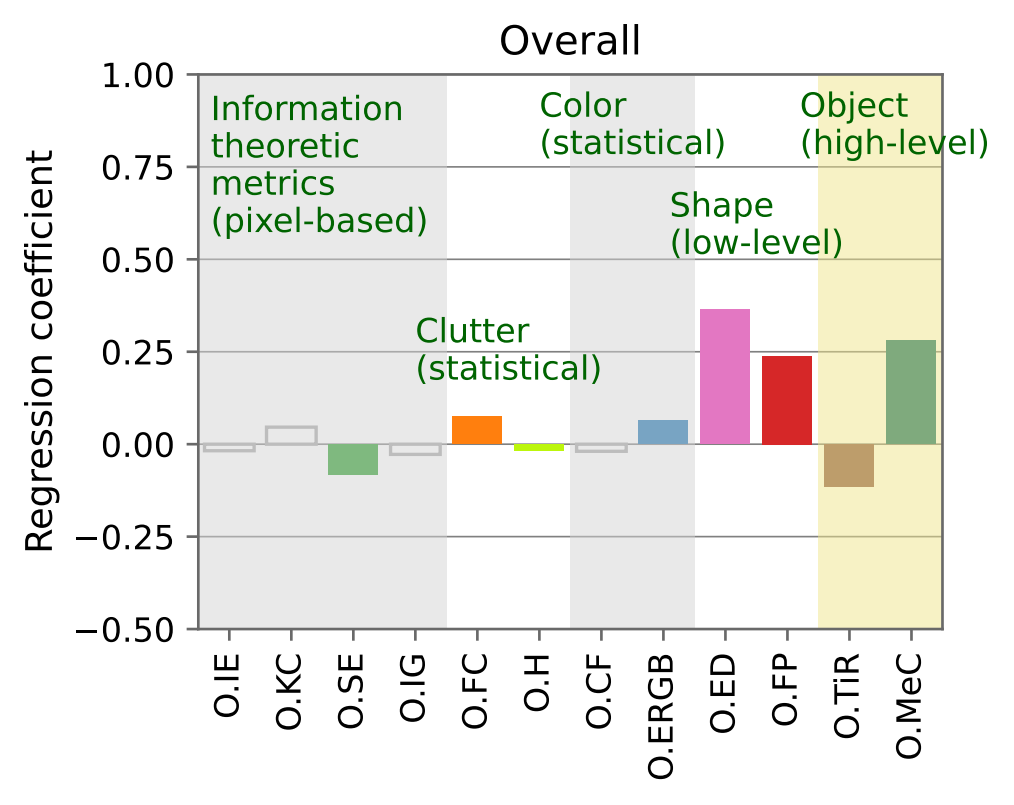}}
    \caption{\textbf{Factoring visual complexity.}
    The \rvision{relative} contribution of each image metric is represented by the magnitude of its corresponding regression coefficients, \rvision{modeled using PLS}, with the significant ones highlighted in solid color (metrics significant at the 0.05 level). }
    \label{fig:factorization}
\end{figure}

\autoref{fig:factorization} presents the resulting PLS coefficients, with the metrics grouped from left to right in shaded zones by the five factor categories defined in~\autoref{tab:ovc}. Solid-colored bars represent significant metrics ($p<0.05$), while those outlined in gray wireframes are not significant. The number of significant metrics indicates that all five categories originally derived from natural scene perception are applicable to visualization contexts. While only one metric is significant in each of the information-theoretic, clutter, and color categories, all metrics in the shape-based and object-based categories are significant. The $R^2$ value reflects the proportion of variance in perceived VC explained by the model---the higher the value, the better the model fit. In this case, $R^2$=0.41 indicates an overall modest fit across all images.

\textbf{(1) The influence of generic pixel-based information-theoretic metrics and color metrics on perceived VC is minimal.} The pixel-based information-theoretic metrics, O.SE ($coe=-0.08, p=0.02$), is the only significant coefficient at the $5\%$ level. O.IE ($coe=-0.02, p=0.87$), O.KC ($coe=0.05, p=0.21$), and O.IG ($coe=-0.03, p=0.45$) are much less significant in the prediction of VC.

\textbf{(2)} \textbf{Information-theoretic metrics based on specific stimulus and attributes are significant factors.} The pixel-based information-theoretic metrics (O.IE, O.KC, and O.IG) were not the significant main effect, but O.SE using the frequency was. In addition, O.FC in the clutter category and O.ERGB in the color category are also information-theoretic metrics. O.FC adapted subband entropy to statistical distributions of color and texture and O.ERGB uses the original Shannon entropy on color alone. They were both significant main effects with $coe=0.08$ ($p=0.01$) and $coe=0.06$ ($p<0.001$), respectively.

\textbf{(3)} {\textbf{Shape and object metrics contribute most to complexity.}} Metrics capturing shape and object quantity were the strongest predictors of perceived VC. O.ED and O.FP, categorized as shape-based metrics (\autoref{tab:ovc}), reflect low-level features following Oliva et al.~\cite{olivia2004identifying}, while O.MeC represents a high-level feature that estimates the number of objects based on color count~\cite{greene2019information}.
\tochange{Overall, metrics that quantify the total number of visual elements, whether shape- or object-based, proved to be significant. Specifically, O.ED ($coe=0.36, p<0.001$), O.FP ($coe=0.24, p<0.001$), and O.MeC ($coe=0.28, p<0.001$) were the top three positive contributors to perceived VC. }

\begin{figure}[!t]
    \centering
    \includegraphics[height=0.6\columnwidth]{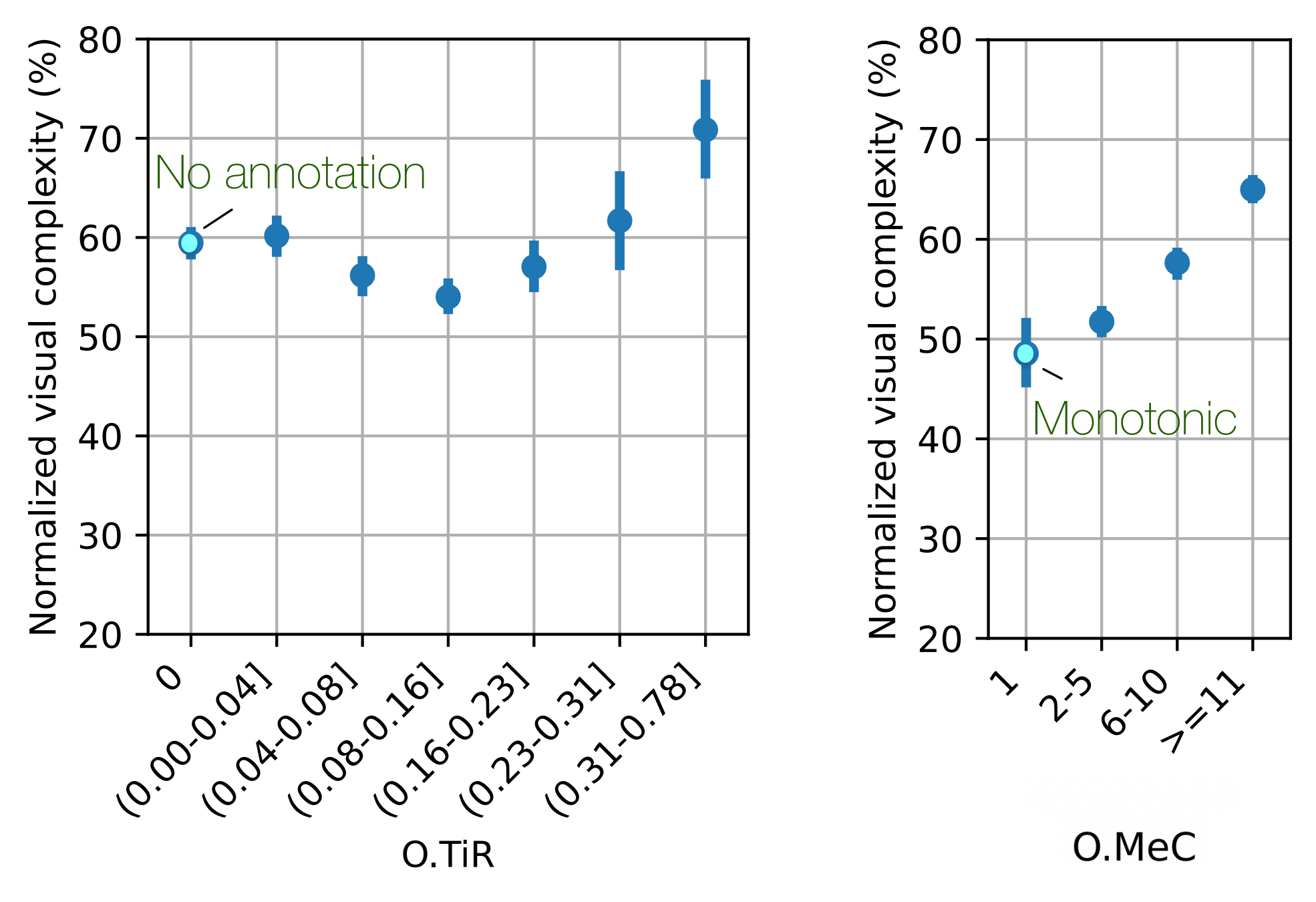}
    \caption{Mean VC scores for visualization images based on the Text-ink-Ratio (O.TiR) and the number of meaningful colors (O.MeC). Error bars represent 95\% confidence intervals.}
    \vspace{-5pt}
    \label{fig:objectmodel}
\end{figure}

\textbf{(4) Text-ink-ratio reduces perceived visual complexity.} O.TiR, which measured the proportion of annotated text within an image, was found to have a negative coefficient (-0.12, p$=$0.005), suggesting that the presence of annotation may in general reduce rather than increase perceived VC. As discussed earlier, annotated text can enhance clarity and aid in interpretation, thereby lowering perceived complexity. To explore this further, we analyzed the trend between O.TiR and perceived VC and observed a non-linear, bell-shaped relationship~(\autoref{fig:objectmodel}). Specifically, VC initially decreases as O.TiR increases, reaching a minimum $0.08-0.16$, and rises again beyond a threshold of approximately $0.23-0.31$. This difference in complexity is statistically significant ($F_{6,1770}=9.8, p<0.001$), indicating a potential optimal range of O.TiR, where a moderate amount of text can reduce perceived complexity by improving clarity, while too little or too much text may have the opposite effect.

{\tochange{\textbf{We also observed that increasing O.MeC was associated with higher perceived VC}~(\autoref{fig:objectmodel}).}} Among the dataset, we found 96 black-and-white or monochrome images with no color variation ($O.MeC=1$). While natural scenes are almost always colorful, visualization images can achieve similar perceptual differentiation through other cues such as texture and shape variation. In general, grayscale or low-color images were perceived as less complex than colored ones, a difference that was statistically significant ($F_{3,1773}=125.0, p<0.001$). For the images with higher O.MeC, we found that color often served as a boundary-defining cue, helping to visually separate elements. This boundary-forming function of color contributed to increased perceived complexity, especially when the colors clearly delineated objects or categories. Notably, this effect was consistent regardless of whether or not the image included textual annotations ($p<0.05$ for both cases).

\paragraph{Summary.} Our analysis shows that \rvision{all variable groups} contribute to higher perceived visual complexity: information-theoretic metrics, greater edge density, more feature points, and a larger number of distinct objects.
\rvision{This result replicates the findings in vision science that visual complexity is likely to be a high-dimensional phenomenon influenced by many factors.} As shown in the three panels of \autoref{fig:teaserVis2025}, the simplest and most complex visualizations differ notably in visual content. Similar patterns are observed in \sm~\autoref{fig:MASSVISandVIS30KExamples} when comparing samples from MASSVIS and VIS30K. Some visualizations, particularly matrix views, appear more complex due to their use of many small points and a wider range of colors. In contrast, simpler images tend to contain fewer elements and use minimal colors. Among these color-rich images, some used discrete colormaps with clearly defined edges, while others employed continuous colormaps such as gradient-style heatmaps. In some cases, these color patterns also reinforced coherent textural structures, further adding to visual complexity. Due to these observations, we next separate color/texture images by continuity to further investigate their relationship with perceived VC.

\section{Comparison to Previous Works}
\label{sec:compareresults} Having established the factors that influence perceived VC through quantitative, metric-based analysis, we applied the same approach to replicate whether these metrics can also account for factors previously studied in the context of node-link diagrams~\cite{purchase2012exploration} and clutter~\cite{rosenholtz2005feature}.

\subsection{Network-based Visual Complexity: Compared to Purchase~\cite{purchase2012exploration}}
\label{sec:purchase}

\begin{figure}[!t]
    \centering
    \includegraphics[width=0.82\columnwidth]{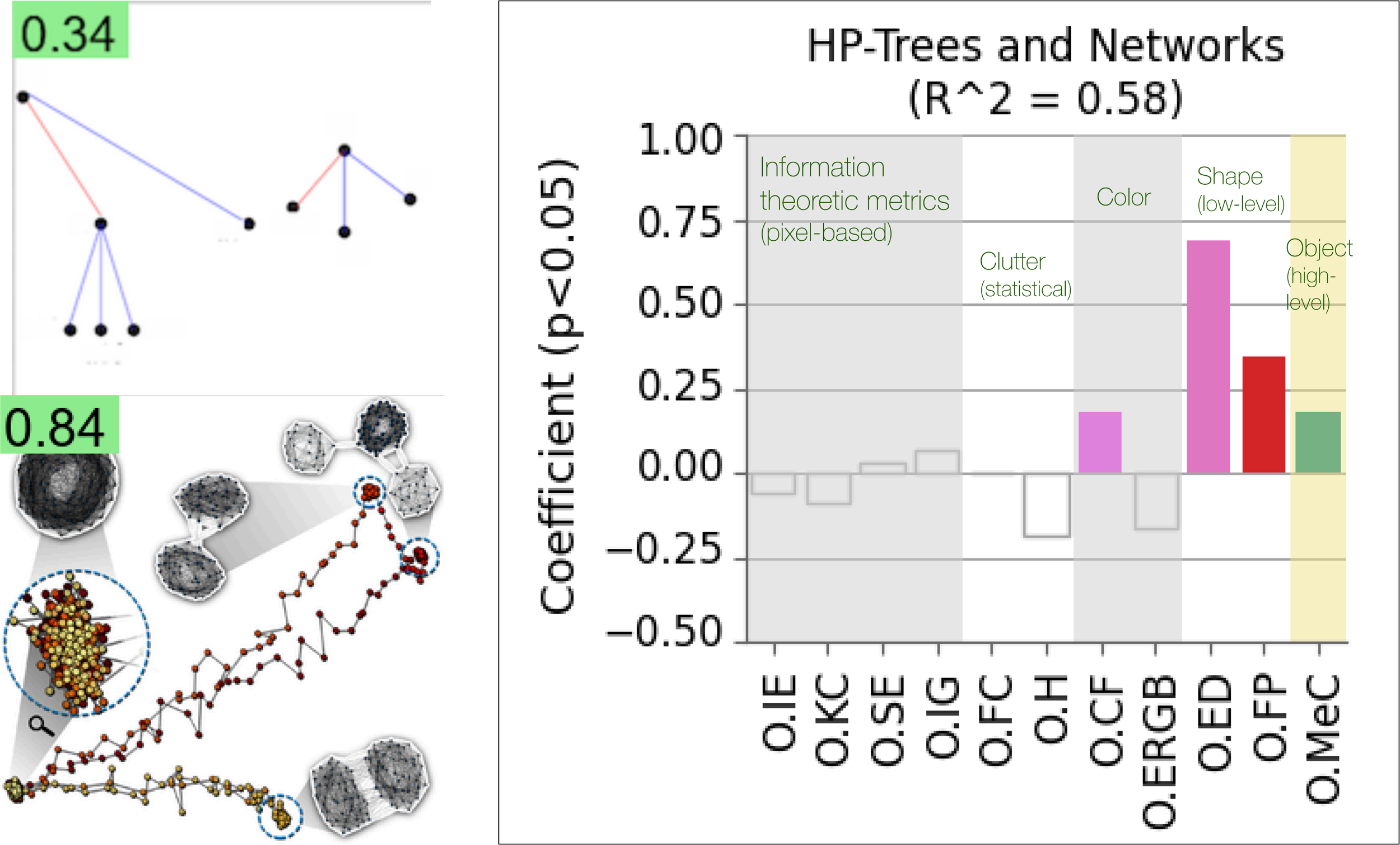}
    \caption{\textbf{Factoring visual complexity of 189 trees and network only} (in  Purchase~\cite{purchase2012exploration} style of direct node-link diagrams).
    The contribution of each image metric is represented by the magnitude of its corresponding regression coefficients, modeled using PLS.
    Only metrics significant at the 0.05 level are shown in solid color.}
    \vspace{-5pt}
    \label{fig:networks}
\end{figure}

There are 189 direct node-link diagrams in our \vcdataset dataset~(\autoref{fig:networks}). We thus studied and compared our factorization analysis to that of Purchase et al.~\cite{purchase2012exploration} to see if we could replicate results in that study. Purchase et al.~\cite{purchase2012exploration} studied network aesthetics by considering four variables: color (or individual RGB components, similar to O.ERGB), object edges (which define boundaries, similar to O.ED), intensity variations (similar to O.CF), and file size (similar to O.KC). Their model explained only $25\%$ of the variance in network visualizations. In contrast, our study employed the same edge detection method but found that O.ED was a statistically significant main effect, suggesting a stronger role for edge-based features in perceived VC.

To replicate and understand this extended large dataset, and compare results with the Purchase et al.\ study on network aesthetics, we curated a subset of our dataset containing network-style visualizations (primarily node-link diagrams, as shown in \sm~\autoref{fig:TreesNetworks}). To better match the original study design, we manually (1) filtered for direct node-link diagram (such that the matrix views were removed); (2) Removed all text and annotations using automated masking followed by a manual cleanup to isolate pure graphical elements given that these texts may activate the edge detector. We refer to this cleaned subset as HP-Trees and Networks. Using our 11 remaining metrics (excluding O.TiR), our PLS regression achieved an $R^2$ of 0.58 indicating a strong fit---substantially higher than the 0.25 reported in Purchase et al.~\cite{purchase2012exploration}. Notably, O.ED ($p<0.001$), O.FP ($p<0.001$), O.MeC ($p=0.003$), and O.CF ($p=0.016$) were key contributors to VC in this subset (\autoref{fig:networks}). We would credit the results to the addition of O.MeC and O.FP, which emerge as main contributors to perceived VC of networks.

\subsection{Information-Theoretic Measure of Color and Texture on Clutter Effect: Compared to Rosenholtz et al.~\cite{rosenholtz2007measuring}}
\label{sec:rosenholtz}

\begin{figure}[!t]
    \centering
    \includegraphics[width=0.82\columnwidth]{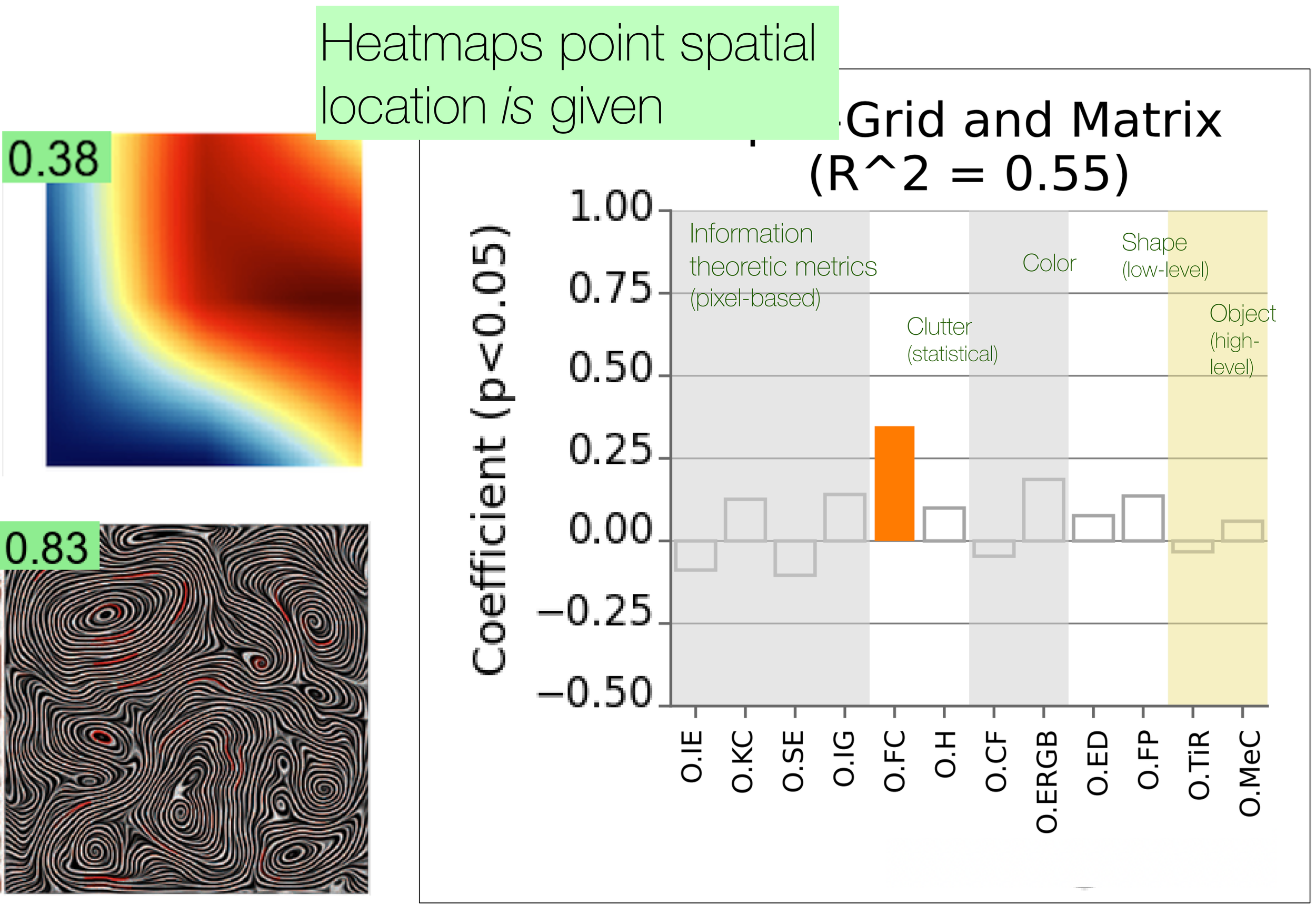}\\
    \vspace{4pt}
    \includegraphics[width=0.82\columnwidth]{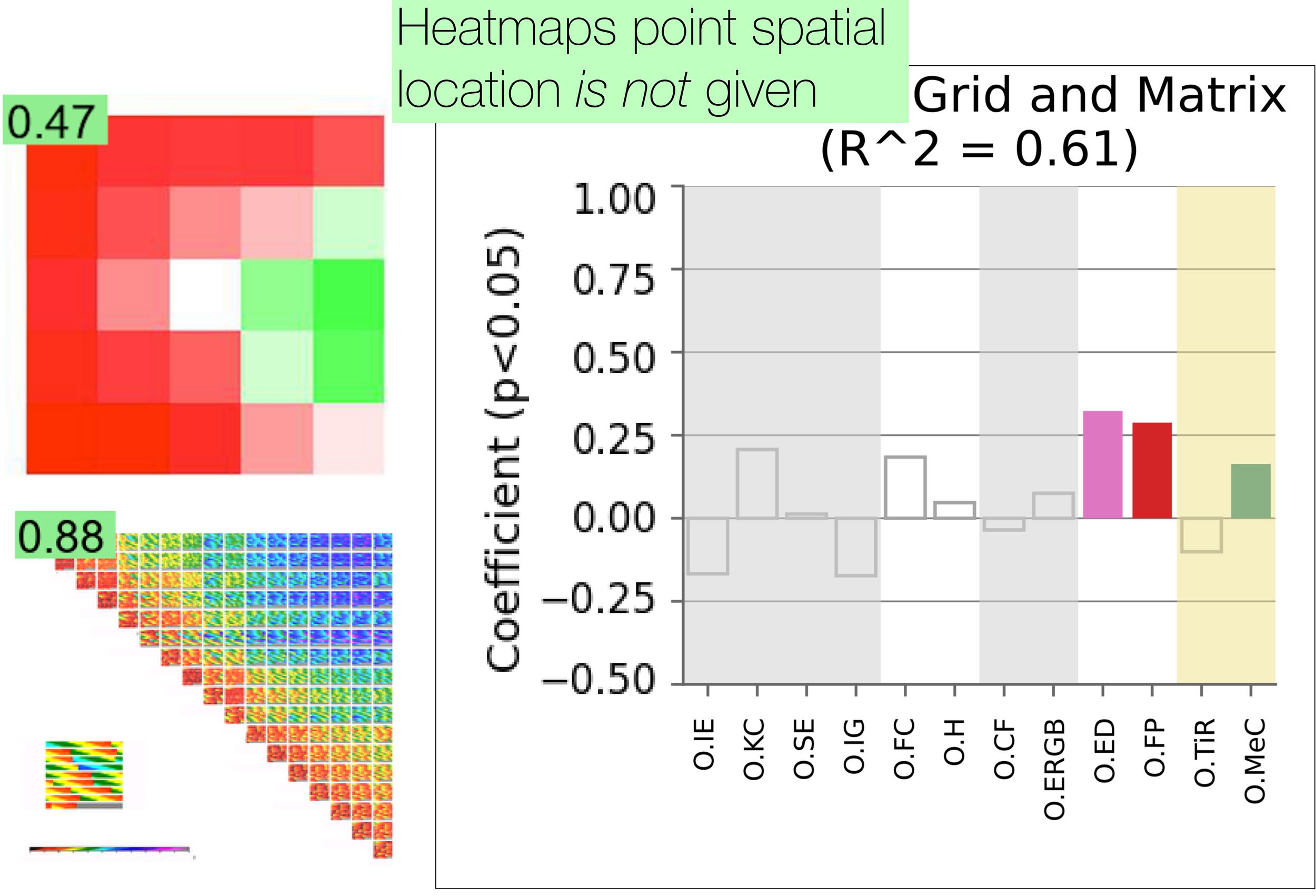}
    \caption{\textbf{Factoring visual complexity of 185 continuous and 126 discrete grid and matrix images}: \rvision{Significant metrics and their contributions to VC quantification vary by whether the visualization is continuous.} \textit{Top}, pixel positions are given and surface are continuous smooth color and texture. O.FC ($p=0.003$) is the only metric significant.
    \textit{Bottom}, the pixel position is discrete, artificially designed and clear boundaries are drawn. Three metrics are significant:
    O.ED ($p=0.003$), O.FP ($p<0.001$), and O.MeC ($p=0.006$).}
    \label{fig:gridmatrix}
\end{figure}

Rosenholtz et al. designed this clutter-based metric to capture color and texture distributions in a natural scene. Our dataset contained 311 `grid and matrix' or `heatmaps'~\cite{borkin2013makes}, of which 185 are spatial, continuous, where spatial location `is given'~\cite{munzner2014visualization}, making them more closely aligned with Rosenholtz’s definition of structured scenes. The rest are discrete, with pixel locations artificially generated---where structured grids and matrix introduce corners and edges. This contrast is evident in our dataset in these two sets of heatmaps (\autoref{fig:gridmatrix}). We also removed all text before modeling these two sets of visualization images to avoid edges and corners confounded by the presence of text.

\autoref{fig:gridmatrix} shows the results. When spatial location \textit{is} given, we found that for visualizations characterized by color and texture, O.FC emerge as the single most informative metric for explaining perceived VC. In contrast, when location is \textit{not} explicitly encoded, shape-based metrics (O.ED and O.FP) and the object-based metric O.MeC are the primary contributors to perceived VC. This result shows that in these discrete grids, edges also act as visual separators, effectively representing the number of visual elements present.

In summary, the continuity-specific results largely mirror the overall findings in terms of dominant contributing metrics, with some variation based on the stimulus type the metric models. What makes this particularly intriguing is that O.FC is highly effective in explaining complexity for continuous heatmap images but had limited explanatory power for artificially generated heatmaps.
\rvision{Given that the coefficients are higher by individual visualization types and continuity, we recommend that type categorization for VC should take into account continuity.}

\section{General Discussion} This section discusses theoretical and design implications of our results. A central problem for the science of visualization is related to perceptual experience as it influences people' understanding about data. One of such high-level perceptual experiences is visual complexity. Our present work contributes to understanding the alignment between human reading and metric measurement.

\subsection{Comparisons to Natural Scene Understanding in terms of Item-Count and Feature Congestion} Multiscale structures emerge in our study. Number of elments operates on object-associated O.MeC and O.TiR, and low-level features such as edges O.ED and corners O.FP, contributing significantly to VC across overall dataset. This result indicates that human perception does not operate on a single scale~\cite{wolfe2021guided}. This high-level perception seems to align to Turing's distinct hypothesis: patterned structures can self-organize through local competitive interactions~\cite{turing2009computing}. As a result, perception can be fast and minimize the need for generic instruction to perform pattern findings. Turing described that people's perceptual computing in this competing mechanism appeared to be more robust to noise, where global structural emergence exhibited topological robustness of insensitivity to variations and noises.

In visualization contexts, global structures formed by color and pattern may support visual Gestalt principles, as captured by the two object-based metrics: O.TiR (related to text) and O.MeC (related to color and object boundaries). These size-scale representations showed that low-level elements also emerge at the level of higher-order textures and objects. This is particularly evident in heatmaps where spatial location is given and aligns with the descriptor attribute O.FC, as demonstrated in our experiments~(\autoref{fig:gridmatrix}). In this context, O.FC represents a meaningful step forward in quantifying human perceptual experiences by treating certain visualization types (here color and texture) as statistical, scene-based representations.

This result may also indirectly support two perspectives. First, while Borkin et al.~\cite{borkin2013makes} treated grid and matrix as a single visualization type to characterize perceptual experiment (memorability), our findings may support that continuity can be considered as a distinguishing factor when we define type. Second, the prolific investigations of information-theoretic measures in data visualizations~\cite{chen2010information, janicke2007multifield, janicke2010salience, yang2000information} may contribute to humans' perceptual experiences---particularly when the underlying information bits are well defined. In our case, the color/texture dimensions in O.FC can be interpreted as encoding structured visual patterns. This result further confirms that continuous color/texture patterns in visualizations is understood differently from discrete grid and matrix.

\subsection{Graphic or Text? The Influence of Semantic Information on Perceived Visual Complexity}

Our findings suggest that the semantic information conveyed by text in visualizations impacts perceived VC differently than graphic elements. We conjectured that explanatory text may reduce perceived VC by enhancing understandability, \eg, Hearst et al.~\cite{hearst2023show} also highlighted the significance of text, considering it as equally important as the visualizations themselves.
\rvision{We found a positive effect of text on VC to a certain extent. In general, explanatory text reduces perceived VC due to its contribution to understandability. However, when the amount of text becomes excessive, it may increase complexity, shifting from aiding comprehension to overwhelming the visual features. Thus using titles and short annotations would be better than long paragraphs.}

\rvision{ Analyzing variations in text layout may reveal new ways to understand perceived VC. Previous eye-tracking behavior data for capturing memorability suggested that people can quickly distinguish text and titles. These text elements always occupied separate areas on the screen, making them easy for the human visual system to decode. In previous studies on three-dimensional scenes, where text overlapped with spatial objects, the list-based approach was often preferred~\cite{chen2004testbed}, highlighting the role of spatial organization in effective information presentation.} \tochange{Other non-data elements beyond annotations may also augment human perceptual experience, although their interaction with graphical elements can be complex. For example, adding a background grid~\cite{stone2008alpha}, incorporating chart-junk~\cite{hullman2011benefitting}, or using subtle contextual cues~\cite{willett2007scented} have all been shown to augment human visualization perceptual experience~\cite{feixas1999information}. The question of how much text is sufficient, and how non-data elements, graphical features, and annotations should appear may need to be personalized to align with the observers' reasoning goals~\cite{pang2025interactive} and the narrative structure of the visualization. }
\subsection{Considerations of Data-Driven Image-based Data Collection}

The use of large, real-world image datasets enabled us to capture a wide range of metric representations. When carefully chosen, a set of metrics can effectively reflect perceptual judgments in other large-scale studies~\cite{kramer2023features}.
\tochange{We thus recommend that progress towards establishing mechanisms underlying VC through quantitative analysis could use large-data collection to address three key challenges: \textit{first}, the development of common and comprehensive image databases to support the assessment of perceptual experiences~\cite{rensink2021visualization}; \rvision{\textit{second}, the design of experiments that directly capture high-level, perceptually driven experiences, such as meaningfulness, understandability, etc;} \textit{third}, the collection and validation of objective metrics capable of explaining those perceptual experiences that would eventually inform design. }

\rvision{Our results revealed that human high-level understanding is supported not only by design-level stimuli (\eg, point and line) but also by efficient object-level representations (\eg, color and text). The implication is to extend the current visualization representation units---whether derived from visual design or cognitive perception---by showing how they are formed by segmenting an image into understandable subsets. Each of these units is mapped to a corresponding data item, enabling fixation allocation and interaction while preserving essential information for the human observer. } Finally, one could also automatically flag overly complex figures for downstream processing, extending the impact of our work into the era of AI-driven data analysis.

\subsection{Limitations and Future Work}

Despite the broader range of metrics and assessments included in this study, limitations remain. \rvision{Experiments like ours are often limited by the datasets used.} For example, one could also validate our outcomes using a small dataset through item response theory for diagnostic efficiency, similar to approaches used in literacy testing~\cite{pandey2023mini}.
\rvision{Here we prioritize explainability based on large data and objective metrics.}
\rvision{In addition, our study used images free of interaction or animation, viewed by the general public rather than domain experts. Also, our evaluation is task-free. One can use our method to incorporate different modalities.} Finally, an exciting direction for future research would be to extend feature congestion modeling~\cite{rosenholtz2024visual} to account for stimulus-specific complexity beyond continuous color-texture stimulus. In principle, metric-based modeling offers the potential for greater precision in capturing a large set of variables of how complexity is perceived.

\section{Conclusion}

Our empirical study systematically examines the relationship between human perception and objective image quality metrics. The \vcdatalink contains visualization stimulus types, human-rated VC scores, and 12 metric measurement scores. Our analyses reveal that the number of elements, whether defined by low-level features like edges and points or high-level boundaries formed by color and text, is a key factor in perceived visual complexity and shapes whether a visualization is perceived as scene-like.

\tochange{Mathematical information-theoretic metrics contribute to perceived VC only when they incorporate high-level structural elements. The strong discriminative power of the feature congestion model is particularly intriguing, as it suggests that probabilistic characteristics may coexist meaningfully within visualizations. If these characteristics can be effectively integrated with scene-structure and information-theoretic metrics, it may be possible to extend feature congestion modeling to a broader range of stimulus types.} As a result, integrating stimulus features, such as set-size, with statistical information-theoretic methods presents a promising pathway for systematically modeling variations of perceived visual complexity by human observers.

Ultimately, such an approach could provide a principled basis for interpreting high-level perceptual experiences across diverse visualization inputs, thereby enhancing the scientific and design potential of complexity quantification. Many aspects of visualization experience and metric behavior extend beyond what is observable in controlled lab-based experiments. This underscores the need to study real-world visualizations used in everyday contexts. To that end, the methodology used in this study, collecting subjective human ratings through large-scale crowdsourcing, offers a scalable way to capture holistic human perception in real-world settings. The complementary insights gained from combining data-driven modeling and experimental methods have the potential to drive transformative advances---an opportunity that the visualization research community is well-positioned to embrace.

\section*{Image Copyrights}

\noindent We, as authors, state that Figures~\ref{fig:overview}, \ref{fig:factorization}, \ref{fig:objectmodel}, \ref{fig:networks}, right pane, and \ref{fig:gridmatrix}, right pane, as well as \autoref{tab:ovc} and~\autoref{tab:svc} are under our own copyright with the permission to be used here. We have also made them available under the \href{https://creativecommons.org/licenses/by/4.0/}{Creative Commons At\-tri\-bu\-tion 4.0 International (\ccLogo\,\ccAttribution\ \mbox{CC BY 4.0})} license and share them at \osflink. All remaining images in the paper are \textcopyright\ IEEE, with permission to be used here.

\acknowledgments{The work was partially supported by the ``Deutsche Forschungsgemeinschaft (DFG, German Research Foundation) under Germany's Excellence Strategy – EXC 2120/1 – 390831618'' through a Visiting Professorship Position, made to Prof. Dr. Jian Chen, via the Cluster of Excellent IntCDC, University of Stuttgart, Germany. \rvision{The crowdsourcing experiment was funded by the Deutsche Forschungsgemeinschaft (DFG, German Research Foundation) – Project-ID 251654672 – TRR 161, with The Ohio State University IRB approval ID: 2022B0363.} \rvision{R.S. Laramee was supported, in part, by funding from the EPSRC (EPSRC UKRI157, APP17227).} We also wish to thank Dr. Ruth Rosenholtz at MIT for sharing the Matlab code for computing Feature Congestion, and colleagues at the Visual Attention Lab at Harvard University, as well as reviewers for their comments. }

\bibliographystyle{abbrv-doi-hyperref}
\bibliography{template}
\clearpage
\appendix

\clearpage
\noindent\begin{minipage}{\columnwidth}
    \vspace{1cm}
    \makeatletter
    \centering
    \sffamily\LARGE\bfseries
    \mytitle\\[1em]
    \large{Additional material}
    \\[1em]
    \makeatother
\end{minipage}
\vspace{1cm}
\setcounter{section}{0}

\section{Reproducibility: \vcdataset: Perceived Visual Complexity Scores, and Objective Image-Quality Metric Databases}
\label{sm:openscience}

Our perceived visual complexity (VC) dataset, \vcdataset, consists of 1,800 visualization images, each annotated with VC scores collected from human participants via the online crowdsourcing platform Prolific. The dataset also includes corresponding objective image-quality metrics and associated metadata. \vcdataset is publicly available through a Google spreadsheet
\vcdatalink. All data analysis code and images can be accessed via Colab on
\vcgoogledrive. \osflink\xspace also has all the analysis results and code. The key columns of \vcdataset include: \begin{enumerate}[label=\textbf{\Alph*}]
    \item
    The \textbf{VIS30K paper DOI} or \textbf{the magazine or web links (column H)}  as a unique identifier to cross-link to other databases such as Vis30K~\cite{chen2021vis30k}, VisPubData~\cite{isenberg2016vispubdata}, KeyVis~\cite{isenberg2016visualization}, and the Practice of Evaluating Visualization{~\cite{isenberg2013systematic,lam2011empirical}}, as well as  MASSVIS~\cite{borkin2013makes}.

    \item
    The \textbf{image name (column C)}, either the  MASSVIS or the VIS30K source image names, as a unique identifier to cross-link to other image datasets, \eg, MASSVIS~\cite{borkin2013makes} and VIS30K~\cite{chen2021vis30kdataset}.

    \item The \textbf{thumbnail of each image (column E)}, either the original image or a sub-panel from the source image, provides a gateway for image annotation and analysis, \eg, through Google CoLab run-time execution \vccolablink.

    \item An indicator \textbf{(column P)} specifying the \textbf{HP-Trees and Networks} visualization images, which are the direct node-link diagrams used in Purchase et al.~\cite{purchase2012exploration}
    and the \textbf{RR-Color and Textured} images,  representing the texture and color stimuli used in Rosenholtz et al.~\cite{rosenholtz2007measuring}
    to support the analysis
    in the main text.

    \item The \textbf{image link (column D)} that points to a web storage address where a full-resolution version is accessible through \vcgoogledrive.
    Please note that all VIS30K image files are copyrighted, and for most the copyright is owned by IEEE. Some have creative commons licenses or are in the public domain. Yet other images are subject to different, specific copyrights as indicated in the figure caption in the paper.

    \item The \textbf{perceived visual complexity (column F)} for the images collected through the crowdsourcing. The interface to collect this data is shown in~\sm~\autoref{fig:interface}.

    \item
    The \textbf{O.MeC score (column BO)} and The \textbf{O.TiR score (column BJ)}, corresponding to the number of color and text-ink-ratio, respectively.
    The Python code for computing these is available in the sub-folders {\textit{Metric.MeaningfulColor(O.MeC)} and \textit{Metric.TexttoInkRatio(O.TiR)}} on \vcgoogledrive.

    \item The other  \textbf{10 image-metrics}, as shown in \textbf{column V to BF}. The Python code for computing these is available in the sub-folder \textit{Metrics.10} on \vcgoogledrive.

\end{enumerate}

\section{Figure Image Courtesy by References}
\label{sm:imageCourtesy}

Since IEEE VIS 2025 has a 2-page citation limits and we used many images from publications, we provide this figure-by-figure citation list to credit the image source. Images in~\autoref{fig:teaserVis2025} are taken from ~\cite{borkin2013makes,cui2011textflow,sanderson2010analysis,gschwandtnei2015visual,jakob2020fluid,yang2004value,huber2005visualizing,otaduy2005haptic,kerr2003thread,hlawatsch2011flow,rufiange2013diffani,tuttle2010pedvis,lekschas2017hipiler,lommerse2005visual,fujiwara2020visual,chen2020co,wilson2005exploring,kondratieva2005application}. Images in~\autoref{fig.TiRMeC} are taken from~\cite{borkin2013makes,botchen2005texture}. Images in~\autoref{fig:metrics} are taken from~\cite{weiss2020revisited, rafiei2005effectively, borkin2013makes}. Images in~\autoref{fig:networks} are taken from~\cite{rafiei2005effectively, liu2018tpflow}. Images in~\autoref{fig:gridmatrix} are taken from~\cite{jakob2020fluid,matvienko2015explicit,zhang2020uncertainty,wattenberg2005note}.

\begin{figure*}[!t]
    \centering
    \includegraphics[width=\textwidth]{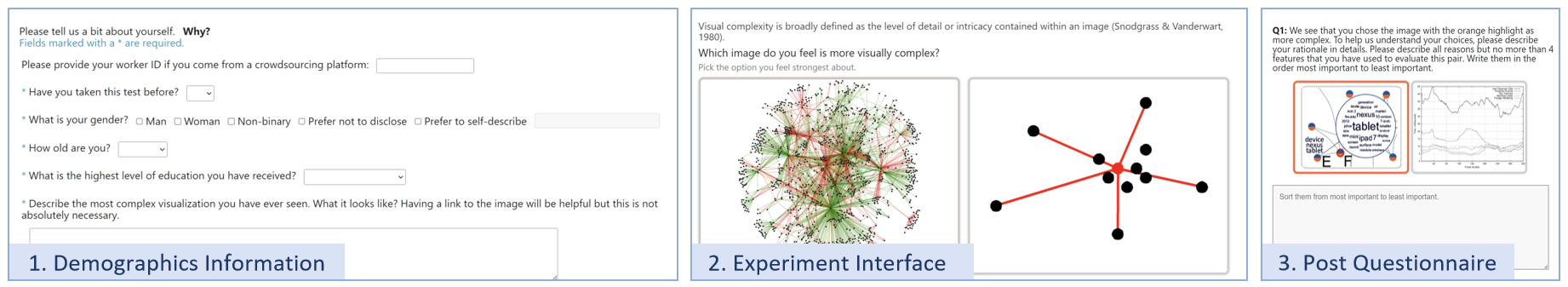}
    \caption{\textbf{Interface for collecting perceived visual complexity.} Participants perform multiple paired comparisons between images, with new pairs assigned using an active sampling method. The collected data were used to assess observer consistency and to compute perceived visual complexity for our quantitative metric analysis (see the main text~\autoref{sec:ParticipantProcedure}).}
    \vspace{-10pt}
    \label{fig:interface}
\end{figure*}

\begin{figure*}[!t]
    \centering
    \includegraphics[width=\textwidth]{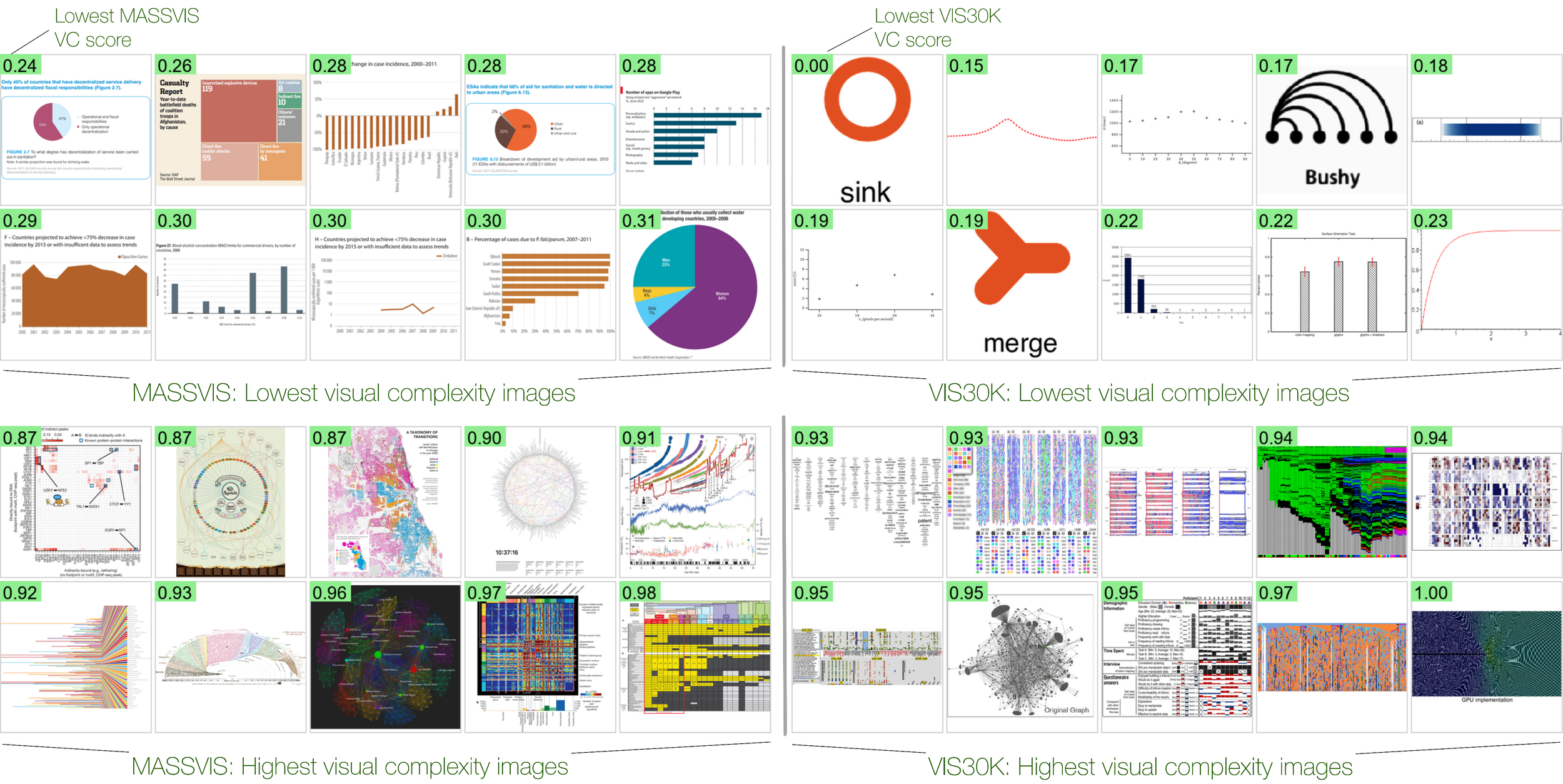}
    \caption{The lowest and highest visually complex images in MASSVIS (left)~\cite{borkin2013makes} and VIS30K (right)~\cite{chen2021vis30kdataset} (see the main text~\autoref{sec:Crowd-sourcing-result}).}
    \label{fig:MASSVISandVIS30KExamples}
\end{figure*}

\begin{figure*}[t!]
    \centering

    \includegraphics[width=\textwidth]{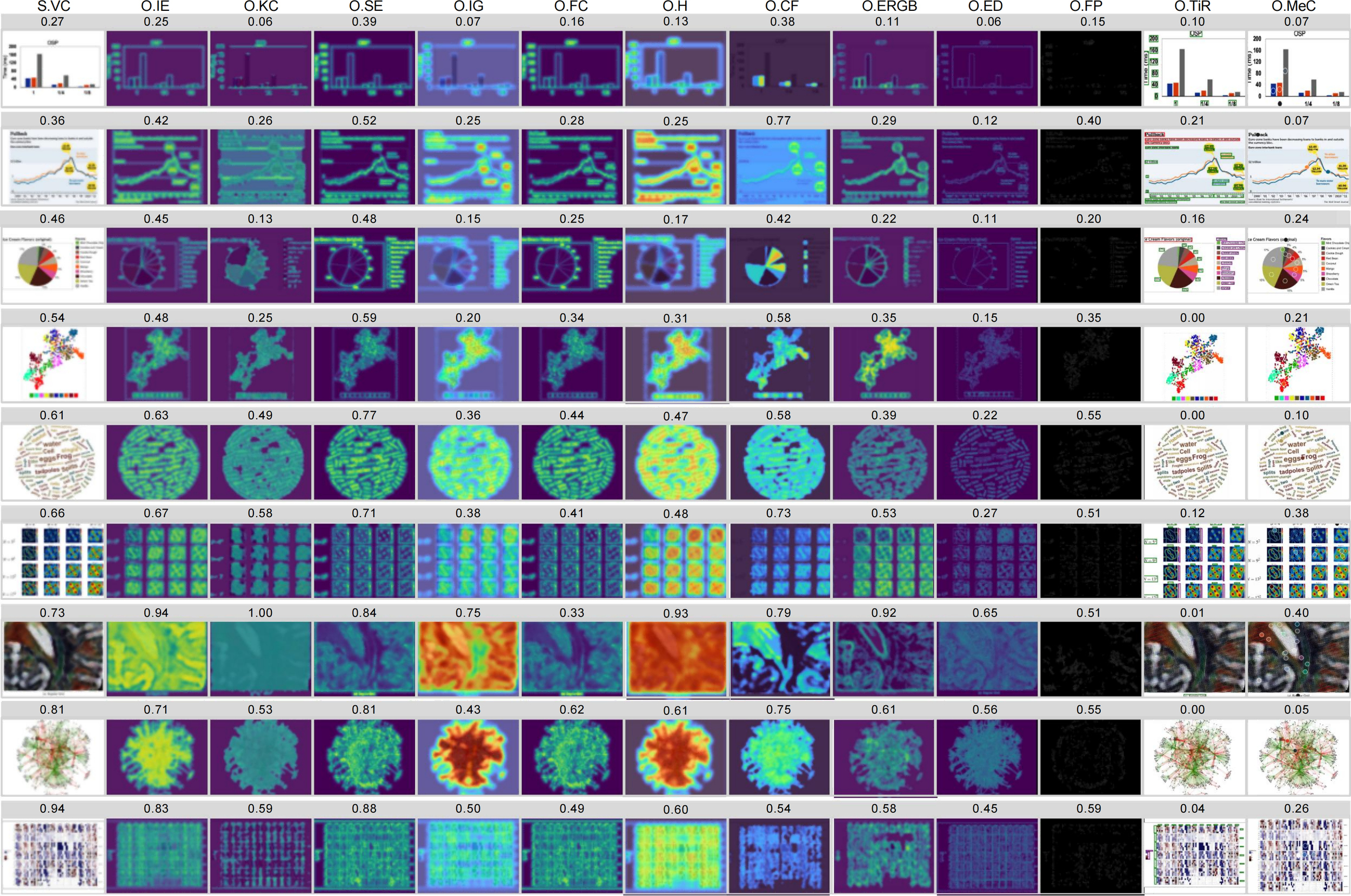}
    \caption{\textbf{Additional examples of image VC and metric scores.} Each row shows the original image followed by visual representations of the 12 objective metrics, along with their corresponding scores. Values indicate normalized percentage scores (see the main text ~\autoref{sec:12metrics}). }
    \label{fig:exampleMetricsVis}
\end{figure*}

Images in~\sm~\autoref{fig:interface} are taken from ~\cite{dinkla2012compressed, sprenger2000h}. Images in~\sm~\autoref{fig:exampleMetricsVis} are taken from ~\cite{dinkla2012compressed,molchanov2018shape,kindlmann2006diffusion,chi2015morphable,xu2010information,setlur2015linguistic,lu2020palettailor,borkin2013makes,wang2020stull}. Images in~\sm~\autoref{fig:MASSVISandVIS30KExamples} are taken from ~\cite{cui2011textflow,sanderson2010analysis,huber2005visualizing,kerr2003thread,gschwandtnei2015visual,kappe2015reconstruction,weigle2005visualizing,balzer2005voronoi,wilson2005exploring,wei2019evaluating,huron2014constructing,Zhao2020PreservingMS,molchanov2018shape,lommerse2005visual,chen2020co,Liu2017PatternsAS,Nguyen2016SenseMapSB}.

\section{\rvision{Addition Experimental Choices and Results}}

\subsection{\rvision{Experimental Data Samples}}
\rvision{The main criteria when we chose the data was diversity of stimuli. It is a large, heterogeneous dataset spanning multiple visualization (node-link diagrams, heatmaps, charts), drawn from real-world sources, in order to increase ecological validity and generalizability of findings. }

\subsection{A Workshop and Pilot Studies}
\label{sec:pilotWork}

There are various approaches to collecting subjective VC experiences, whether dealing with a small set of visualization images or a large set common in vision science. To determine the most suitable method for gathering VC scores, we piloted several approaches including both low-fidelity manual sorting and high-fidelity active sampling-based experiments. This was necessary because a large-scale data gathering of visual experience data for thousands of visualization images had not been established previously in the visualization literature.

The first approach involved asking raters to score a small set of images (about 20-100). For example, He et al.~\cite{he2022beauvis} used 15 images to collect absolute beauvis scores. We tested this in our exploratory workshop but found it difficult to align scores across participants when attempting to rank a larger set of 200 images.

The second approach involved a hierarchical division of images into groups. For example, in the study by Oliva et al.~\cite{olivia2004identifying}, participants divided 100 images into four hierarchies and stated their criteria to categorize natural scenes as `openness' etc. This is also the standard de facto approach to measure other perceptual dimensions from images (\eg, seminal work of texture and visualization image classifications~\cite{rao1996towards, lohse1994classification}). We piloted this approach in our workshop and found it effective for gathering participants' comments to guide subsequent steps. However, similar to the first approach, it presented scalability challenges, limiting its applicability for larger datasets. During the workshop experiment, 46 participants, organized into 20 groups of 2-3 students, were instructed to perform the hierarchical sorting tasks based on perceived visual complexity of the images, defined as ``\textit{the amount of detail or intricacy in images}''~\cite{purchase2012exploration}. These participants were graduate students having domain expertise in machine learning and computer vision, before they took the visualization class; and three participants had extensive industrial experience. Participants were asked to document the factors that influenced their sorting decisions.

Participants on day 1 sorted 20 images, with all groups completing the task in an average of 20 minutes. The scalability challenge arrived on day 2 for sorting 50 images, with the same instruction. Participants described the sorting task as "overwhelming." No group was able to finish within 1.5 hours, and only $50\%$ of the groups managed to complete the task within 2 hours. First, participants noted that increasing image set made any linear sorting impractical; and two groups utilized a 2D Cartesian coordinate system to organize VC clusters. Second, their comments suggested the benefits of text display: without the knowledge of the data source and differences between MASSVIS and VIS30K, 11 groups pointed out that some (MASSVIS) images `\textit{looked nice}' and these (VIS30K) \textit{show sophisticated scientific phenomena}. `\textit{Text (annotated) greatly improved clarity; thus I rated the image with lower complexity}'. Participants also commented that `\textit{After reading the texts, I know where to look.}' Others mentioned object set-size and said `\textit{there were not many items in the image; thus these (MASSVIS) images were simpler.}' Third, half of the groups commented that ``\textit{these (VIS30K scientific visualization) images were difficult to understand without text or legend showing what an image represents.}'' These subjective experiences led us to exclude 3D surface and volume rendering visualizations from our formal experiment, as they were challenging for participants and were likely to be rated more complex.

The third approach is active-sampling based, similar to how game competition assigns competing players~\cite{bradley1952rank} to score human players on a global scale. For example, both Bradley-Terry ranking algorithm~\cite{bradley1952rank} and Microsoft's TrueSkill 2\texttrademark~\cite{minka2018trueskill, nagle2020predicting} enabled scoring thousands of images in computer vision~\cite{saraee2020visual}. We also tried this for our pilot study with 200 images and found that this approach was reliable. However, scaling up this approach to 1,800 images with one-time sampling would require a significantly larger number of trials, making it nearly infeasible in terms of cost and time.

\begin{figure*}[!t]
    \centering
    \begin{subfigure}{0.3\textwidth}
        \centering
        \includegraphics[height=0.7\textwidth]{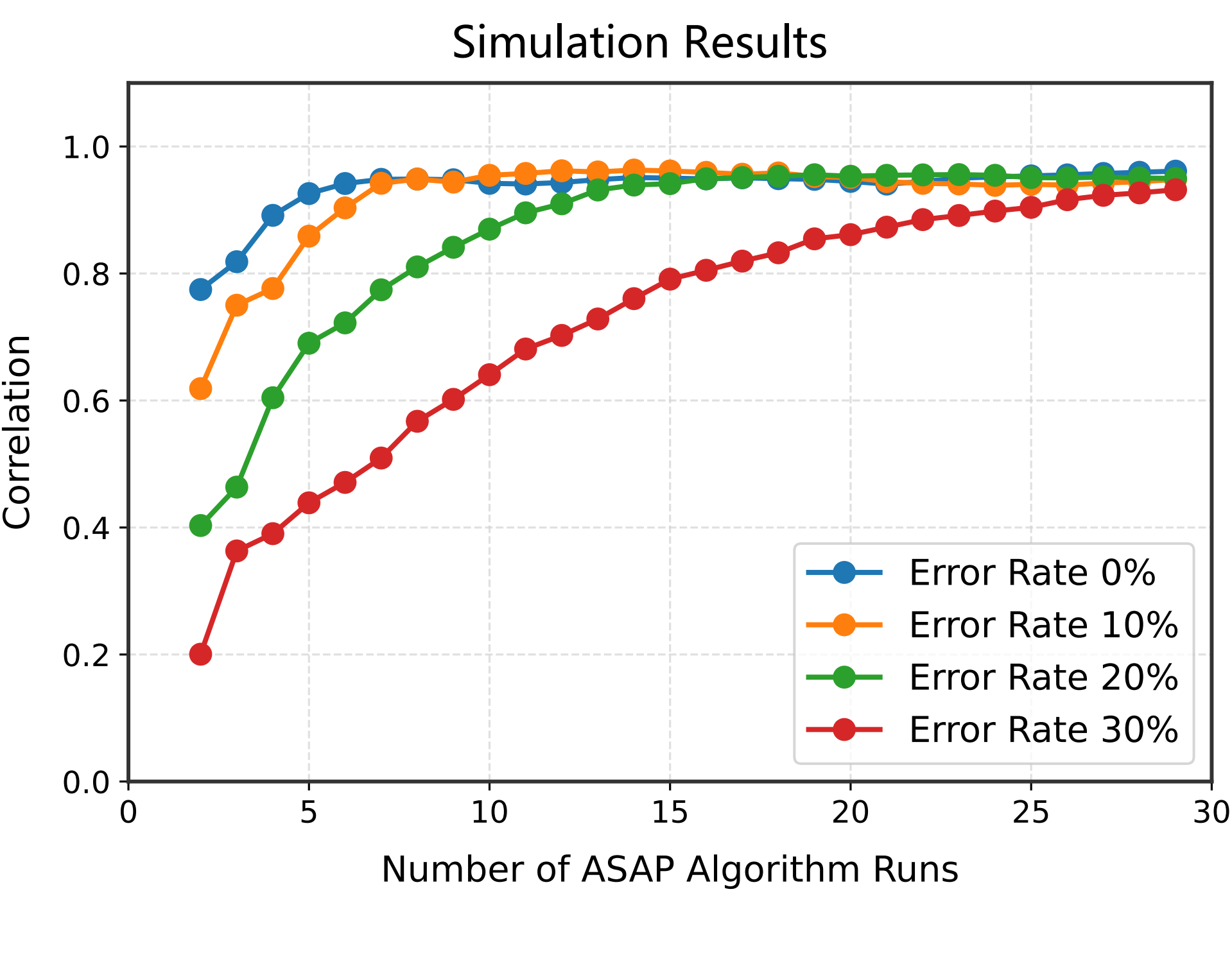}
        \caption{Simulation of noise input.}
        \label{fig:asap:noise}
    \end{subfigure}
    \begin{subfigure}{0.3\textwidth}
        \centering
        \includegraphics[height=0.7\textwidth]{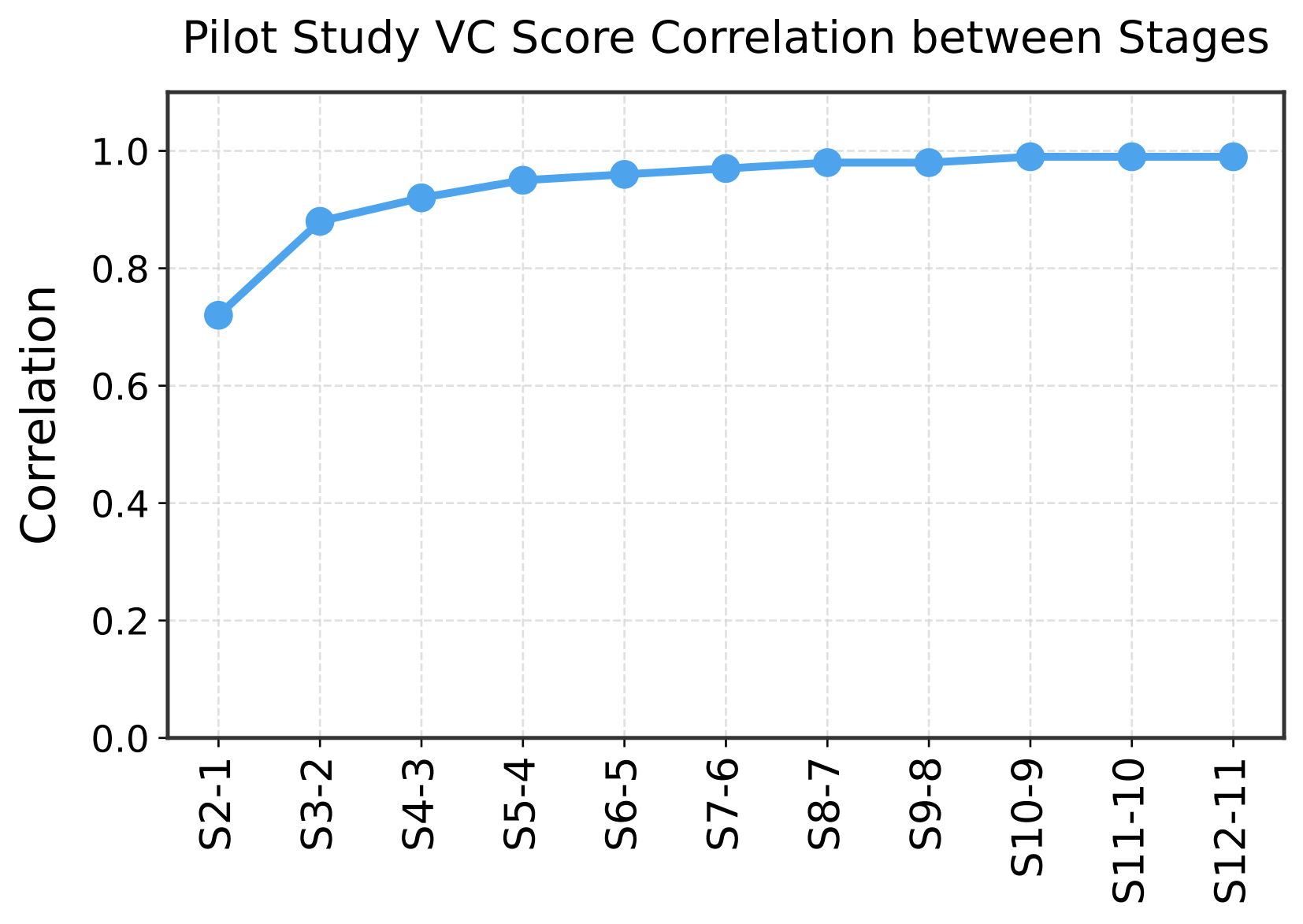}
        \caption{Pilot study reliability.}
        \label{fig:asap:stages}
    \end{subfigure}
    \begin{subfigure}{0.3\textwidth}
        \centering
        \includegraphics[height=0.7\textwidth]{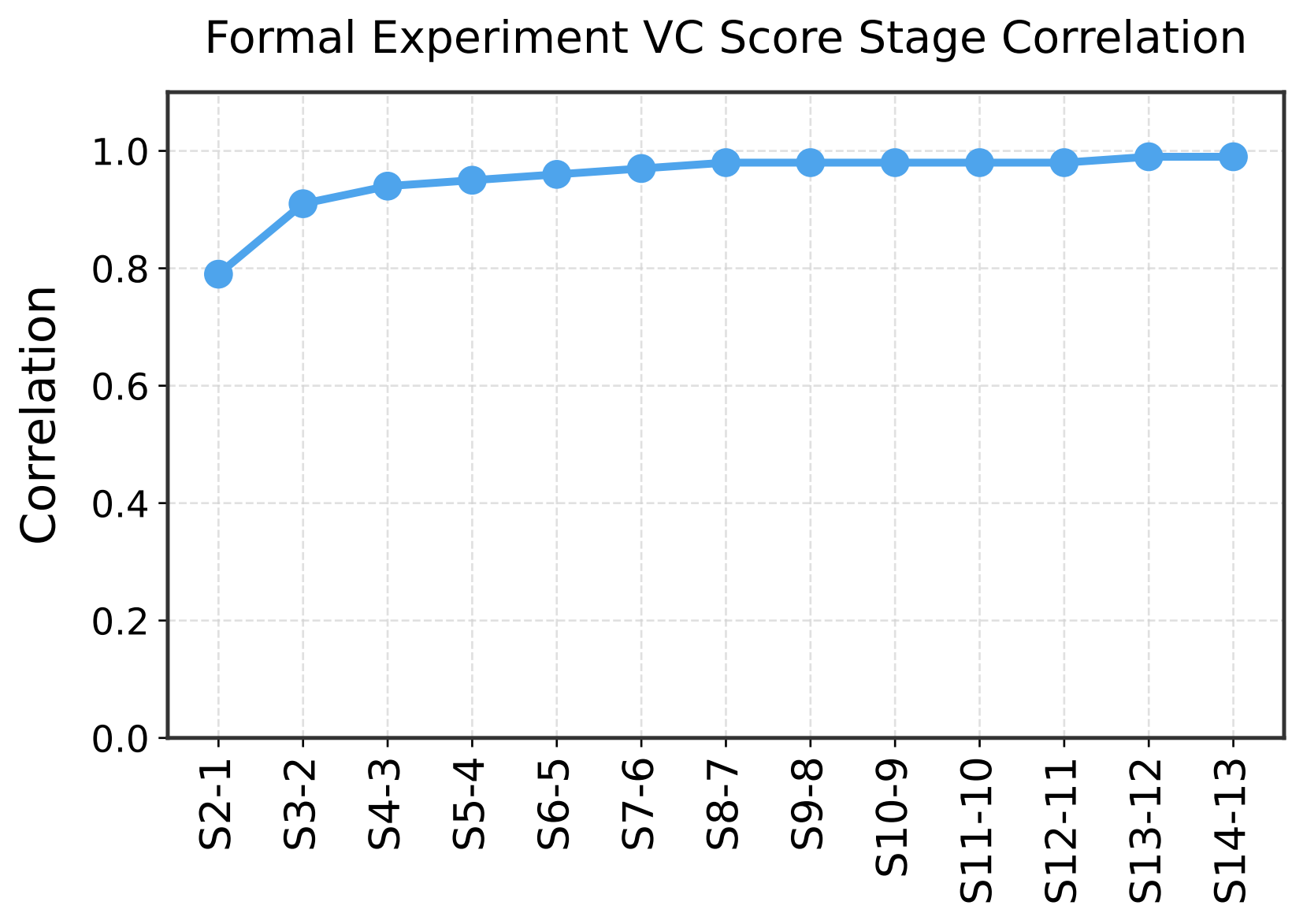}
        \caption{Formal study reliability.}
        \label{fig:asap:correation}
    \end{subfigure}
    \caption{\textbf{Reliability of the Mikhailiuk et al.'s multi-stage active sampling algorithm~\cite{mikhailiuk2021active}}. This figure validates the active sampling algorithm and estimates the number of stages required for our data collection.
    \textbf{(a). Robustness to human input noise.} We simulated user error by randomly flipping 0\%, 10\%, 20\%, and 30\% of comparison outcomes, leading the algorithm to recommend a different set of images in the next phase. Despite the noise, the algorithm remained robust---at noise levels $\leq$ 20\%, the correlation between noisy and original image scores remained above 0.9 and stabilized within 13 stages.
    \textbf{(b). Pilot test of algorithmic reliability (N=90 images).} In a 12-stage pilot, 72 different participants (6 per stage) ranked 90 images. The plot shows high correlations between image scores across adjacent stages, indicating convergence.
    \textbf{(c). Reliability in the formal full study (N=1,800 images).} Our stopping rule required a correlation $>0.95$ across three consecutive stages. After the $13^{th}$ stage, the VC score correlation with the previous stage reached 0.98. Our data collection stopped at the $14^{th}$ stage, and results from this stage were used for analysis.
    \textbf{Observations.} The results indicated that the scores stabilized by the $10^{th}$ stage after approximately 890 comparisons. Comparisons generated by the active sampling algorithm were randomly distributed across 6 participants per stage (see the main text~\autoref{sec:pilotSummary}). }
    \vspace{-10pt}
    \label{fig:validationExps}
\end{figure*}

We finally used a multi-stage active sampling algorithm~\cite{mikhailiuk2021active}, which reduces the number of comparisons by 10 fold while maintaining the same level of reliability. In each phase, the algorithm optimizes the information obtained to adapt the algorithms to assign an absolute score.~\autoref{fig:validationExps} shows the validation of this approach.

\subsection{\rvision{Partial Least Square (PLS) for Variable Selection: Additional Discussion}}
\label{sec:PLS}

\rvision{PLS is a relatively new method and is gaining interests in choosing variables in many scientific domains such as bio-informatics, chemistry and neuroscience~\cite{krishnan2011partial}. It is becoming widely used mainly due to scientists' ability to collect large datasets with many variables and PLS's ability to handle a large number of variables. The process of choosing the most useful variable is called variable selection, where variables in our case are the objective image metrics (predictors) and the response is the collected VC scores via the crowdsourcing experiment.} Unlike principal component regression, which maximizes variance in predictors, PLS explicitly maximizes the covariance between predictors and the target variable (VC score), allowing us to extract latent components that are optimally aligned with perceptual responses. This made PLS an appropriate tool to retain and analyze all metrics jointly while still yielding interpretable component structures.

\rvision{\textbf{Our calculation.} We assessed the appropriate number of PLS components to retain by running models with 1 to 10 components and selecting the number where the regression $R^2$ began to plateau. Based on this procedure across different experiments, we selected 5 components for all final analyses to ensure a balance between model complexity and interpretability. For the bootstrapping procedure to assess the reliability of each metric’s contribution, we resampled the data with replacement across multiple iterations to estimate the distribution of the PLS regression coefficients, from which we computed standard errors and approximate p-values. This enabled us to identify statistically significant and stable metric variables. Lastly, to aid interpretation, we complemented coefficient analysis with individual metric vs. VC plots (\eg, \autoref{fig:objectmodel} and extended in the supplemental) that visualize how VC scores vary across the value range of each feature.}

\subsection{Verbal Reasoning}
\label{sec:verbal}
\tochange{While the main text focuses on quantitative analysis, we also conducted an additional analysis of the 698 self-reported verbal reasoning responses collected from the post-questionnaire. In these responses, participants explained their rationale for two randomly selected trials. The words and their frequencies offer insights into the cognitive processes involved, helping to construct vocabularies that represent semantic categories relevant to the perception of visual complexity.

This approach aligns with prior work in high-order perception research, such as studies on memorability~\cite{kramer2023features} and attitude~\cite{ashokkumar2021social}.
\autoref{tab:keywordCountTop10} shows the proportion of words grouped by semantic category, ordered by descending word frequency.
\autoref{tab:vcDimensionNames} summarizes the resulting comments into the categories. The most frequently cited themes were clarity and readability ($47.3\%$) and color and contrast ($39.7\%$), followed by information density ($34.1\%$), number of elements ($33.0\%$), abstractness and familiarity ($25.9\%$), visual clutter ($16.8\%$), interconnectedness ($7.4\%$), and beautifulness ($2.6\%$).}

The quantitative results also align with our metric-based findings, supporting the existence of a hierarchy of perceptual features that influence visual complexity. Interconnectedness (7.4\%) reflects participants’ attention to the relationships between visualization elements, with frequently used terms such as "connection" (9 mentions) and "interact" (6 mentions), primarily associated with node-link diagrams.

\rvision{The high-level attributes of clarity and readability associated with verbal reasoning may explain that some high-level details could not be explained by the PLS models. For communication purposes, the familiarity, structure, surprise~\cite{sartas2025complexity}, distinctiveness~\cite{kyle2025scene, kyle2023characterising}, and aesthetic appeal, are important. Furthermore, we also did not measure \textit{correctness}, (\ie, the representations to use visualization that represent key knowledge in order to access meaning) and \textit{semantic distance} (the goodness-of-fit between representation and its meaning). Our image-based metrics do not represent this set of high-level attributes, but our method can capture these to further examine complexity.}

\section{Additional Implications on Using Complexity to Support Design}

Ideally, this project outcome would support new design.
\textit{Declutter} removes any additional ink not needed: grid lines, labels, colors, and 3D effects are taken as redundant visual elements~\cite{ajani2021declutter}. The color removal in their decluttering experiment was appropriate for conditions where colors were not used to encode data. Subsequently, the authors added highlight colors and text annotations to convey key takeaways from the figures. In doing so, the purpose of the images shifted from classic rhetorical visualizations to communicative tools, integrating textual explanations with semiotic marks. The authors observed a weak improvement in perceived VC, a finding that is consistent with the insights from our inquiry. On the other hand, human perception of ``redundancy'' depends on the expertise of the viewer: People with less experience would benefit from more data-relevant elements.

Our regression model revealed that metrics associated with edge density and feature points, along with color count, play a dominant role in determining perceived complexity, which is in close alignment with the work of Oliva et al.~\cite{olivia2004identifying}. For visualization applications, extensive studies of visual abstraction and suggestive contours, which are driven by perceptual principles~\cite{viola2017pondering, decarlo2023suggestive}, could potentially help reduce VC.

\begin{table*}[!t]
    \centering
    \small
    \caption{A comprehensive literature review of terms, factors, their definitions, and real-world applications reported to influence visual complexity. This review served as the foundation for summarizing subjective factors in~\autoref{tab:svc} and guiding the selection of objective metric measurements in~\autoref{tab:ovc} (see the main text~\autoref{sec:relatedwork}).}
    \label{table:svccomplete}
    \label{tab:paperlist}
    \fontsize{6.5}{6.5}\selectfont \begin{tabular}{|p{1.5cm}|p{2cm}|p{5.5cm}|p{1cm}|p{5.5cm}|}
        \toprule
        \textbf{Category} &\textbf{Factor} & \textbf{Definition} & \textbf{Source} & \textbf{Goal} \\
        \midrule
        &Entropy & The amount of information required to encode the distribution of pixel values. & ~\cite{saraee2020visual,ryan2018glance} & To measure the information content and ``visual richness'' of an image.\\
        &Compression ratio & The size of the compressed image divided by the size of the original raw image. & ~\cite{saraee2020visual, corchs2014no} & To measure image complexity.\\
        &Noise & The unwanted variations and features present in the image. & ~\cite{le2012representing} & To propose a 3D feature space to represent visual complexity of images.\\
        Information-theoretic       &Quantity of information & The number of units of information on the screen. & ~\cite{miniukovich2018visual} & To measure the facets of visual complexity for GUIs.\\

        &Meaningfulness & The meaningfulness of the stimuli. & ~\cite{olivia2004identifying} & To judge visual complexity of images.\\
        &Understandability & The difficulty of semantically understanding a scene. It relates to two aspects: quantity of elements and relationships between elements. & ~\cite{corchs2016predicting} & To evaluate real world image complexity.\\

        & Prototypicality & Compliance  with past experience or habit. & ~\cite{miniukovich2018visual} & To quantify interface visual complexity.\\
        &Cognitive load & The cognitive effort required for interaction with the interface. &~\cite{harper2009toward} & Definition of visual complexity as an implicit measure of cognitive load.\\

        &Perceivability of detail & Limitations of human perception requiring more effort to perceive details make interfaces seem more complex. & ~\cite{miniukovich2018visual} & To measure the facets of visual complexity for GUIs.\\

        \hline
        & Clutter & The number and complexity of items or their representation or organization. & ~\cite{rosenholtz2007measuring} & To formalize the concept of visual clutter.\\
        &  & The amount of distractors and focal points competing for attention. & ~\cite{olivia2004identifying}  & To evaluate clutter measurement methods of perceived scene complexity.\\
        Clutter      & Feature congestion & The energy or strength of edges, color/luminance variations, and orientations within a local neighborhood. & ~\cite{saraee2020visual} & To calculate and measure visual clutter and complexity of local area.\\
        &Visual clutter & Influences of search performance of target. & ~\cite{miniukovich2018visual} & To quantify interface visual complexity.\\
        \hline

        &Colorfullness & The variety and diversity of colors present in the region. & ~\cite{rosenholtz2007measuring} & To evaluate visual clutter.\\
        & Color variability & The amount of color variation across a region quantified by summing the euclidean distances between the color of each pixel and the mean color of the region. & ~\cite{rosenholtz2007measuring} & To measure attributes for visual clutter.\\
        &   & The number of  dominant colors, perceived color depth and idiosyncratic color  preferences. & ~\cite{miniukovich2018visual} & To quantify interfaces.\\
        & Colors & Number of unique RGB colors. & ~\cite{purchase2012exploration} & To measurement attribute.\\
        & Color regions & The numbers of regions of different color/texture that can be segmented. & ~\cite{corchs2016predicting} & To indicate varied colors/textures and visual complexity.\\
        & Color harmony & The coordination of colors, luminance, and hues in the image. & ~\cite{corchs2016predicting} & To evaluate the color balance of an image and reflect visual complexity.\\
        Color     & Color coherence & The level of continuous and homogeneous intensity of color signal.& ~\cite{iliyasu2013mining} & To measure the chromatic contributions to visual complexity and model system representing human visual system's assessment or evaluation of visual complexity.\\
        & Luminance variations & The amount of local luminance variation across a region quantified by summing the luminance contrast between each pixel and its neighbors. & ~\cite{rosenholtz2007measuring} & To evaluate visual clutter.\\
        & Gradient strength & The average magnitude of graylevel gradients, indicating the overall level of contrast and transitions in the image. & ~\cite{mario2005image} & To derive a human criterion-free metric for image complexity.\\
        & Saturation complexity & The vividness of a color. & ~\cite{fernandez2019visual} & To analyze the effectiveness and informativeness of the HSV channels in predicting visual complexity.\\
        & Hue complexity & The color type. & ~\cite{fernandez2019visual} &  \\
        & Value complexity & The brightness of a color. & ~\cite{fernandez2019visual} &  \\
        &   Lighting complexity & The complexity of lighting including the contrast of light and shadow and richness of shadow effects. & ~\cite{ramanarayanan2008dimensionality} & To measure the perceived visual complexity in a computer graphics scene.\\

        \hline
        &  Symmetry & The degree to which one half of the image predicts the other half. & ~\cite{olivia2004identifying} & \\
        & & The similarity  of an object reflection. & ~\cite{miniukovich2014quantification} & To quantify interface visual complexity.\\
        & Organization & The spatial structure, regularity, and placement of items. & ~\cite{olivia2004identifying} & To reflect how the combination of elements affects visual complexity.\\
        & Spatial Frequency & The rate of change or repetition of image patterns across space. & ~\cite{corchs2016predicting} & To indicate the number of cycles or repetitions per unit visual angle or image width. \\
        &  Variety of visual form & The number of visual features like colors, shapes, sizes, textures used to represent information. & ~\cite{miniukovich2018visual} &         GUIs.\\
        &  Spatial organization & Repetition and regular positioning of visual elements simplifies the perceived information. & ~\cite{miniukovich2018visual} &
        GUIs.\\
        &  Ease of grouping & The similarity within a  visual group, and the difference between one visual group and  other groups. & ~\cite{miniukovich2014quantification} & To quantify interface visual complexity.\\
        &  Grid & Regular repetition of similar structural elements. & ~\cite{miniukovich2014quantification} & To quantify interface visual complexity.\\

        &  Edges & The contours and boundaries in an image that exhibit significant graylevel changes. & ~\cite{mario2005image} & Photos $\&$ images. \\
        Shape
        and
        Structure      &  Edge density & The ratio between the number of edge pixels and the total number of pixels in an image. & ~\cite{mario2005image} & To measure image complexity.\\
        & & The area of the image occupied by edges. & ~\cite{purchase2012exploration} & To measure visual complexity.\\
        &  Edge congestion & The density of edges. & ~\cite{miniukovich2018visual} & To quantify interface visual complexity.\\
        &  Structure & The visual patterns and edges present in the image. Images with more edges and patterns are considered to have higher structure. & ~\cite{le2012representing} & To propose A 3D feature space to represent visual  complexity of images.\\
        &      Compactness & The ratio of squared perimeter to area. & ~\cite{attneave1957physical} & To judge complexity of shapes.\\
        &  Turns & The number of sides or turns in the shape. & ~\cite{attneave1957physical} & To judge complexity of shapes.\\
        &  Skeletons & The internal structure of the shape. & ~\cite{sun2021curious} & To measure shape complexity by explaining why a shape has the external features. \\

        &  Texture variation & A higher contrast, correlation, energy and lower homogeneity indicate more texture variation and complexity. & ~\cite{corchs2016predicting} & To predict visual complexity across different image categories.\\
        &   Material complexity & The richness, variety, and intricacy of materials and textures. & ~\cite{ramanarayanan2008dimensionality} & To measure the perceived visual complexity in a computer graphics scene.\\

        &  Co-occurrence matrix & Symbiotic relationship between different texture. &~\cite{iliyasu2013mining} & To evaluate visual complexity.\\

        \hline
        &  Number of objects & The number of objects or components that make up a visual scene. & ~\cite{olivia2004identifying} & To evaluate clutter.\\
        &  Object categories & The number of unique object categories in a scene. &~\cite{chai2010scene} & To account for the effect of perceptual grouping. This means that similar objects can be perceived as a group, reducing the perceived complexity. \\
        Object-based     &  Element relationships & More complex interactivity and relationships between elements in a scene. & ~\cite{heft2000evaluating} & To evaluate scene complexity.\\
        &  Element heterogeneity & Diversity and heterogeneity of elements. & ~\cite{snodgrass1980standardized} & To evaluate visual complexity of image.\\
        &  Element diversity & To represent variety of visual elements present in the image including objects, textures, and corners, etc. & ~\cite{le2012representing} & To propose a 3D feature space based on structure, noise, and diversity.\\
        \bottomrule
    \end{tabular}

\end{table*}

\begin{figure*}[!t]
    \centering {\includegraphics[width=\textwidth]{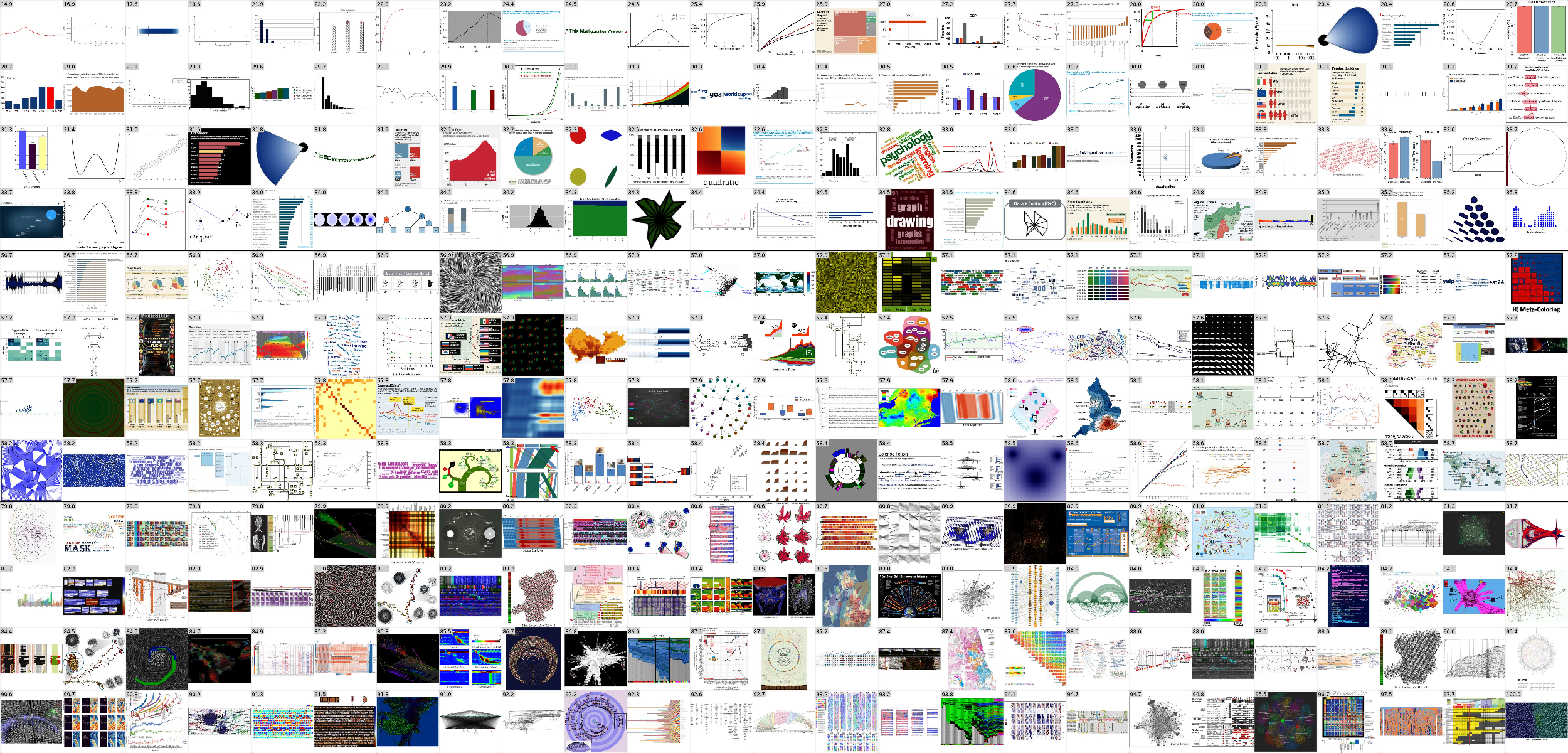}}

    \caption{\textbf{A 300 Example image set with visual complexity scores.} From 1,800 images,~\textit{Top:} The least visually complex visualization images from our experiment. \textit{Middle:} The images around the median visual complexity scores.
    \textit{Bottom:} The most visually complex images from our experiment (see the main text ~\autoref{sec:Crowd-sourcing-result}). }
    \label{fig:VCExamplesFirstSet}
\end{figure*}

\begin{figure*}[!t]
    \centering
    \centering
    \fbox{
    \includegraphics[width=0.85\textwidth]{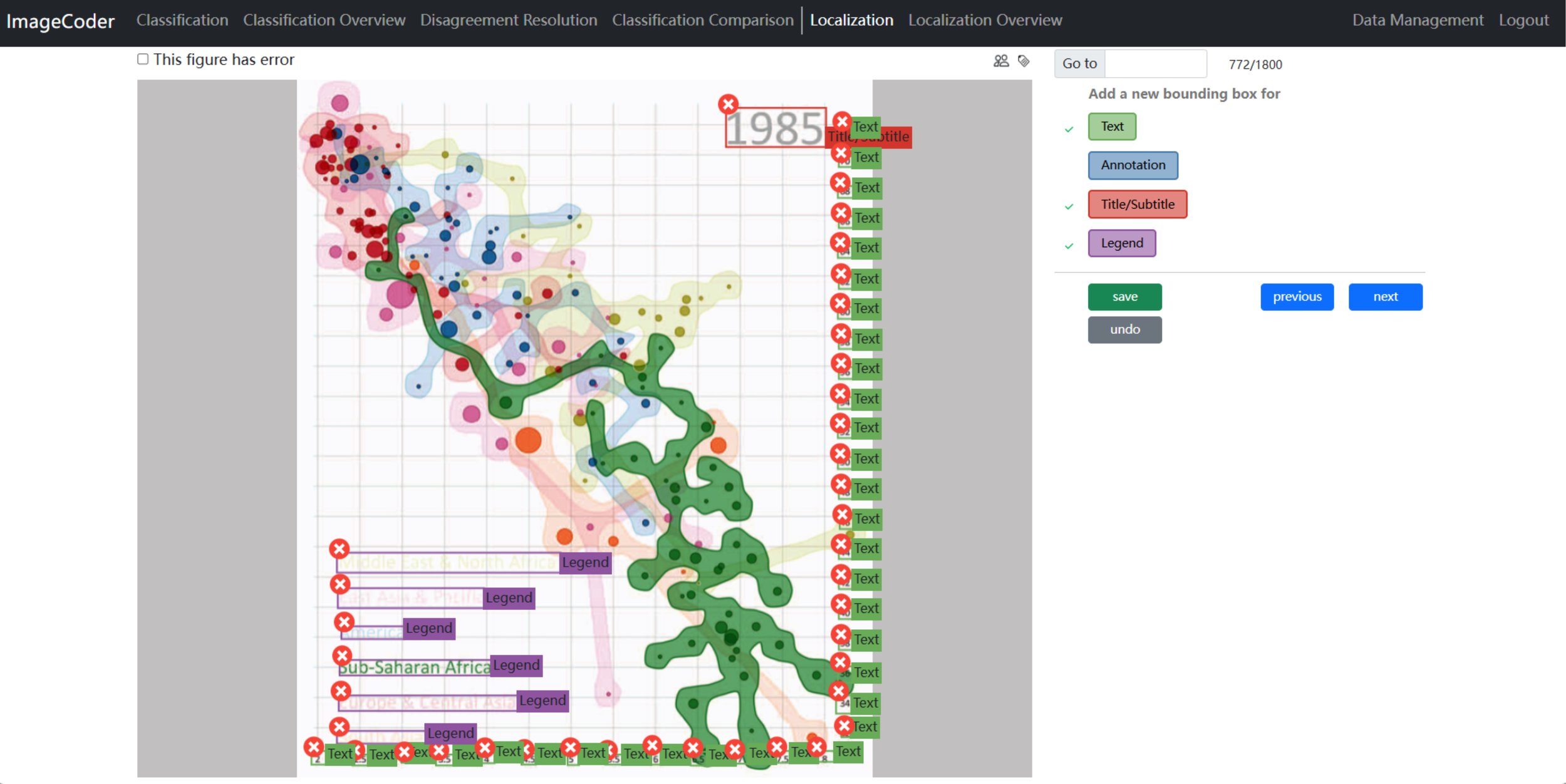} }
    \caption{ \textbf{Interface for curating the text labels.} Here, our labels separate `annotation' from other texts such as `Title/subtitle', and `legend' (see main text \autoref{sec:objMetrics}).}
    \label{fig:OTiRInterface}
\end{figure*}

\begin{figure*}[!t]
    \centering
    \setlength{\picturewidth}{0.3\textwidth}
    \setlength{\pictureheight}{3cm}

    \begin{subfigure}{\picturewidth}
        \centering
        \includegraphics[height=\pictureheight]{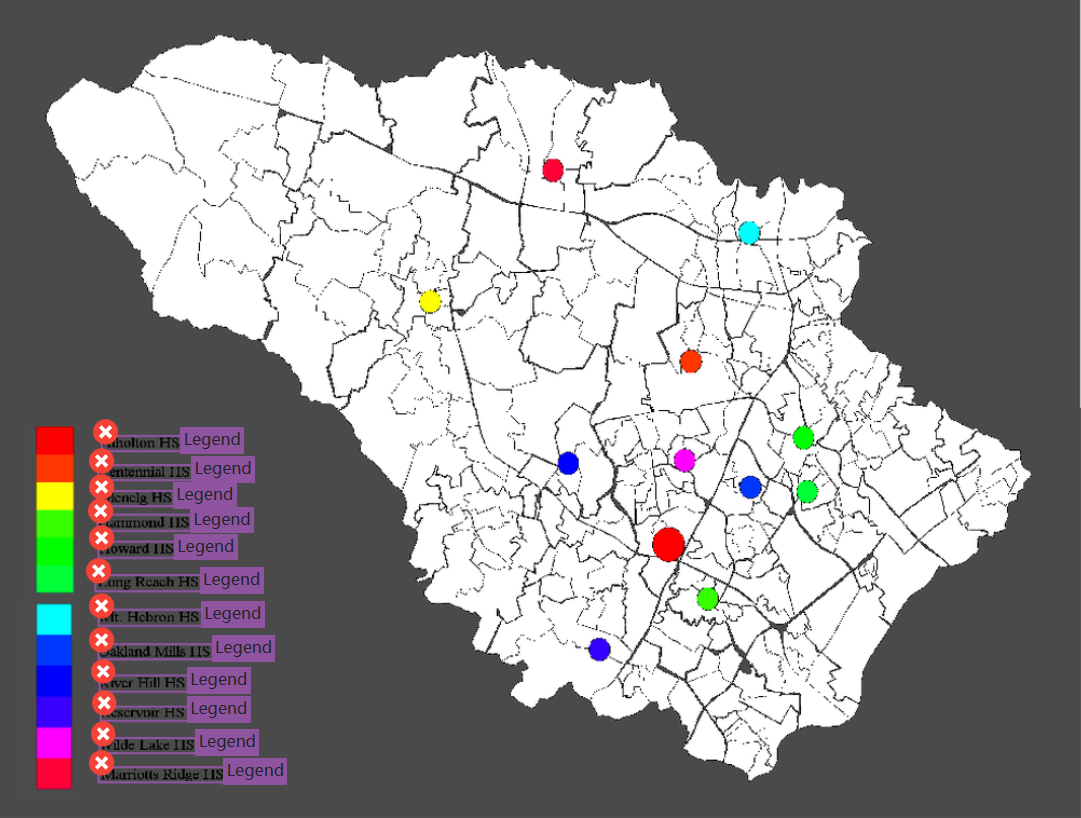}
        \caption{O.TiR=1.97\%}
    \end{subfigure}
    \begin{subfigure}{\picturewidth}
        \centering
        \includegraphics[height=\pictureheight]{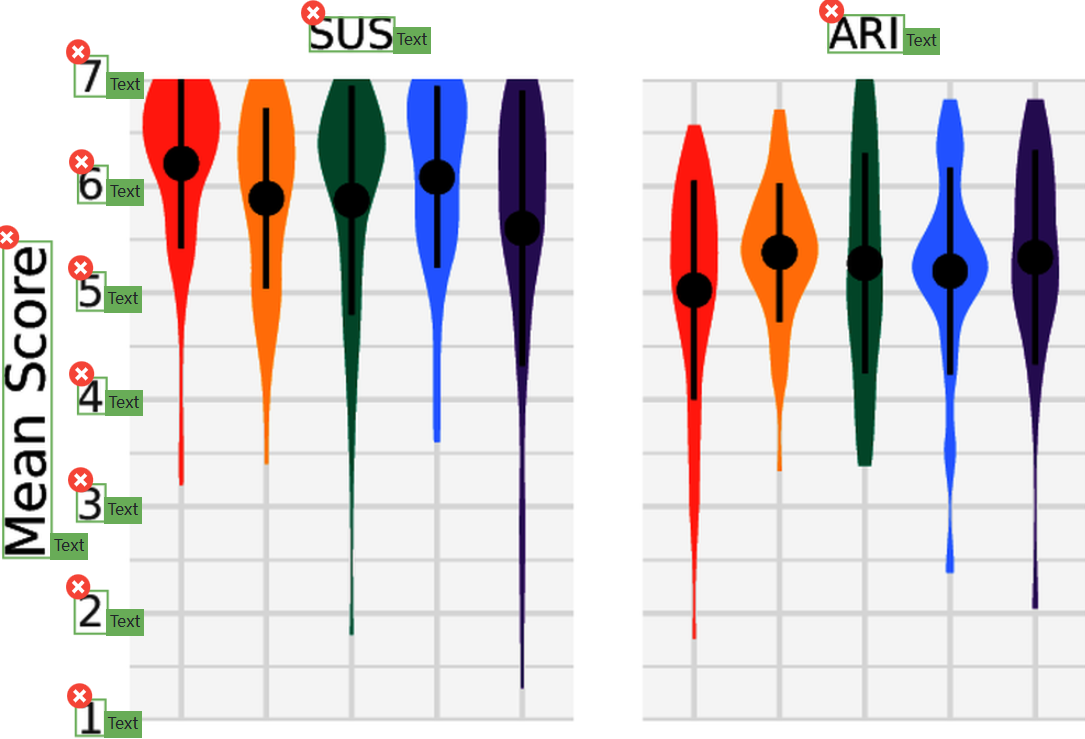}
        \caption{O.TiR=3.69\%}
    \end{subfigure}
    \begin{subfigure}{\picturewidth}
        \centering
        \includegraphics[height= 2cm]{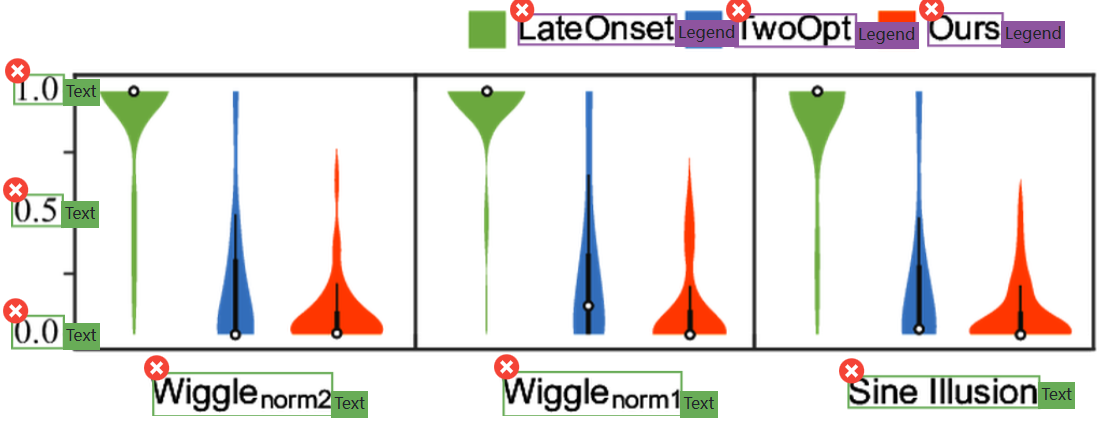}
        \caption{O.TiR=8.51\%}
    \end{subfigure}

    \begin{subfigure}{\picturewidth}
        \centering
        \includegraphics[height=\pictureheight]{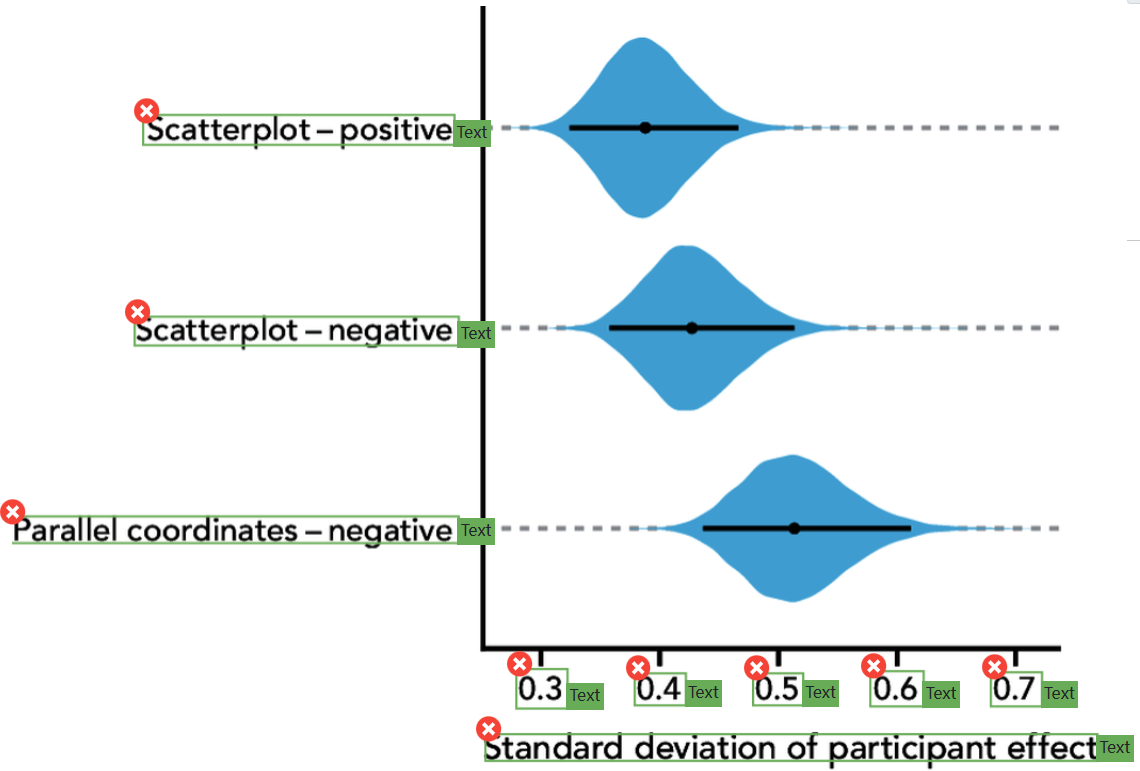}
        \caption{O.TiR=6.71\%}
    \end{subfigure}
    \begin{subfigure}{\picturewidth}
        \centering
        \includegraphics[height=\pictureheight]{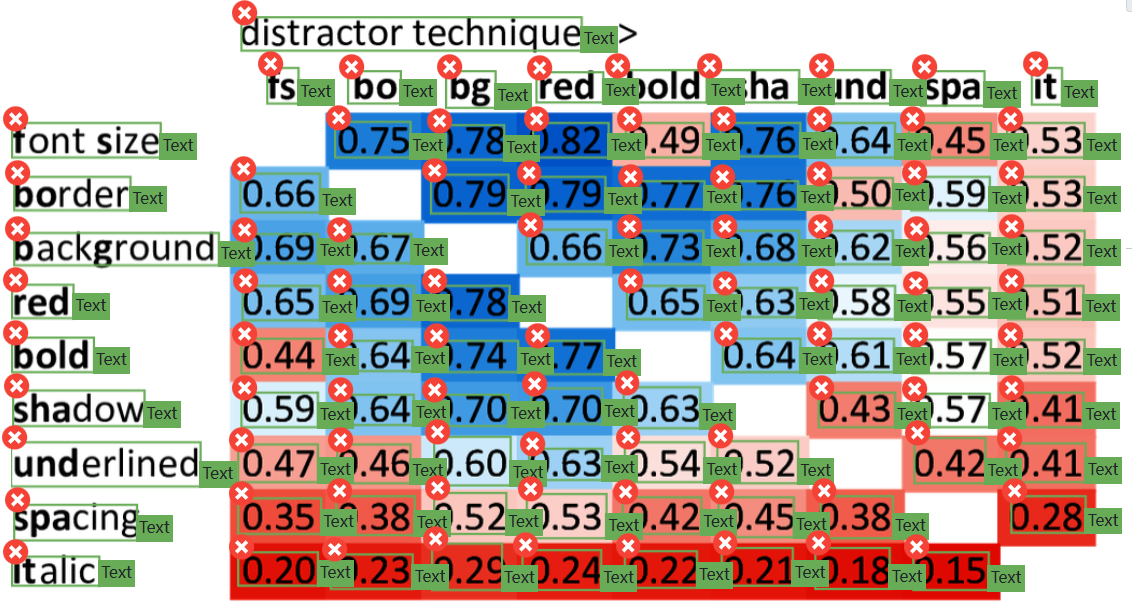}
        \caption{O.TiR=41.39\%}
    \end{subfigure}
    \begin{subfigure}{\picturewidth}
        \centering
        \includegraphics[height=\pictureheight]{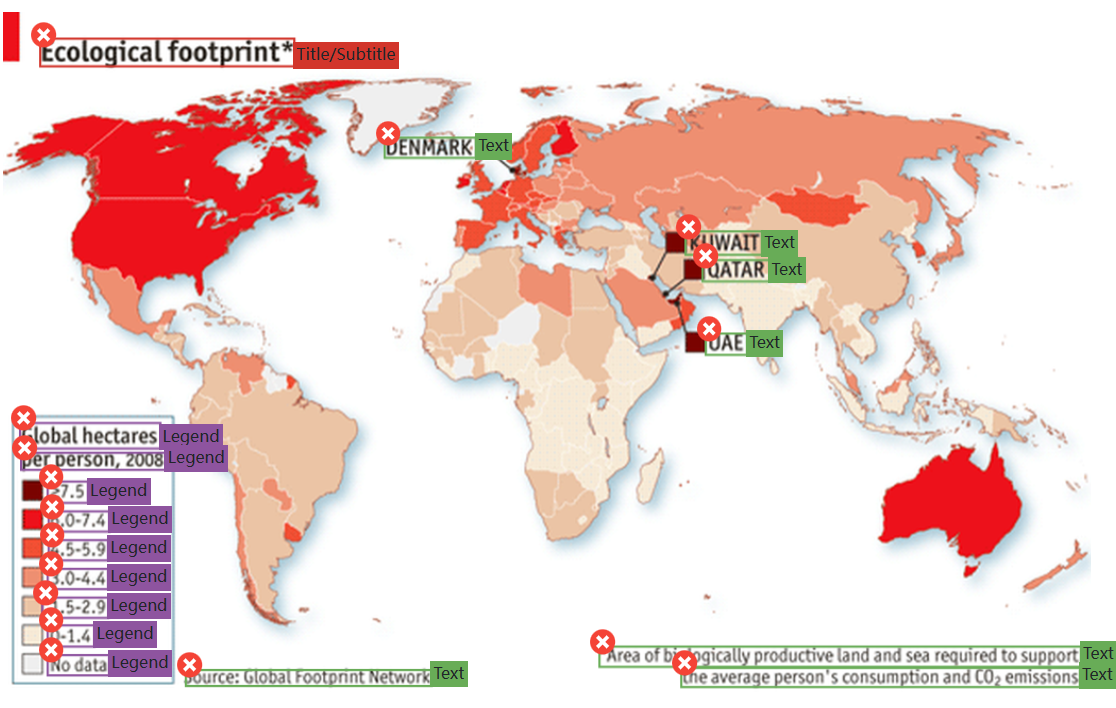}
        \caption{O.TiR=6.54\%}
    \end{subfigure}

    \begin{subfigure}{\picturewidth}
        \centering
        \includegraphics[height=\pictureheight]{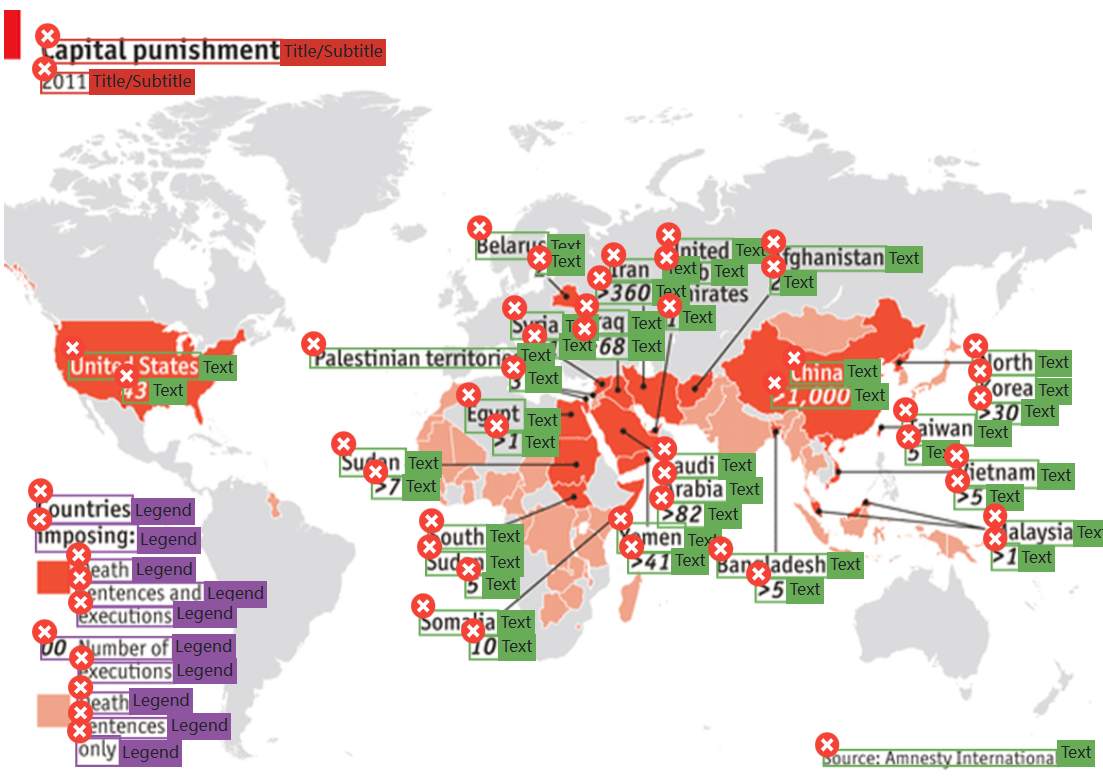}
        \caption{O.TiR=10.99\%}
    \end{subfigure}
    \begin{subfigure}{\picturewidth}
        \centering
        \includegraphics[height=\pictureheight]{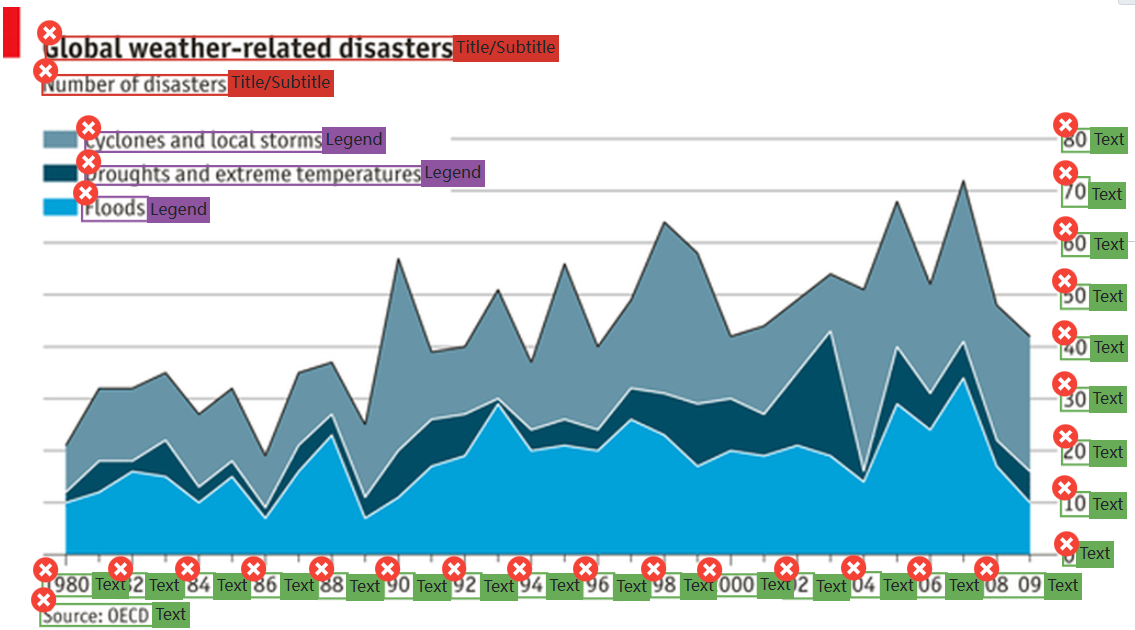}
        \caption{O.TiR=6.94\%}
    \end{subfigure}
    \begin{subfigure}{\picturewidth}
        \centering
        \includegraphics[height=\pictureheight]{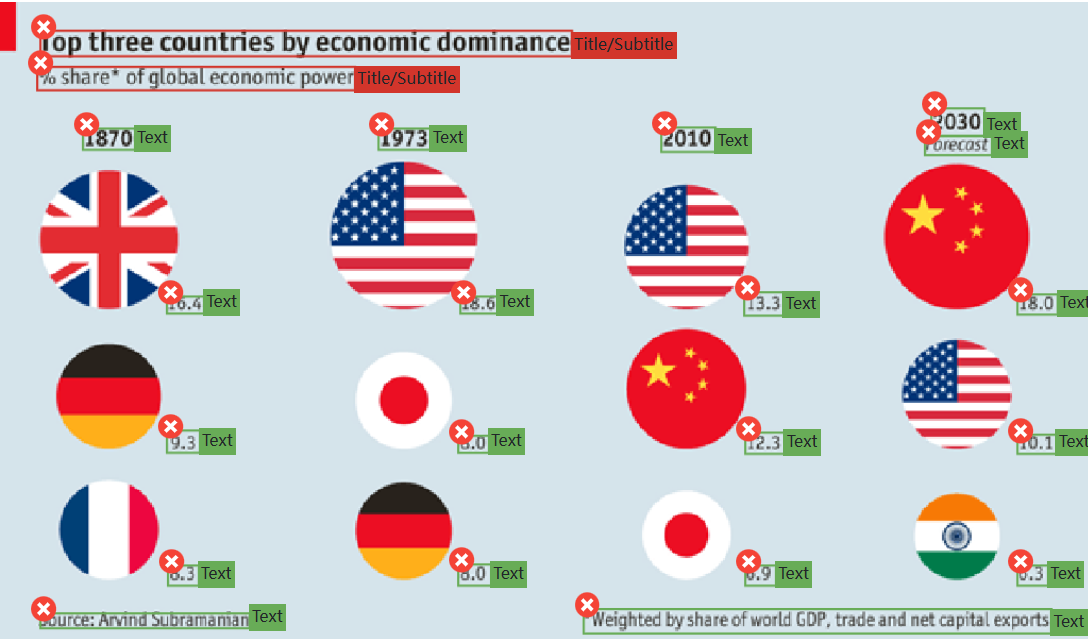}
        \caption{O.TiR=7.21\%}
    \end{subfigure}

    \begin{subfigure}{\picturewidth}
        \centering
        \includegraphics[height=\pictureheight]{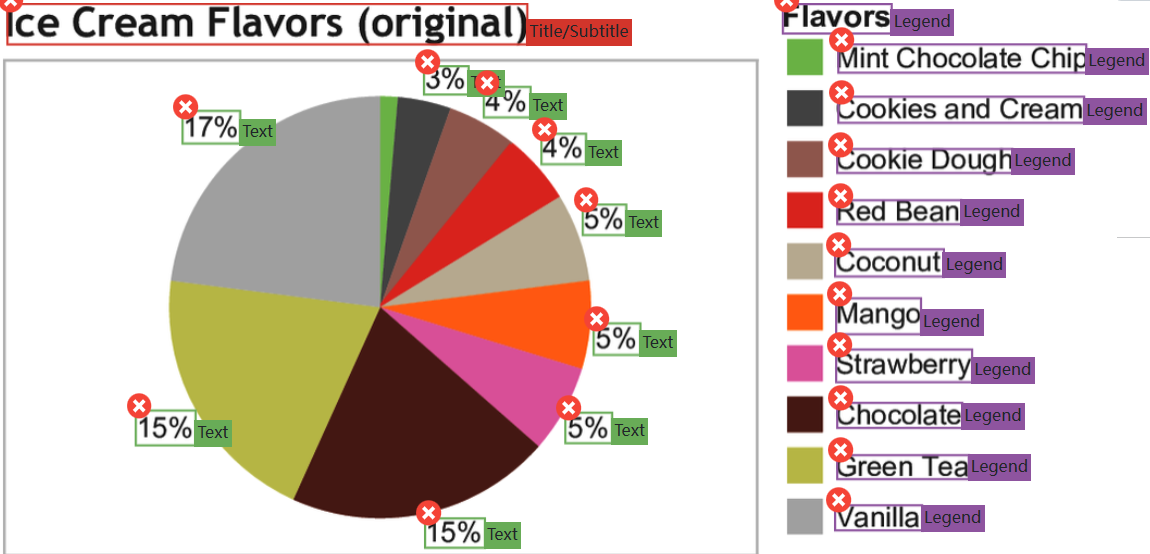}
        \caption{O.TiR=12.35\%}
    \end{subfigure}
    \begin{subfigure}{\picturewidth}
        \centering
        \includegraphics[height=\pictureheight]{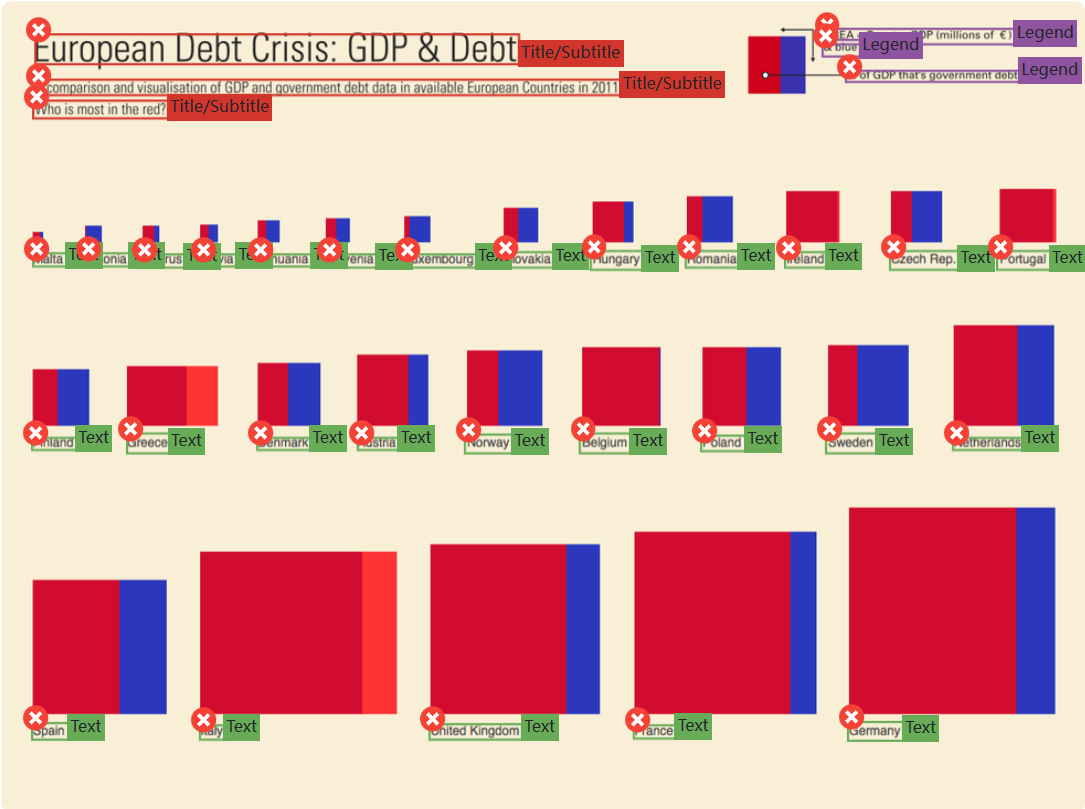}
        \caption{O.TiR=5.88\%}
    \end{subfigure}
    \begin{subfigure}{\picturewidth}
        \centering
        \includegraphics[height=\pictureheight]{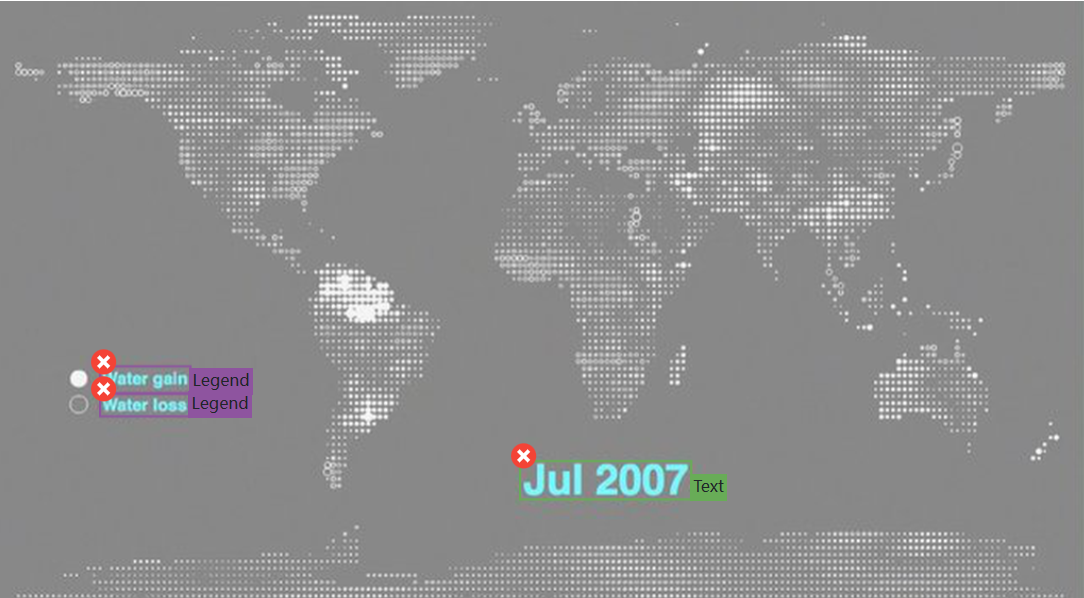}
        \caption{O.TiR=1.68\%}
    \end{subfigure}

    \begin{subfigure}{\picturewidth}
        \centering
        \includegraphics[height=\pictureheight]{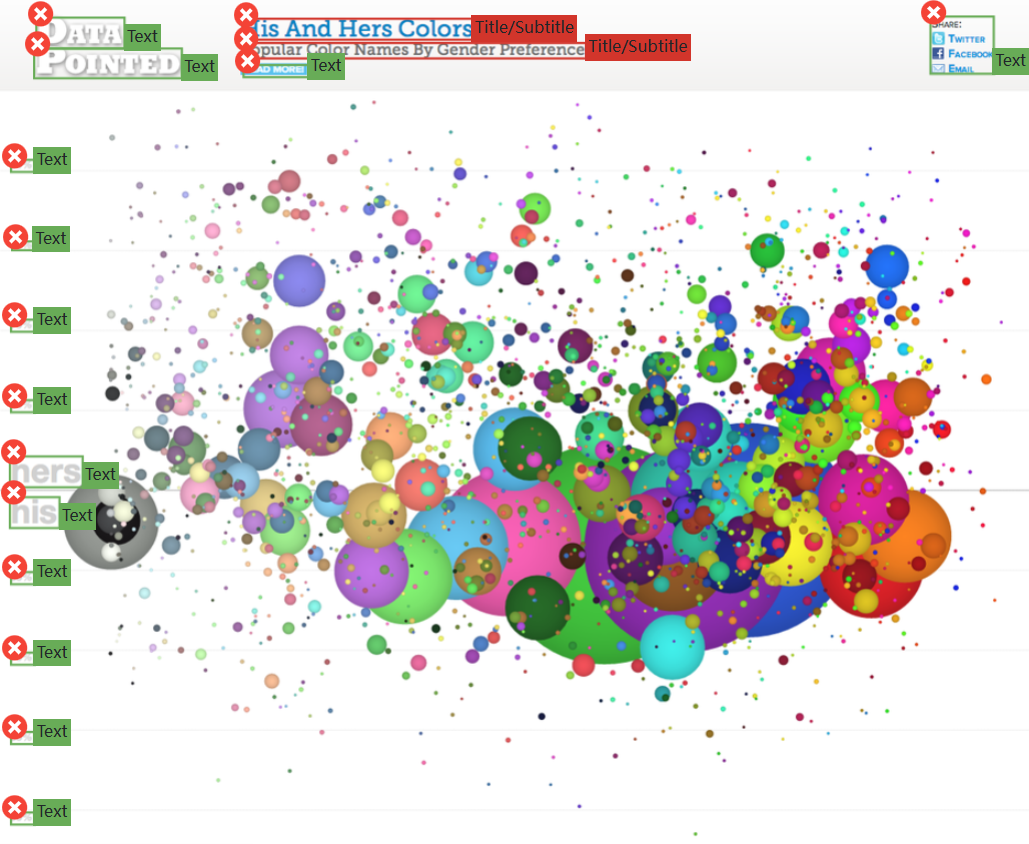}
        \caption{O.TiR=3.02\%}
    \end{subfigure}
    \begin{subfigure}{\picturewidth}
        \centering
        \includegraphics[height=\pictureheight]{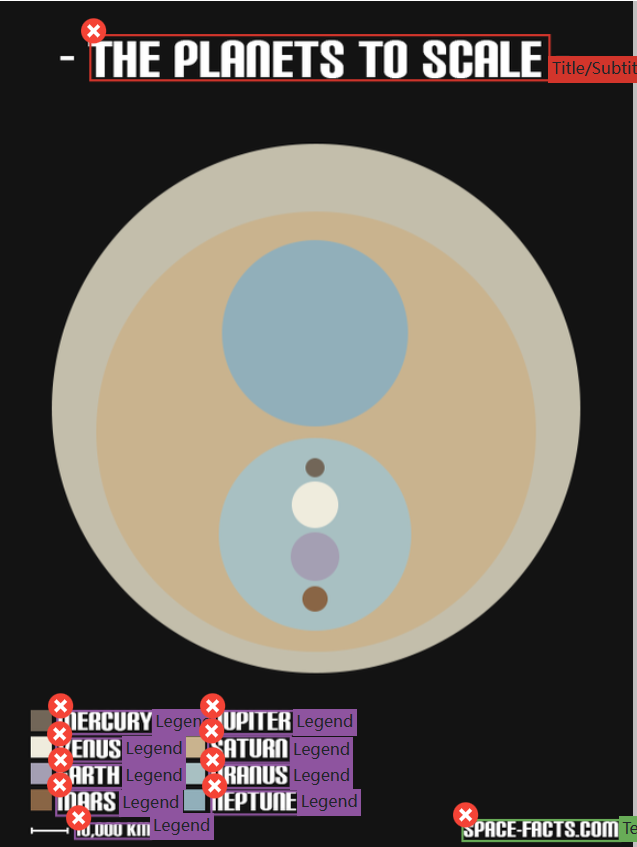}
        \caption{O.TiR=7.49\%}
    \end{subfigure}
    \begin{subfigure}{\picturewidth}
        \centering
        \includegraphics[height=\pictureheight]{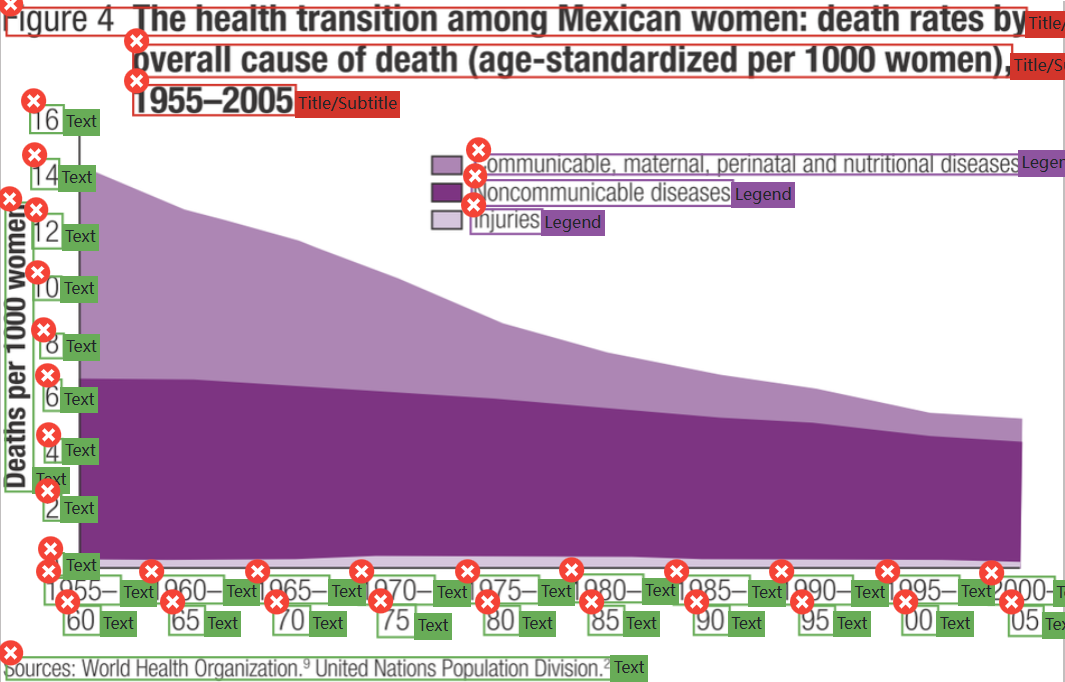}
        \caption{O.TiR=19.57\%}
    \end{subfigure}

    \begin{subfigure}{\picturewidth}
        \centering
        \includegraphics[height=\pictureheight]{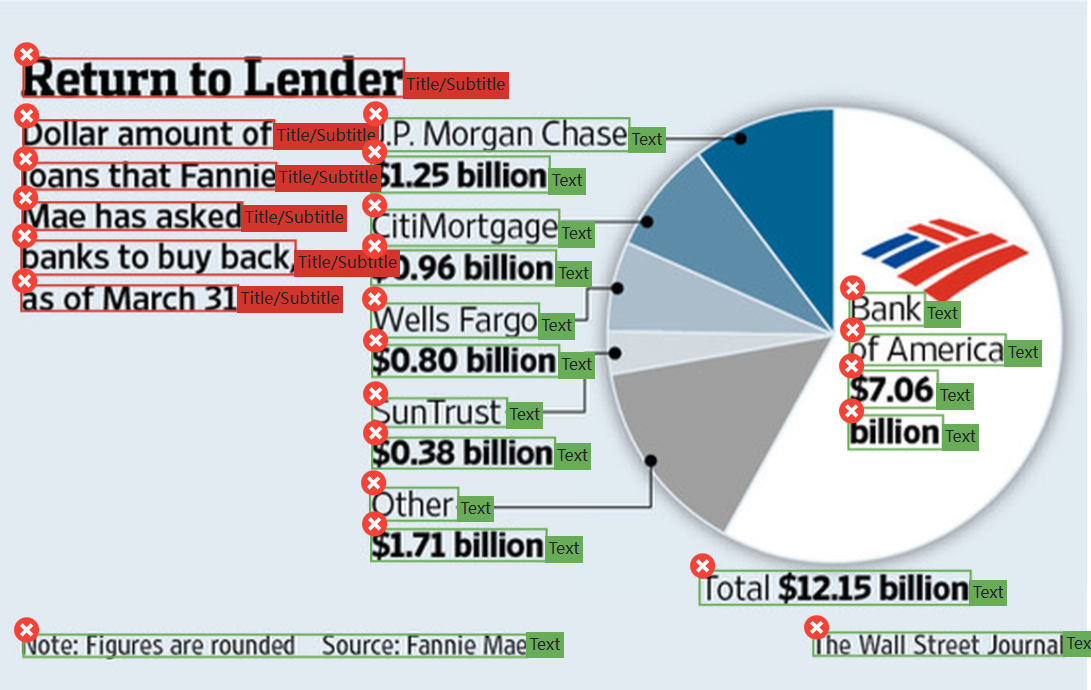}
        \caption{O.TiR=19.52\%}
    \end{subfigure}
    \begin{subfigure}{\picturewidth}
        \centering
        \includegraphics[height=\pictureheight]{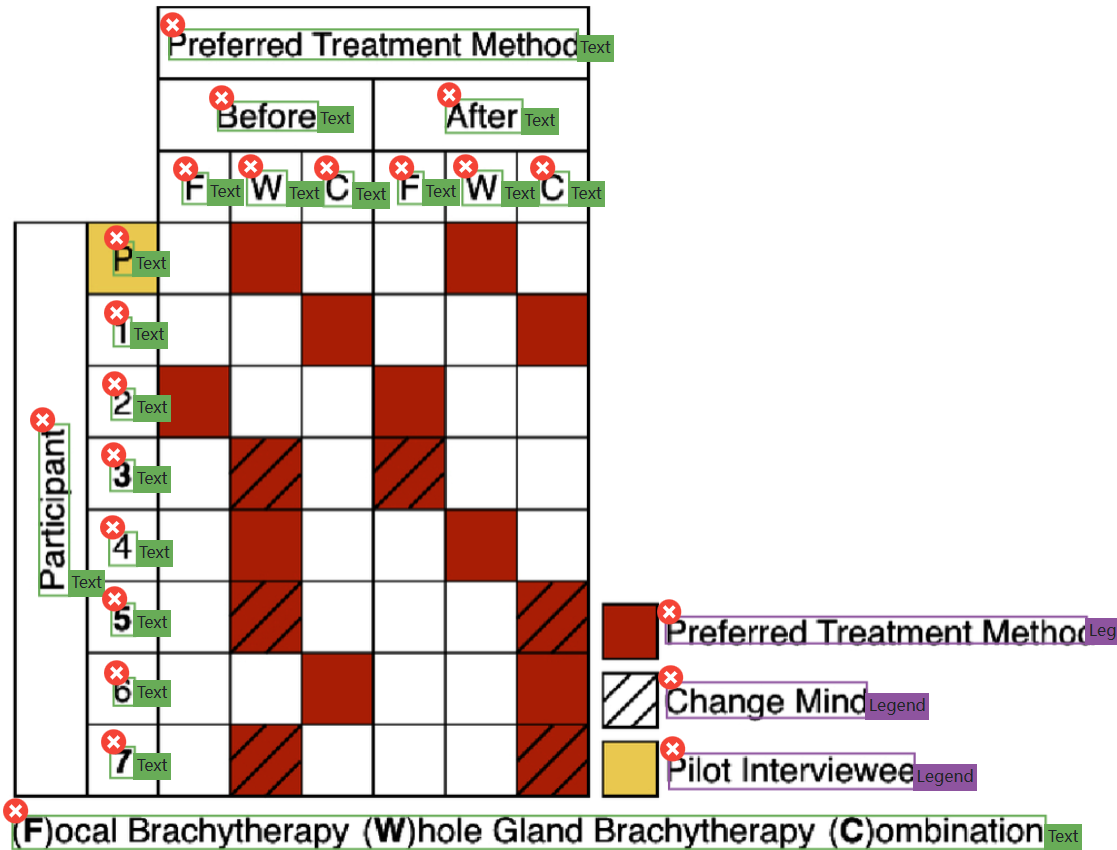}
        \caption{O.TiR=10.41\%}
    \end{subfigure}
    \begin{subfigure}{\picturewidth}
        \centering
        \includegraphics[height=\pictureheight]{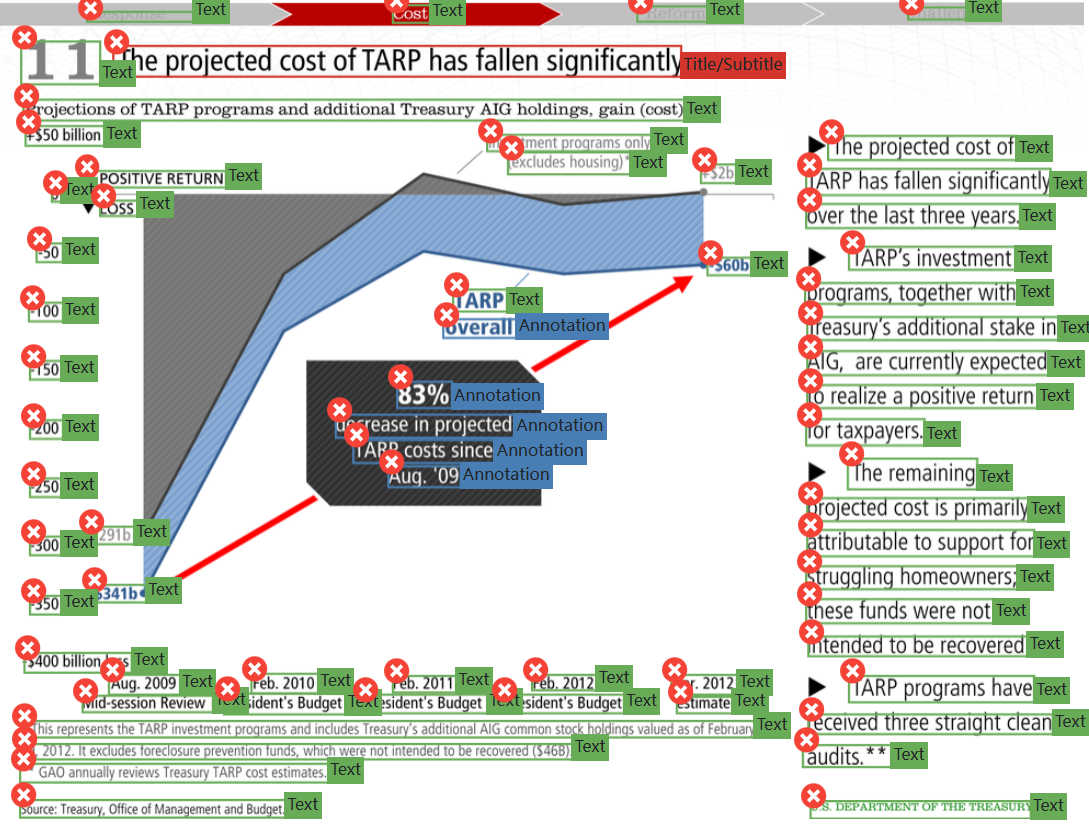}
        \caption{O.TiR=23.07\%}
    \end{subfigure}

    \caption{\textbf{O.TiR examples.} Each subcaption shows its O.TiR value. Additional results for computing O.TiR for all images in \vcdataset are available in the corresponding Google Sheet \vcdatalink, under the tab \texttt{O.TiR} (see main text \autoref{sec:objMetrics}).}
    \label{fig:tir_bbox_examples}
\end{figure*}

\begin{figure*}[!t]
    \centering
    \begin{subfigure}[b]{\textwidth}

        \includegraphics[width=\textwidth]{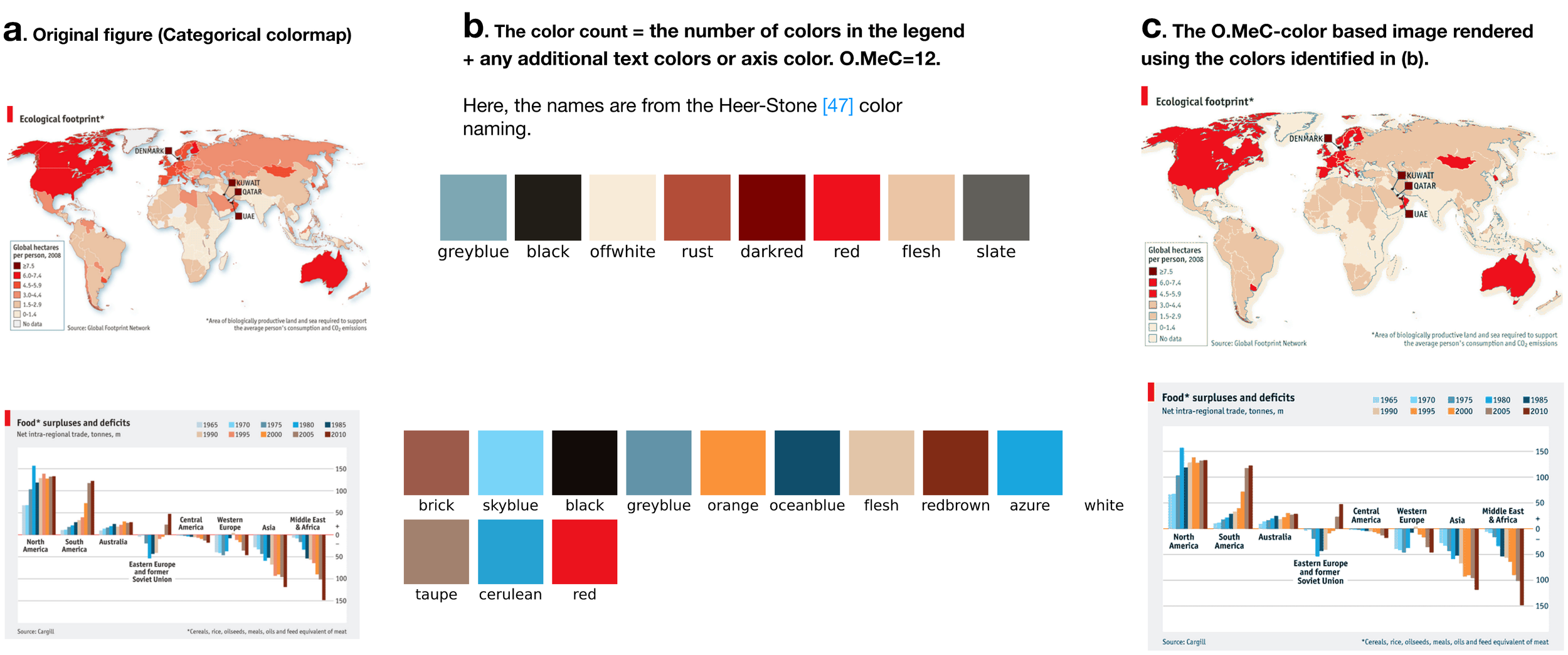}
    \end{subfigure}

    \hrulefill \vspace{10pt}
    \begin{subfigure}[b]{\textwidth}
        \includegraphics[width=\textwidth]{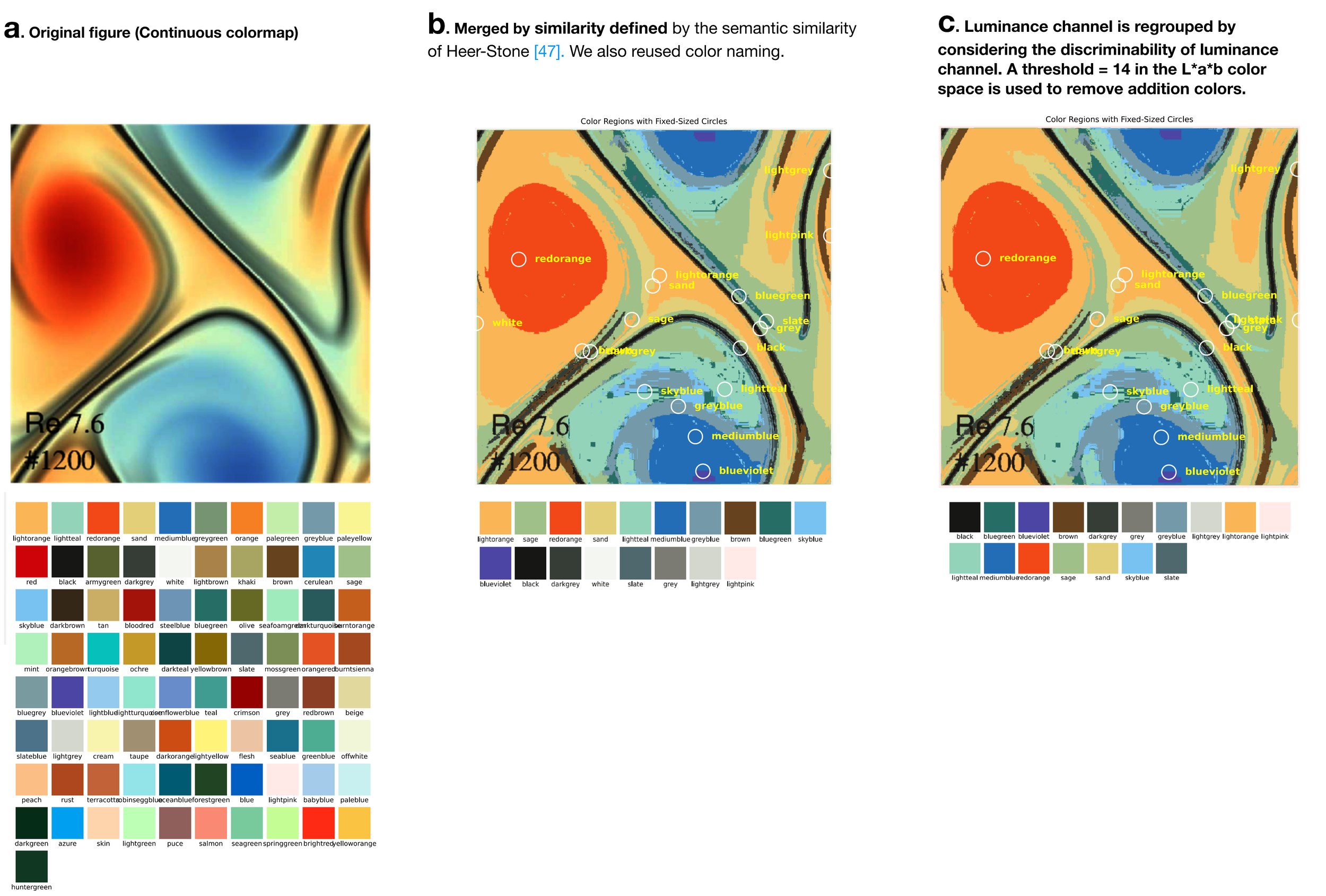}
    \end{subfigure}
    \caption{\textbf{The O.MeC metric computational workflow.} \textit{Top}: categorical color map processing. \textit{Bottom}: continuous color map processing. Step-by-step results for computing O.MeC on both continuous and discrete color images in \vcdataset are available in the corresponding Google Sheet \vcdatalink, under the tab \texttt{O.MeCResults} (see main text \autoref{sec:objMetrics}). }
    \label{fig:ProcessingMeCWorkflow}
\end{figure*}

\begin{figure*}[!t]
    \centering
    \setlength{\pictureheight}{3.5cm}

    \begin{subfigure}{0.32\textwidth}
        \centering
        \includegraphics[height=\pictureheight]{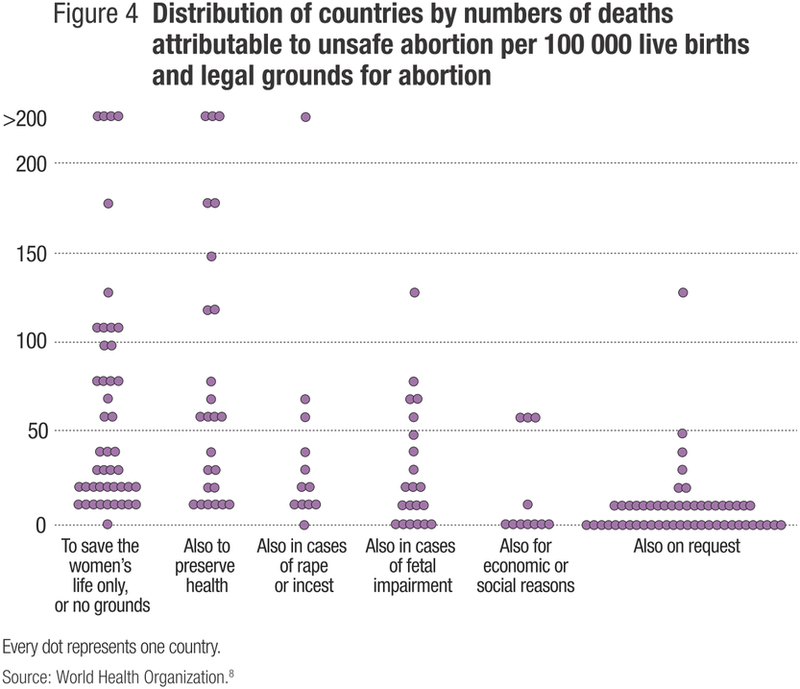}
        \caption{O.MeC=2}
        \label{MEC:fig1}
    \end{subfigure}
    \begin{subfigure}{0.32\textwidth}
        \centering
        \includegraphics[height=\pictureheight]{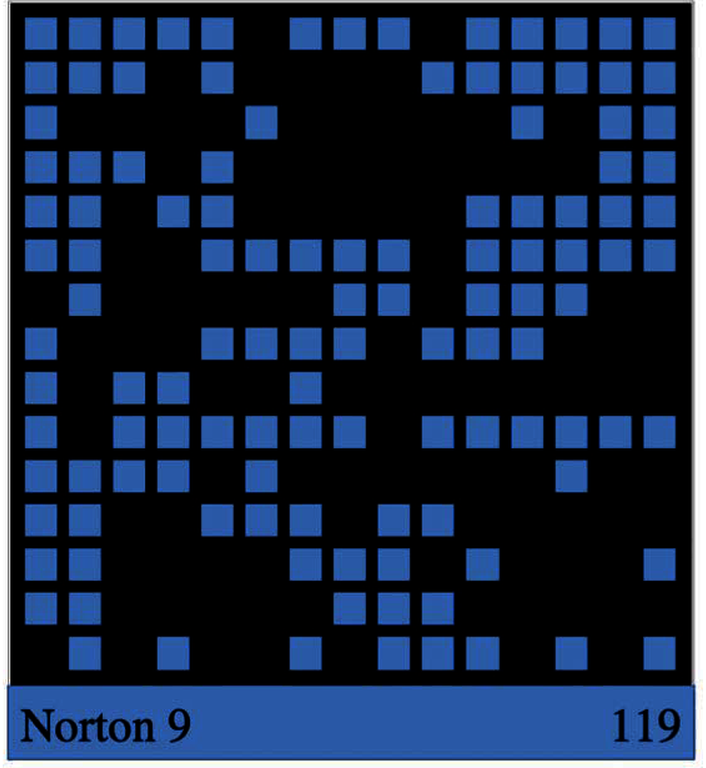}
        \caption{O.MeC=2}
        \label{MEC:fig2}
    \end{subfigure}
    \begin{subfigure}{0.32\textwidth}
        \centering
        \includegraphics[height=\pictureheight]{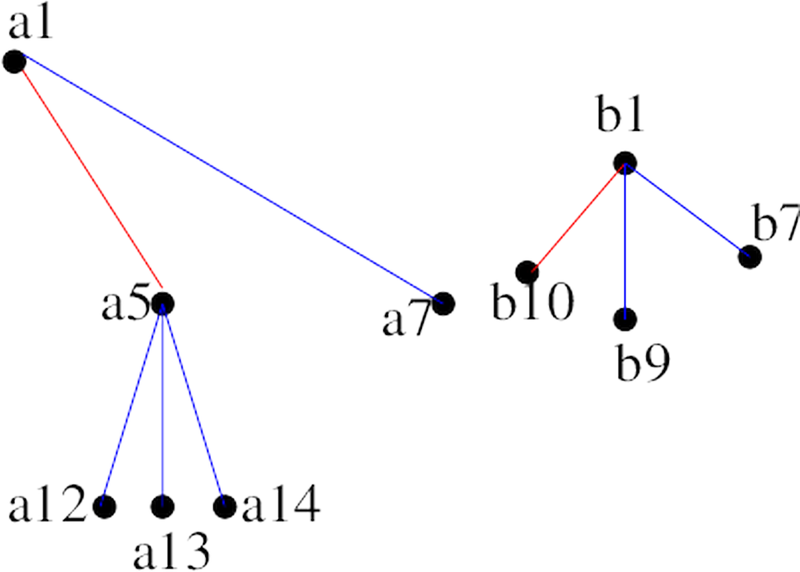}
        \caption{O.MeC=3}
        \label{MEC:fig3}
    \end{subfigure}

    \begin{subfigure}{0.32\textwidth}
        \centering
        \includegraphics[height=\pictureheight]{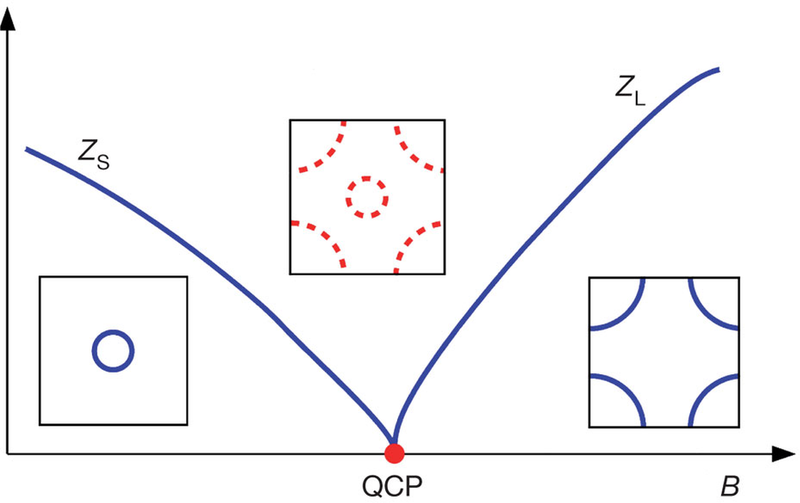}
        \caption{O.MeC=3}
        \label{MEC:fig4}
    \end{subfigure}
    \begin{subfigure}{0.32\textwidth}
        \centering
        \includegraphics[height=\pictureheight]{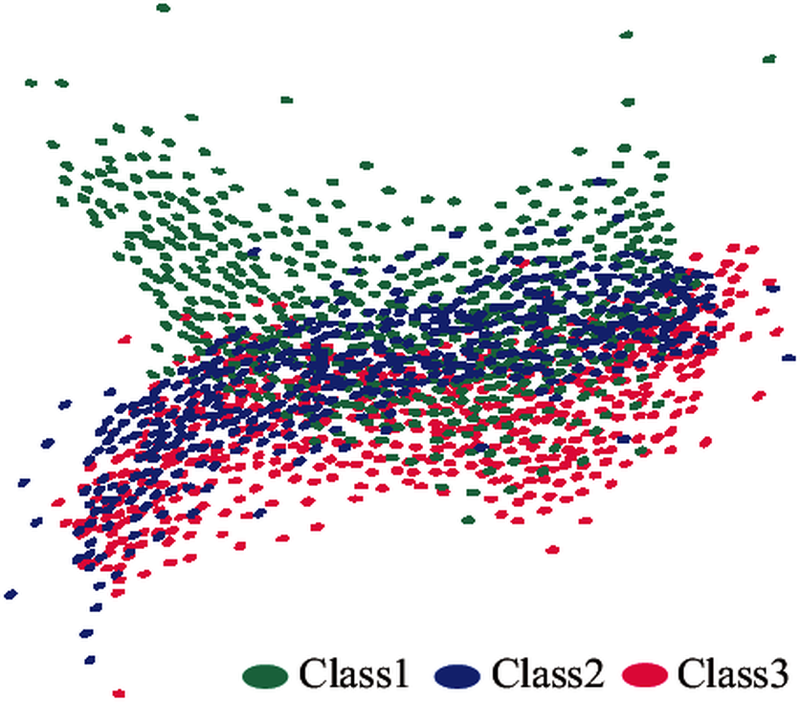}
        \caption{O.MEC=4}
        \label{MEC:fig5}
    \end{subfigure}
    \begin{subfigure}{0.32\textwidth}
        \centering
        \includegraphics[height=\pictureheight]{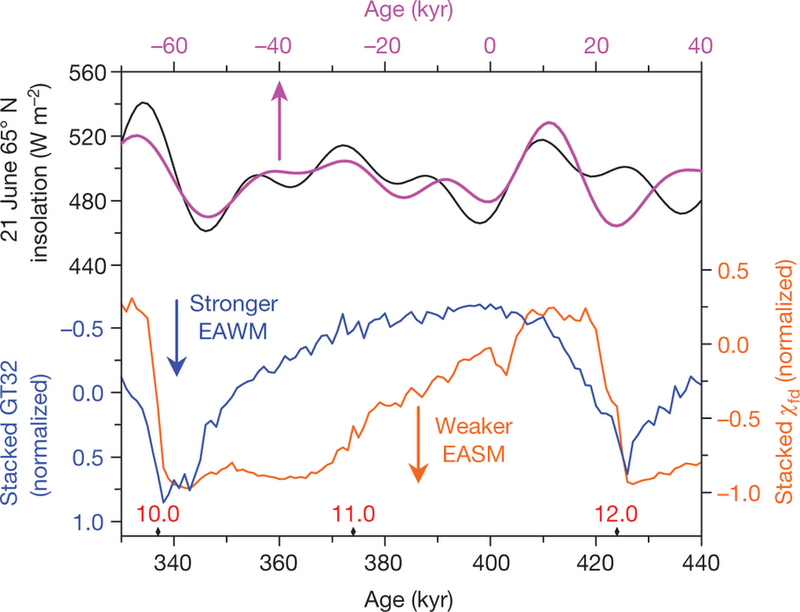}
        \caption{O.MeC=5}
        \label{MEC:fig6}
    \end{subfigure}

    \begin{subfigure}{0.32\textwidth}
        \centering
        \includegraphics[height=\pictureheight]{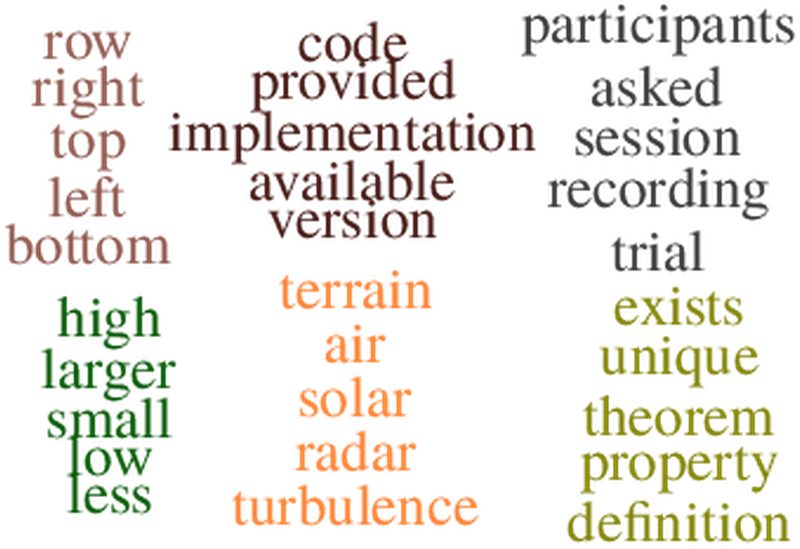}
        \caption{O.MeC=6}
        \label{MEC:fig7}
    \end{subfigure}
    \begin{subfigure}{0.32\textwidth}
        \centering
        \includegraphics[height=\pictureheight]{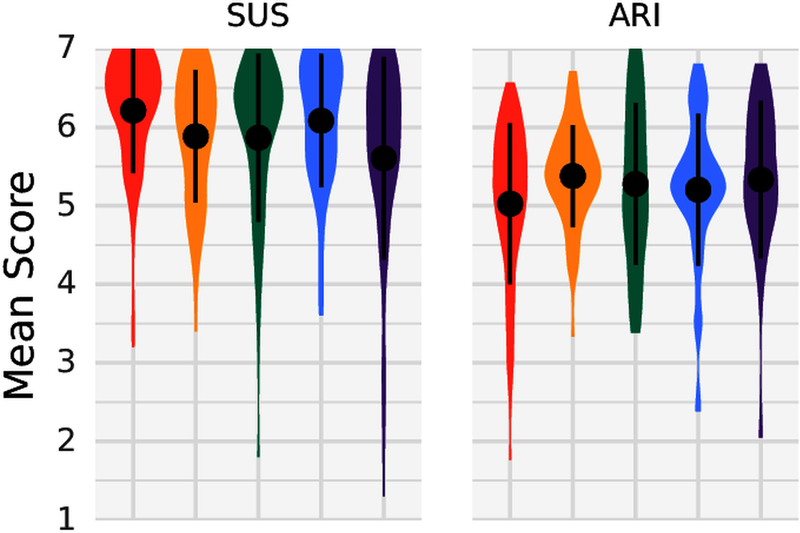}
        \caption{O.MeC=7}
        \label{MEC:fig8}
    \end{subfigure}
    \begin{subfigure}{0.32\textwidth}
        \centering
        \includegraphics[height=\pictureheight]{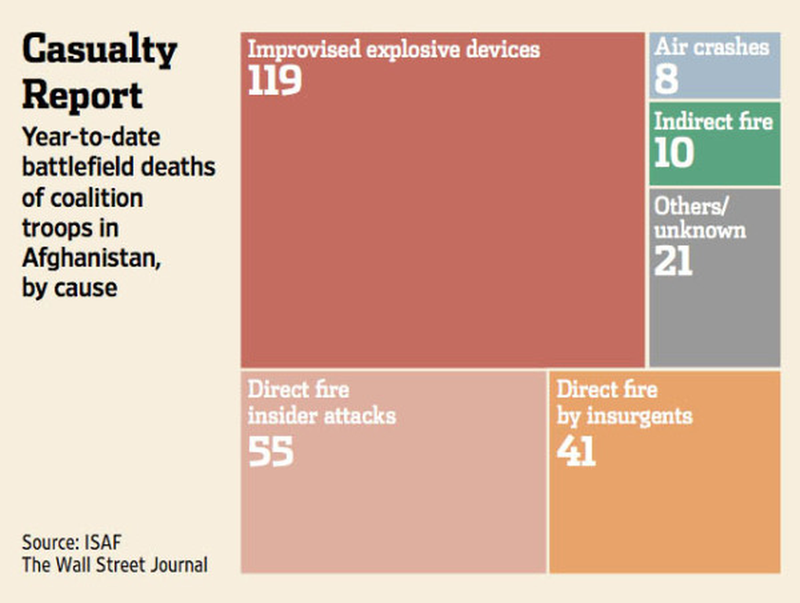}
        \caption{O.MeC=9}
        \label{MEC:fig9}
    \end{subfigure}
    \begin{subfigure}{0.32\textwidth}
        \centering
        \includegraphics[height=\pictureheight]{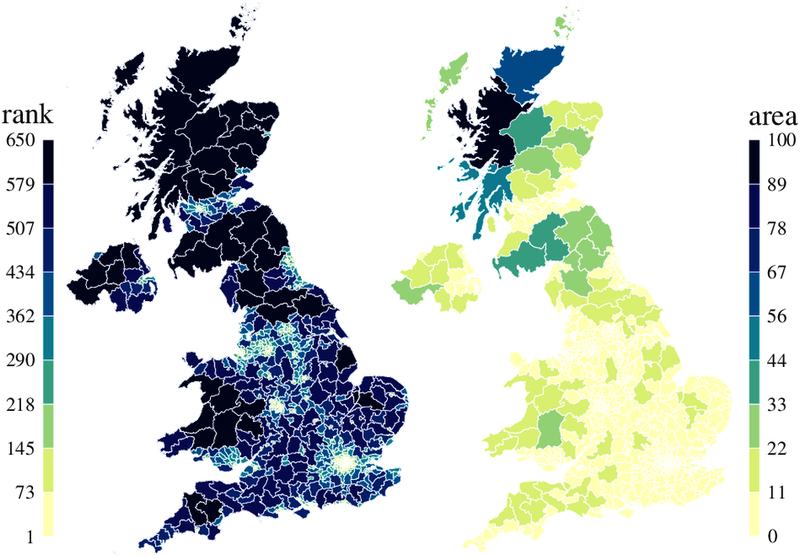}
        \caption{O.MeC=10}
        \label{MEC:fig10}
    \end{subfigure}
    \begin{subfigure}{0.32\textwidth}
        \centering
        \includegraphics[height=\pictureheight]{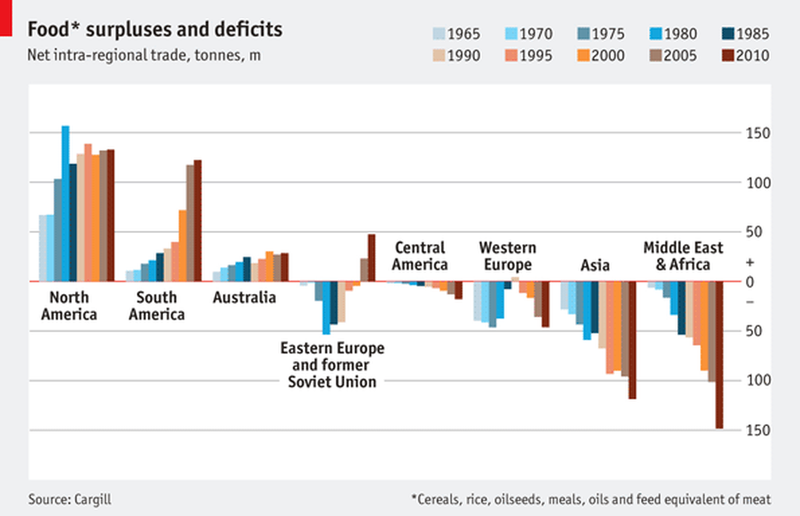}
        \caption{O.MeC=13}
        \label{MEC:fig11}
    \end{subfigure}
    \begin{subfigure}{0.32\textwidth}
        \centering
        \includegraphics[height=\pictureheight]{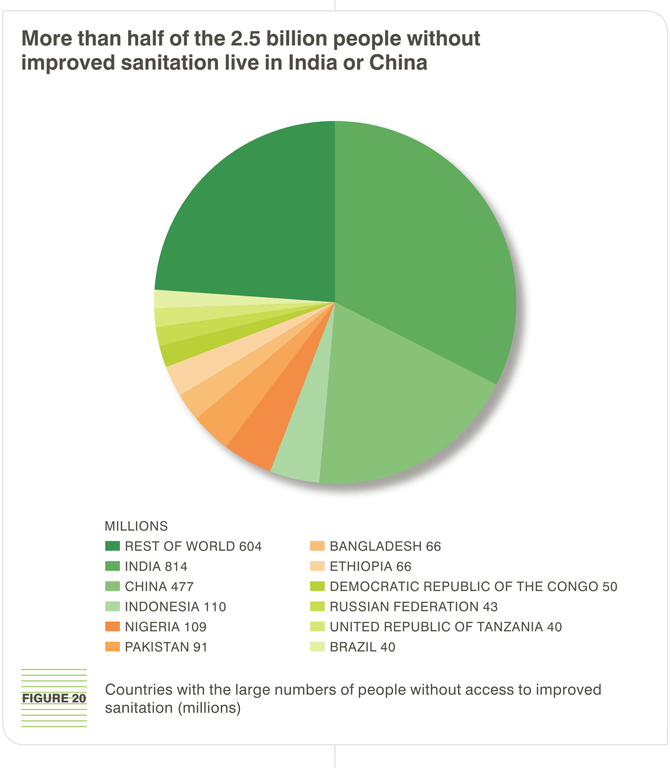}
        \caption{O.MeC=14}
        \label{MEC:fig12}
    \end{subfigure}

    \caption{More O.MeC examples for \textbf{discrete} representations. Subcaptions contain the O.MeC number (see main text \autoref{sec:objMetrics}).}
    \label{fig:discrete_mec}
\end{figure*}

\begin{figure*}[!t]
    \centering
    \setlength{\pictureheight}{3cm}
    \begin{subfigure}{0.32\textwidth}
        \centering
        \includegraphics[height=\pictureheight]{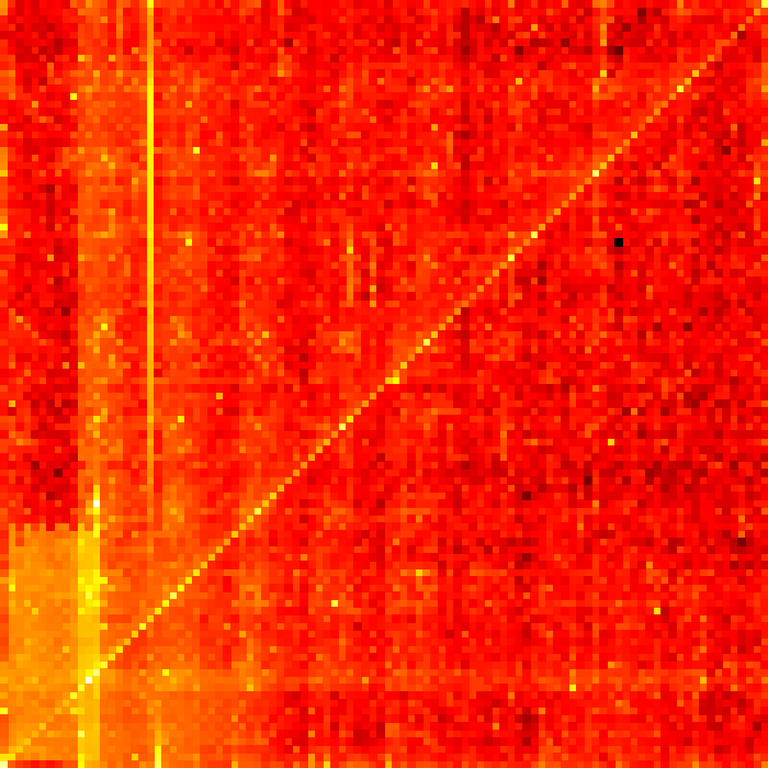}
        \caption{O.MeC=4}
        \label{MEC_c:fig1}
    \end{subfigure}
    \begin{subfigure}{0.32\textwidth}
        \centering
        \includegraphics[height=\pictureheight]{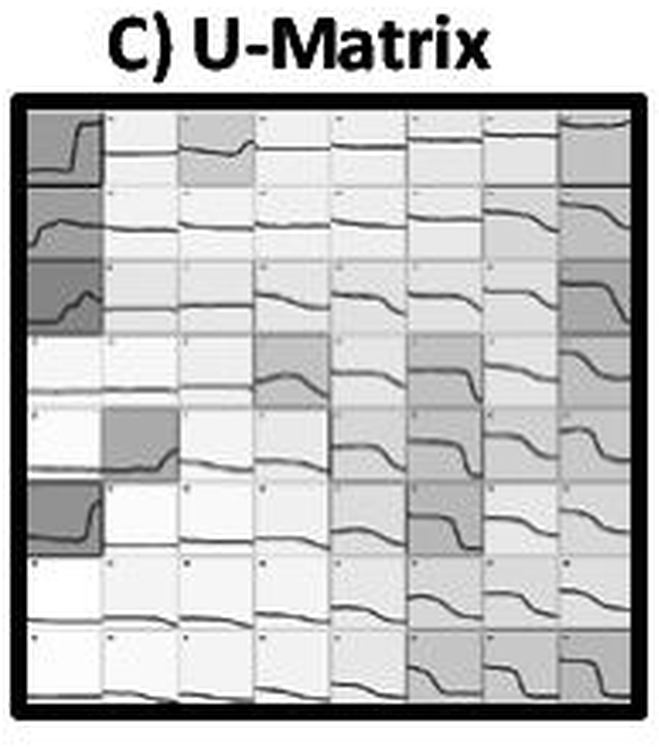}
        \caption{O.MeC=5}
        \label{MEC_c:fig2}
    \end{subfigure}
    \begin{subfigure}{0.32\textwidth}
        \centering
        \includegraphics[height=\pictureheight]{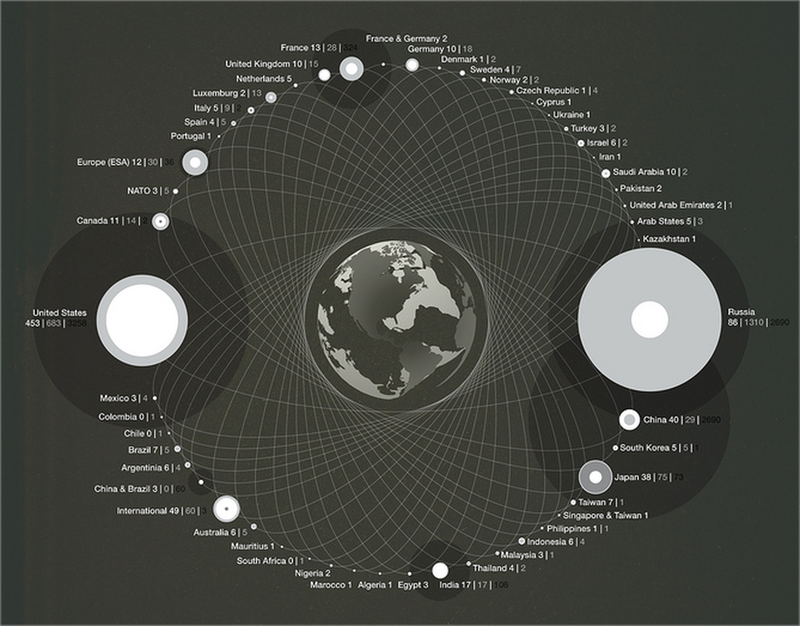}
        \caption{O.MeC=7}
        \label{MEC_c:fig3}
    \end{subfigure}
    \begin{subfigure}{0.32\textwidth}
        \centering
        \includegraphics[height=\pictureheight]{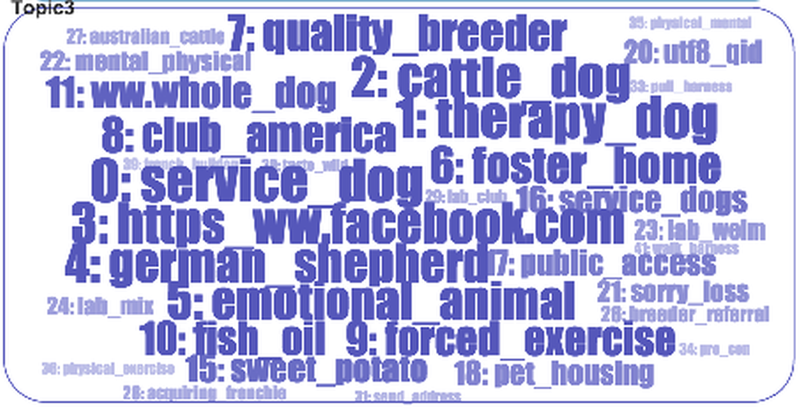}
        \caption{O.MeC=8}
        \label{MEC_c:fig4}
    \end{subfigure}
    \begin{subfigure}{0.32\textwidth}
        \centering
        \includegraphics[height=\pictureheight]{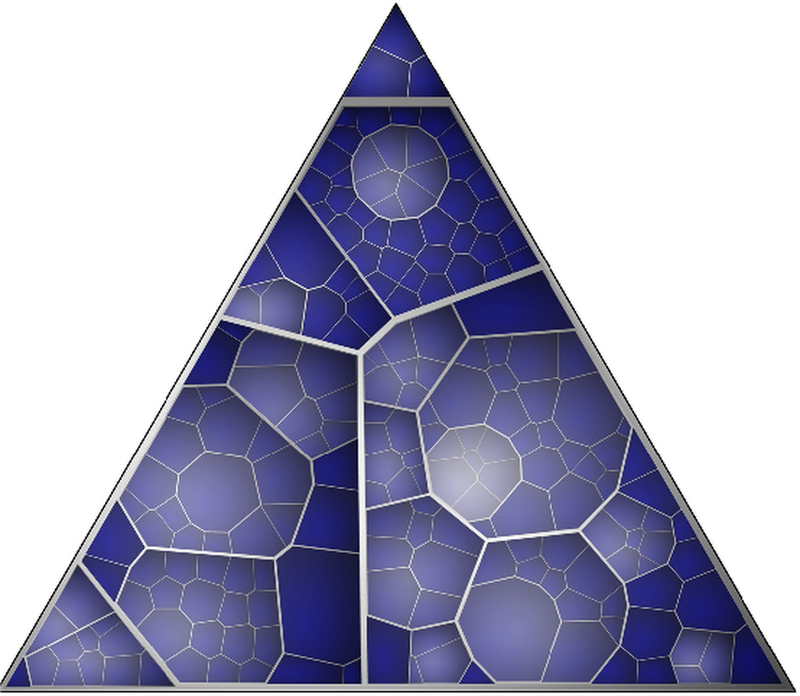}
        \caption{O.MeC=11}
        \label{MEC_c:fig5}
    \end{subfigure}
    \begin{subfigure}{0.32\textwidth}
        \centering
        \includegraphics[height=\pictureheight]{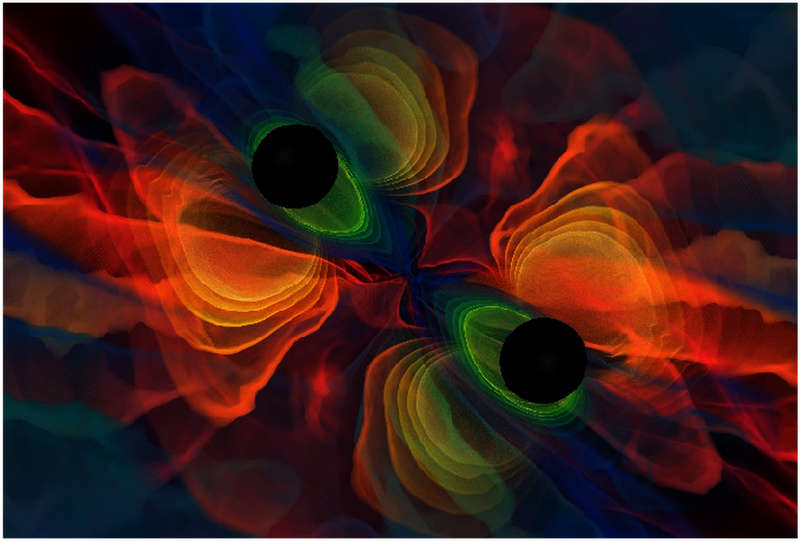}
        \caption{O.MeC=14}
        \label{MEC_c:fig6}
    \end{subfigure}
    \begin{subfigure}{0.32\textwidth}
        \centering
        \includegraphics[height=\pictureheight]{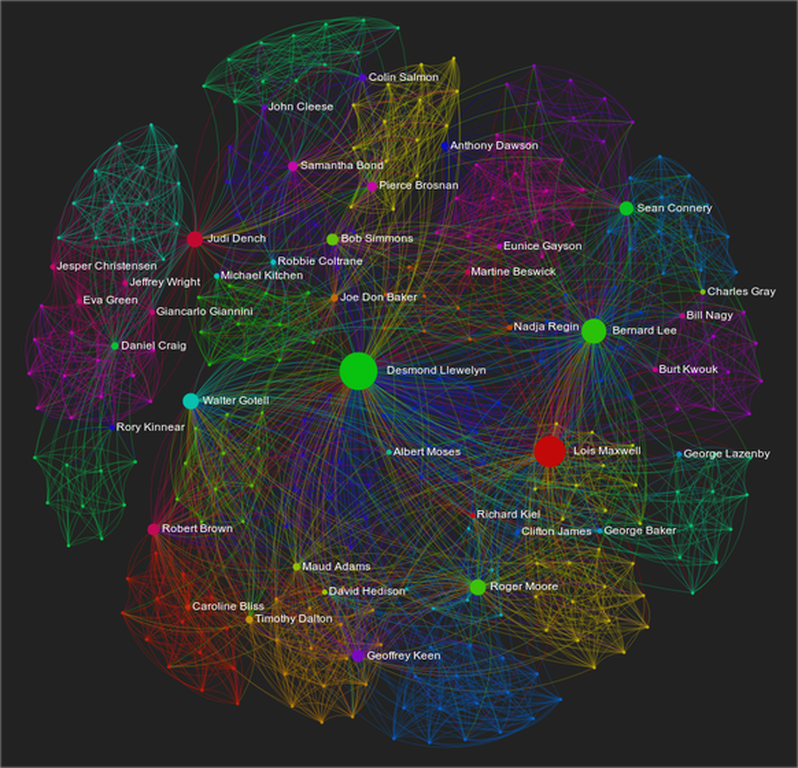}
        \caption{O.MeC=15}
        \label{MEC_c:fig7}
    \end{subfigure}
    \begin{subfigure}{0.32\textwidth}
        \centering
        \includegraphics[height=\pictureheight]{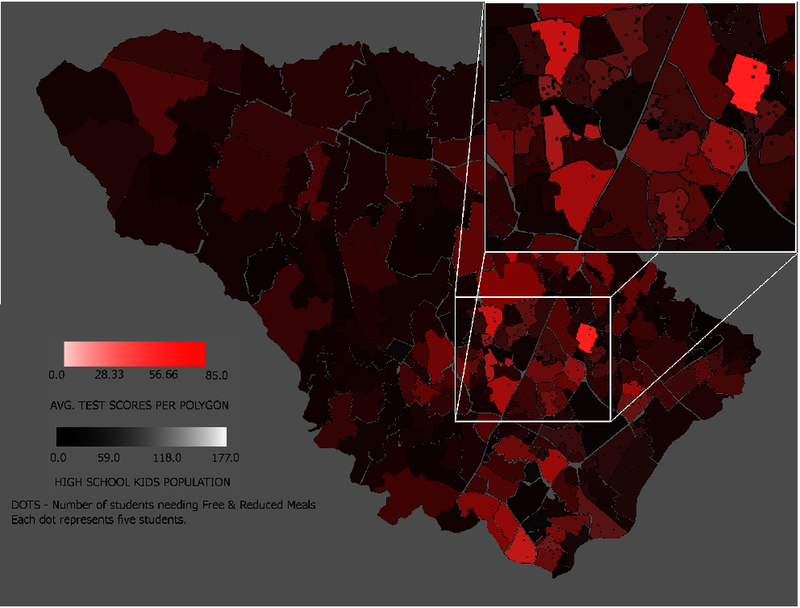}
        \caption{O.MeC=15}
        \label{MEC_c:fig8}
    \end{subfigure}
    \begin{subfigure}{0.32\textwidth}
        \centering
        \includegraphics[height=\pictureheight]{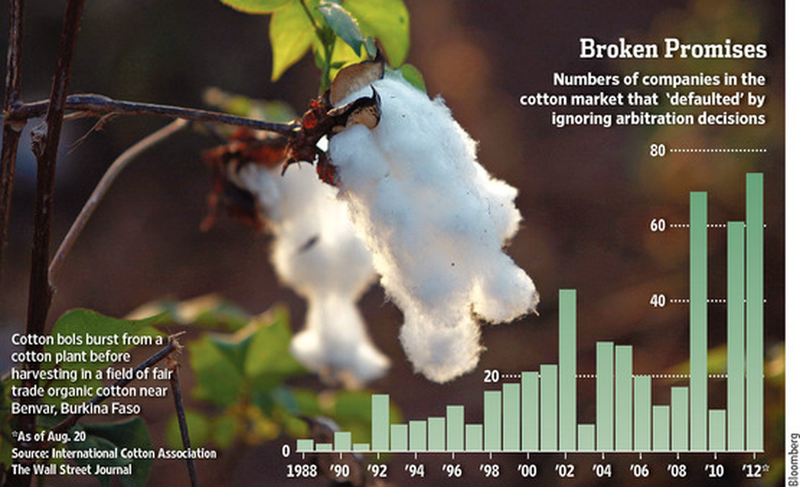}
        \caption{O.MeC=16}
        \label{MEC_c:fig9}
    \end{subfigure}
    \begin{subfigure}{0.32\textwidth}
        \centering
        \includegraphics[height=\pictureheight]{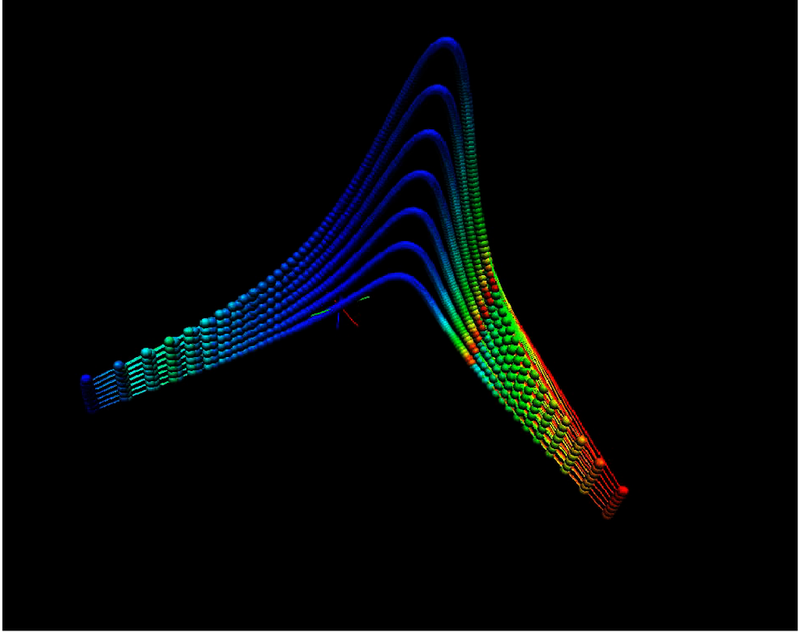}
        \caption{O.MeC=18}
        \label{MEC_c:fig10}
    \end{subfigure}
    \begin{subfigure}{0.32\textwidth}
        \centering
        \includegraphics[height=\pictureheight]{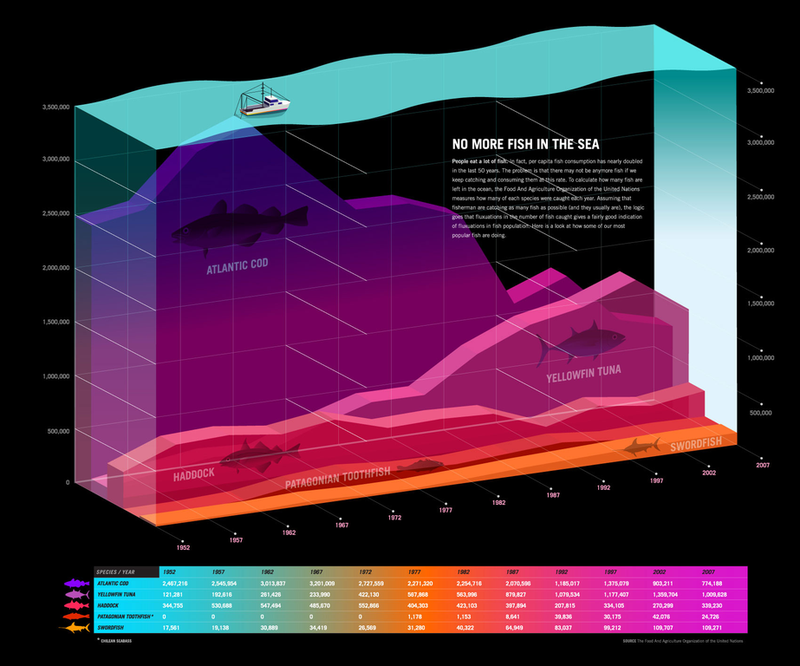}
        \caption{O.MeC=22}
        \label{MEC_c:fig11}
    \end{subfigure}
    \begin{subfigure}{0.32\textwidth}
        \centering
        \includegraphics[height=\pictureheight]{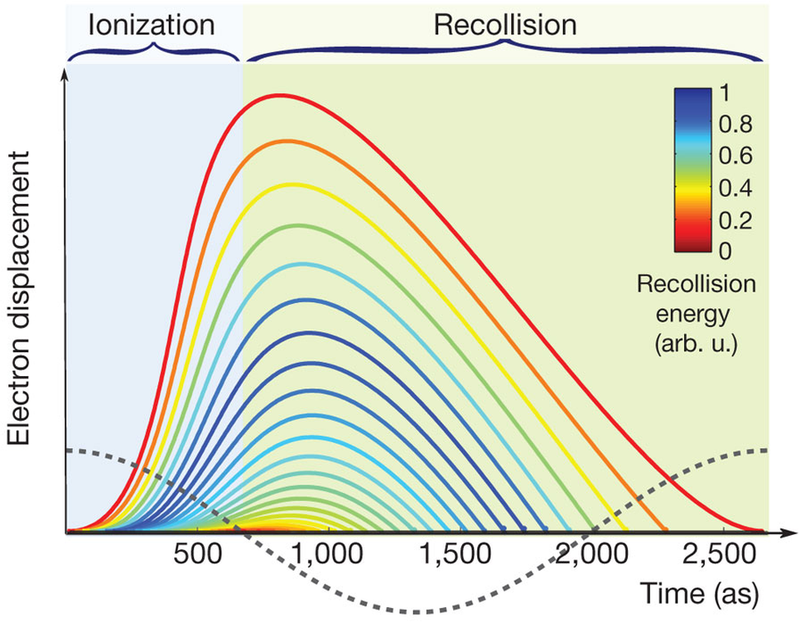}
        \caption{O.MeC=24}
        \label{MEC_c:fig12}
    \end{subfigure}

    \caption{More O.MeC examples for \textbf{continuous} representations. Subcaptions contain the O.MeC number (see main text \autoref{sec:objMetrics}). }
    \label{MEC_c_c}
\end{figure*}

\begin{figure*}[!t]
    \centering
    \includegraphics[width=0.9\columnwidth]{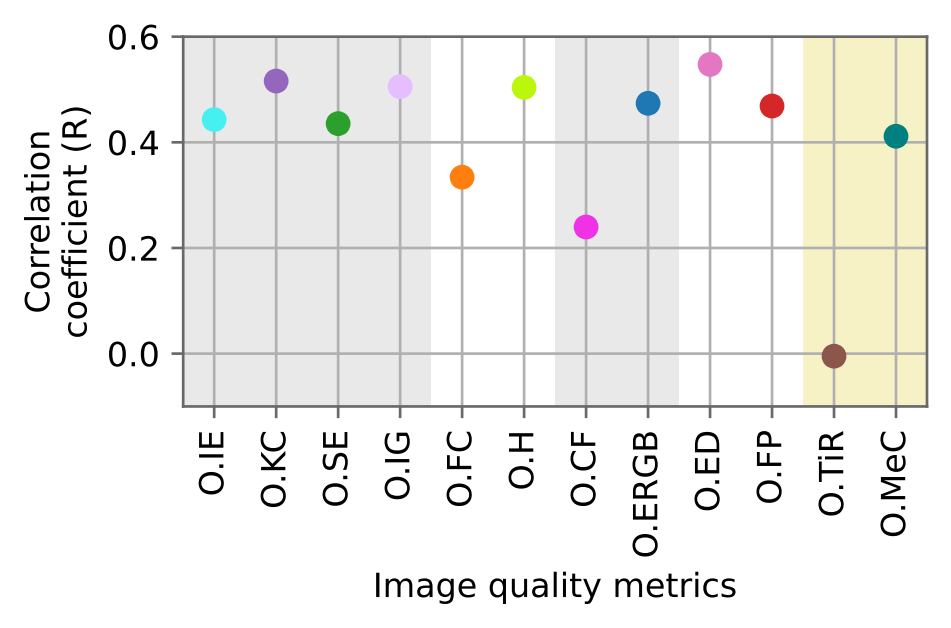}
    \caption{\textbf{Single variable analysis: correlation between each metric and perceived VC.} Pearson correlation coefficients ($r$) were used to assess the linear relationships between each objective metric and perceived VC. The associated effect size is typically categorized as small ($r\ge0.10$), medium ($r\ge0.30$), and large ($r\ge0.50$).
    \textbf{Observation.} All but TiR metrics were positively correlated, with most showing moderate to strong correlations. Between the two object-based metrics, O.MeC demonstrated a moderate correlation with perceived VC ($r=0.41$), whereas O.TiR showed a negative but statistically insignificant correlation with perceived VC ($r=-0.01, p=0.56$) (see the main text ~\autoref{sec:12metrics}). }
    \label{fig:corrFeatureVC}
\end{figure*}

\begin{figure*}[!t]
    \centering
    \includegraphics[width=\textwidth]{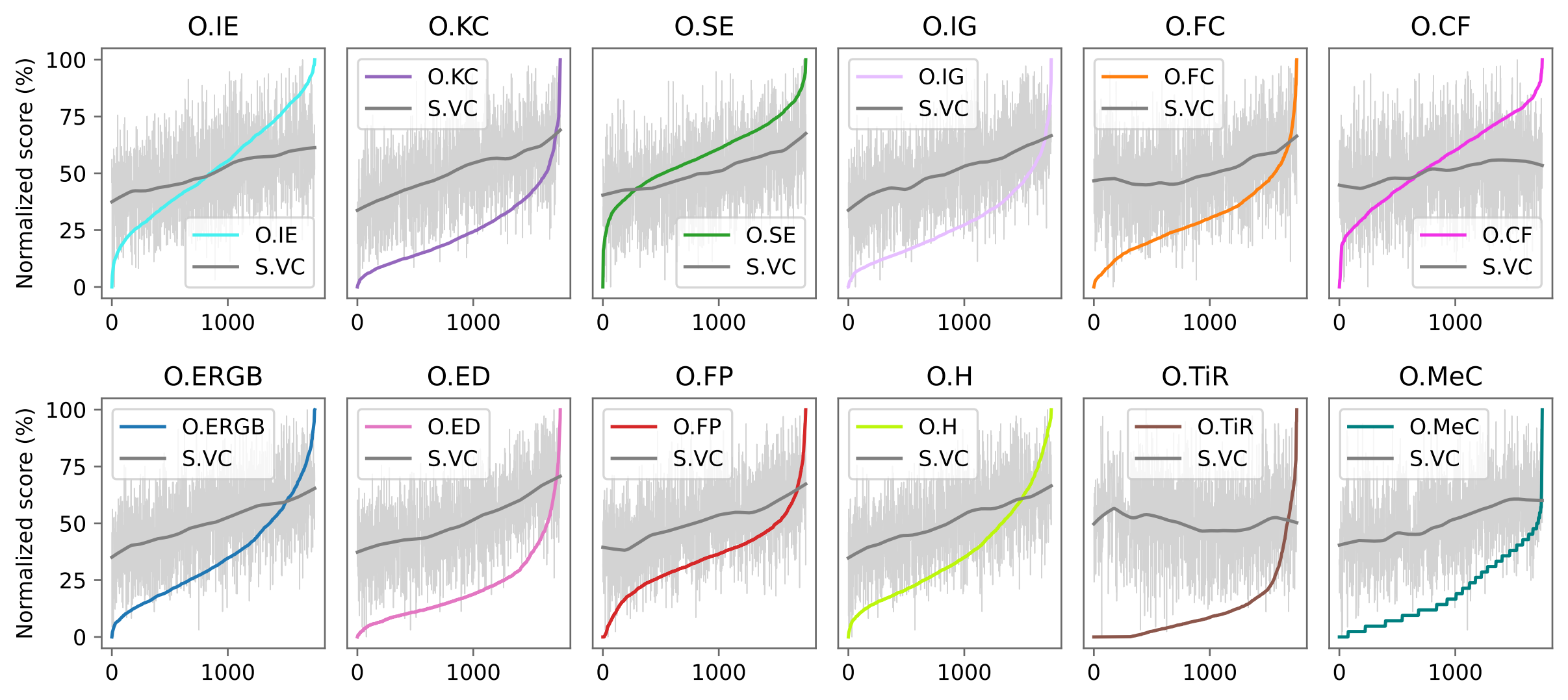}\caption{\textbf{Metrics vs. perceived VC.} The colored lines show the metric values in ascending order for the 1800 images, while the light-gray lines in the background are perceived VC scores, overlaid by a locally weighted smoothing line in gray (see the main text ~\autoref{sec:12metrics}). }
    \label{fig:visComplexityByType}
\end{figure*}

\begin{figure*}[!t]
    \centering
    \setlength{\pictureheight}{4.5cm}

    \begin{subfigure}{0.23\textwidth}
        \centering
        \includegraphics[height=\pictureheight]{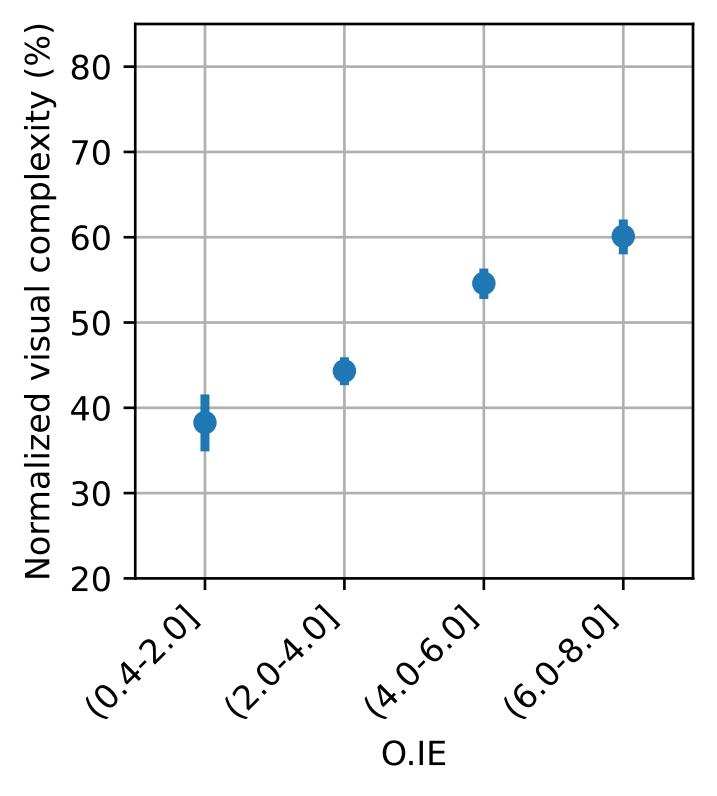}
    \end{subfigure}
    \begin{subfigure}{0.23\textwidth}
        \centering
        \includegraphics[height=\pictureheight]{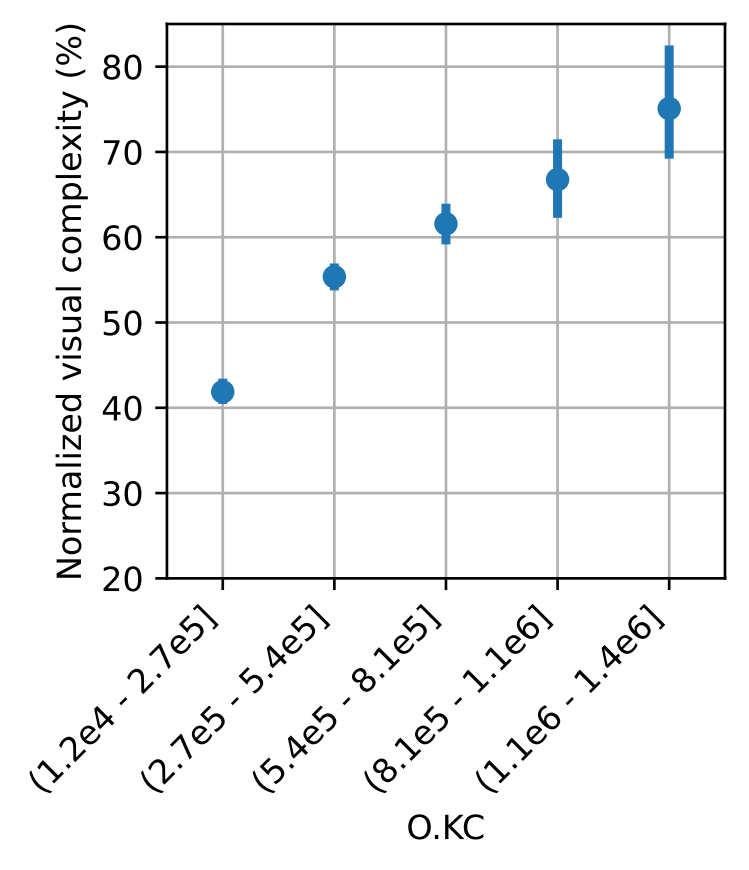}
    \end{subfigure}
    \begin{subfigure}{0.23\textwidth}
        \centering \includegraphics[height=\pictureheight]{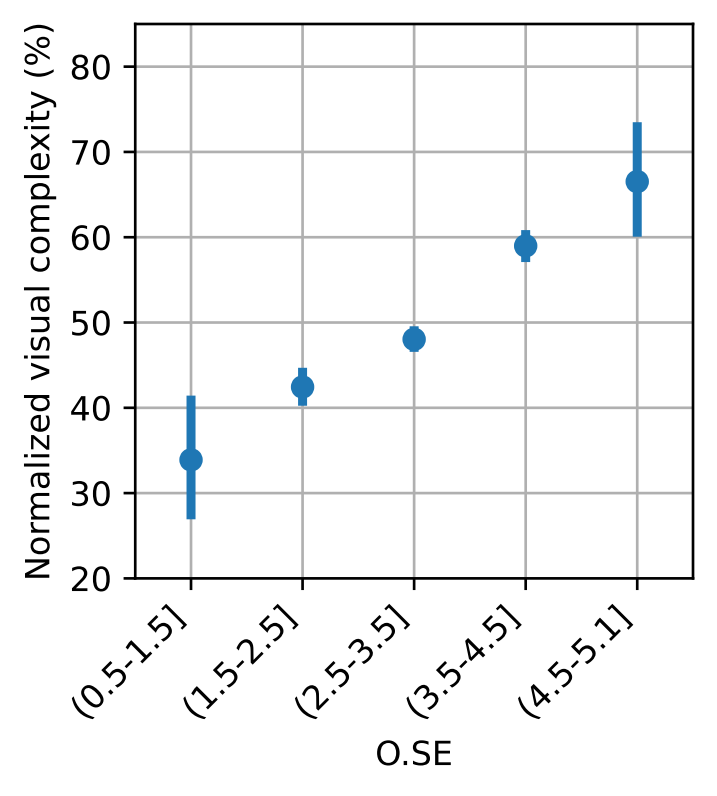}
    \end{subfigure}
    \begin{subfigure}{0.23\textwidth}
        \centering
        \includegraphics[height=\pictureheight]{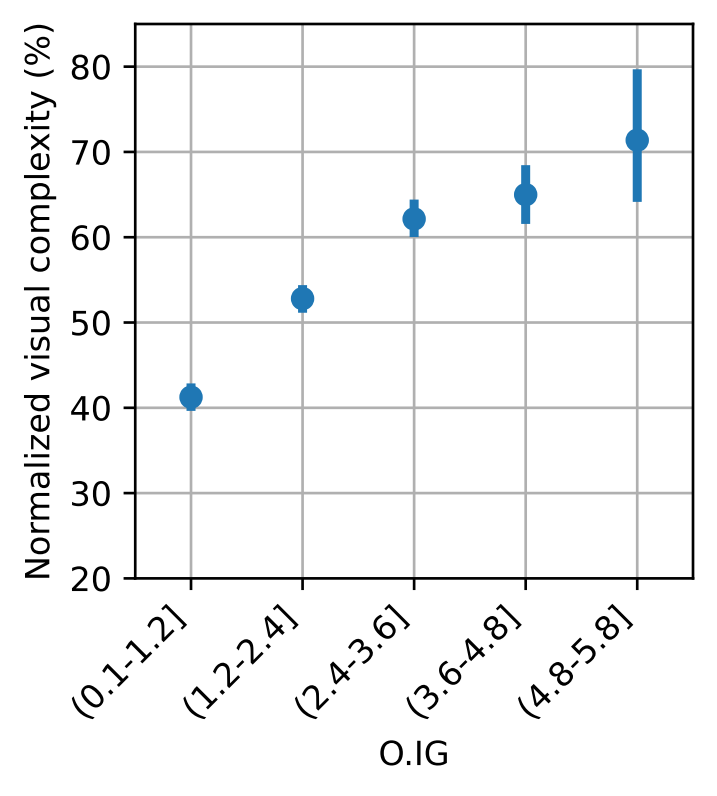}
    \end{subfigure}
    \begin{subfigure}{0.23\textwidth}
        \centering
        \includegraphics[height=\pictureheight]{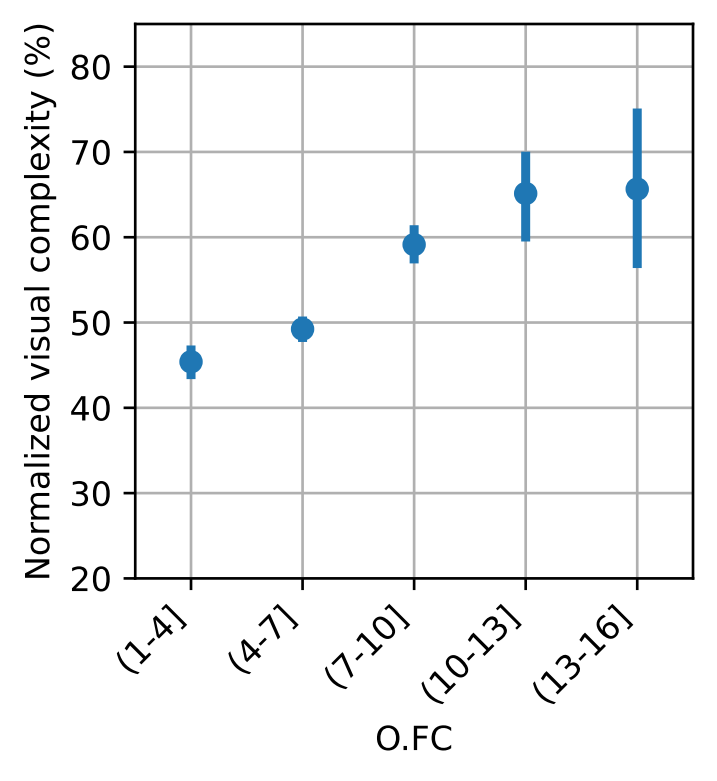}
    \end{subfigure}
    \begin{subfigure}{0.23\textwidth}
        \centering
        \includegraphics[height=\pictureheight]{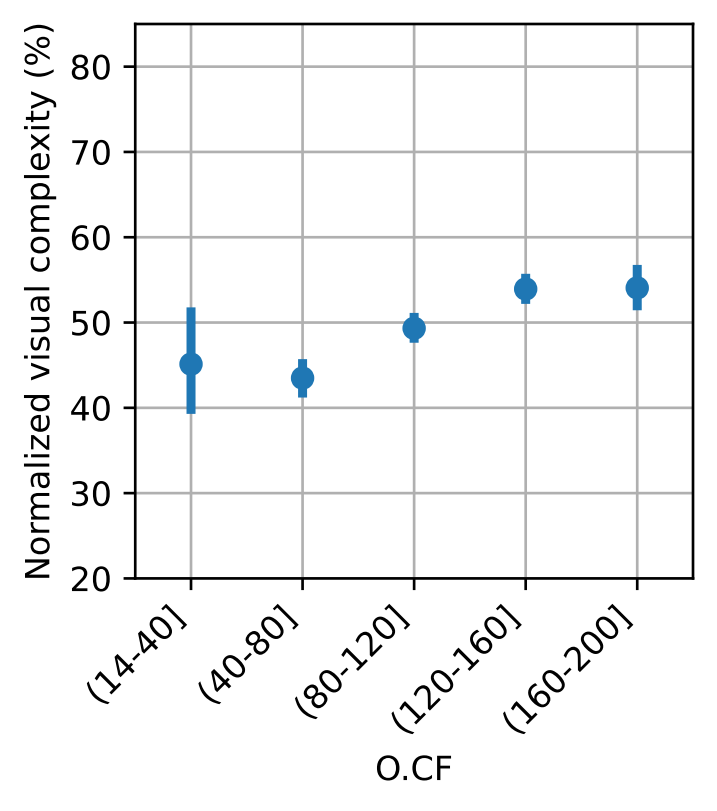}
    \end{subfigure}
    \begin{subfigure}{0.23\textwidth}
        \centering
        \includegraphics[height=\pictureheight]{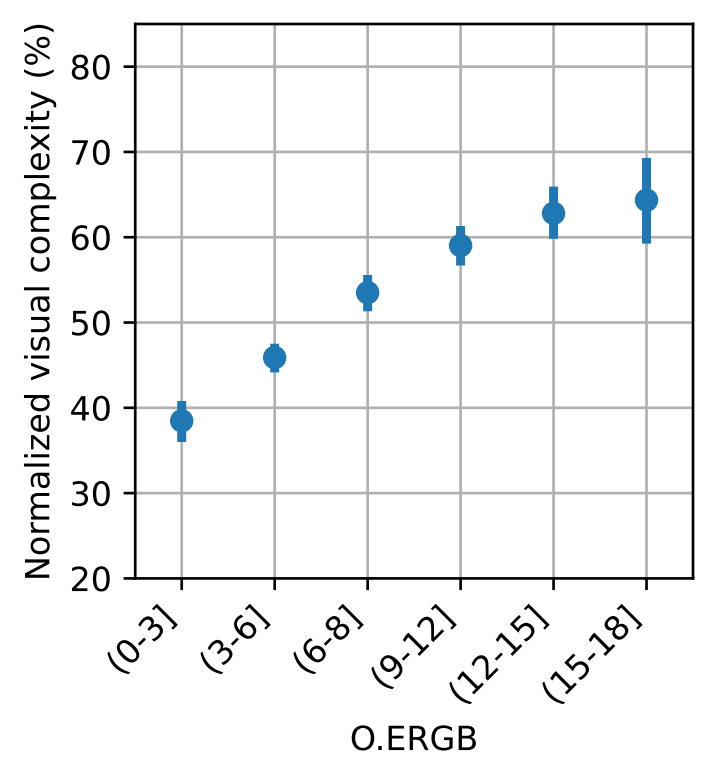}
    \end{subfigure}
    \begin{subfigure}{0.23\textwidth}
        \centering
        \includegraphics[height=\pictureheight]{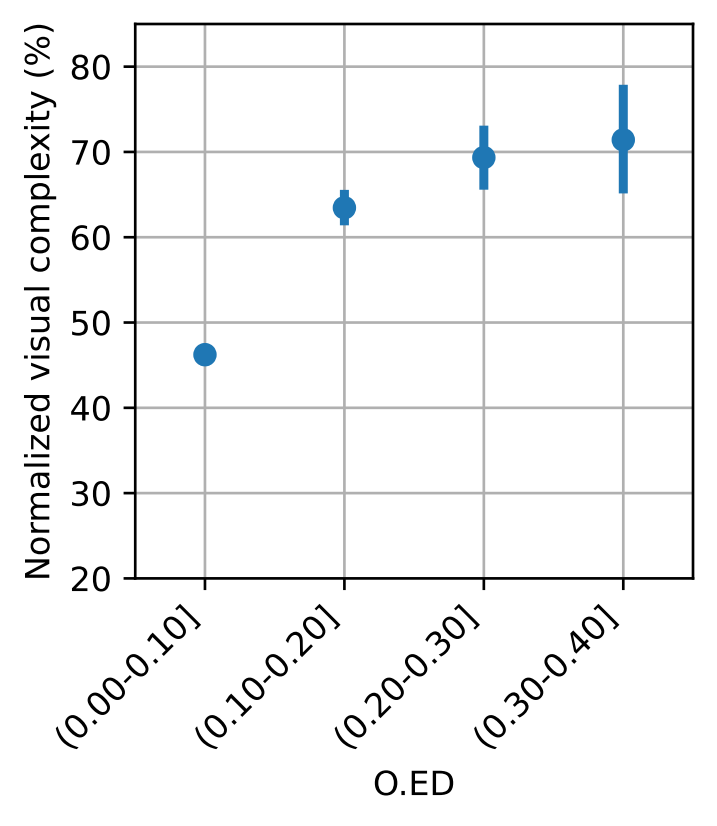}
    \end{subfigure}

    \begin{subfigure}{0.23\textwidth}
        \centering \includegraphics[height=\pictureheight]{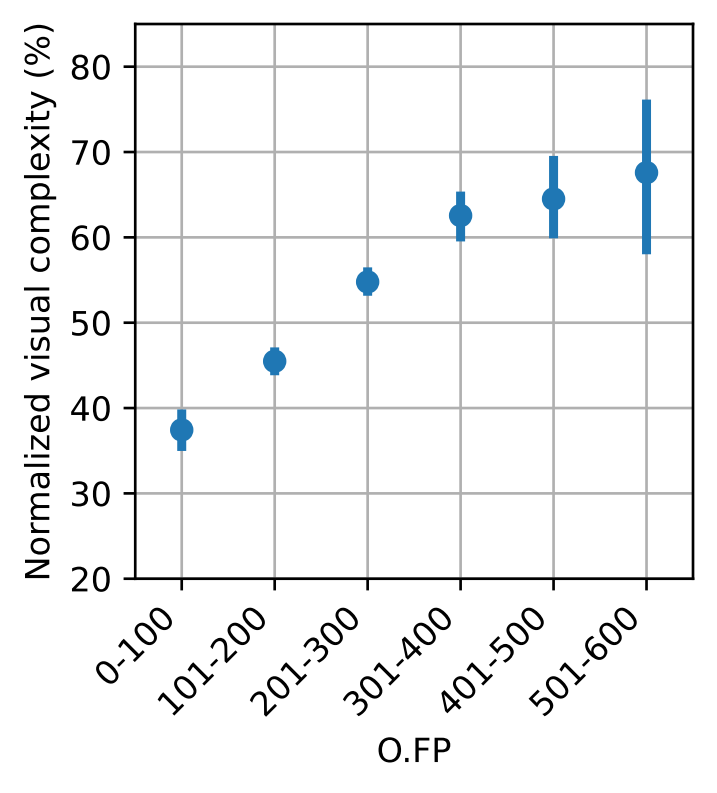}
    \end{subfigure}
    \begin{subfigure}{0.23\textwidth}
        \centering

        \includegraphics[height=\pictureheight]{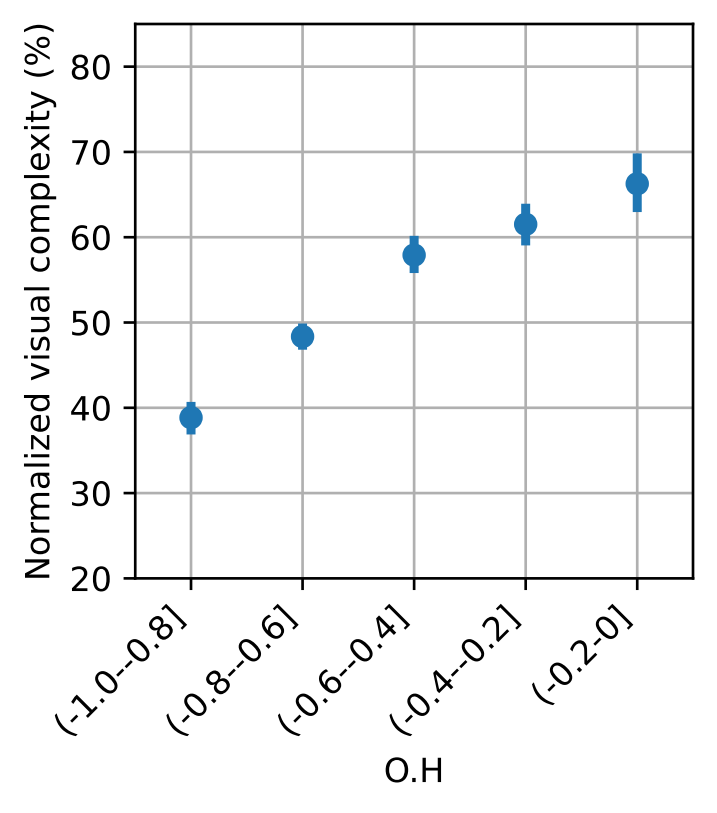}
    \end{subfigure}
    \caption{\rvision{\textbf{Mean visual complexity (VC) across for visualization of all 10 objective metrics.} The O.TiR and O.MeC metric results are in the main text~\autoref{fig:objectmodel}. Error bars represent $95\%$ confidence intervals. These correlations should be interpreted with caution. VC is a high-level phenomenon, so a single-variant analysis like this may not fully account for the variations observed in VC (see the main text ~\autoref{sec:12metrics}).} }
    \label{fig:singleVariableAnalysis}
\end{figure*}

\clearpage
\begin{table*}[!t]
    \centering
    \caption{\textbf{Metric correlation analyses.} P-values for between-metric Pearson correlations. O.TiR has large p-values with most other metrics, indicating non-significant correlations. O.MeC exhibits non-significant correlations with O.TiR and O.FC. In all other cases, $p\leq0.05$ indicates significant correlations (see the main text \autoref{sec:12metrics} and \sm~\autoref{fig:metricsCorrelationfig}).} \begin{tabular}{lrrrrrrrrrrrr}
        \toprule
        {} & \textbf{O.IE} & \textbf{O.KC} & \textbf{O.SE} & \textbf{O.IG} & \textbf{O.FC} & \textbf{O.H} & \textbf{O.CF} & \textbf{O.ERGB} & \textbf{O.ED} & \textbf{O.FP} & \textbf{O.TiR} & \textbf{O.MeC} \\
        \midrule
        O.IE    &  &        &        &        &        &        &        &        &        &        &        &        \\
        O.KC    & $<$0.001 &  &        &        &        &        &        &        &        &        &        &        \\
        O.SE    & $<$0.001 & $<$0.001 &  &        &        &        &        &        &        &        &        &        \\
        O.IG    & $<$0.001&$<$0.001&$<$0.001&  &        &        &        &        &        &        &        &        \\
        O.FC    &$<$0.001&$<$0.001&$<$0.001&$<$0.001& &        &        &        &        &        &        &        \\
        O.H     &$<$0.001&$<$0.001&$<$0.001&$<$0.001&$<$0.001&&        &        &        &        &        &        \\
        O.CF    &$<$0.001&$<$0.001&$<$0.001&$<$0.001&$<$0.001&$<$0.001& &        &        &        &        &        \\
        O.ERGB  &$<$0.001&$<$0.001&$<$0.001&$<$0.001&$<$0.001&$<$0.001&$<$0.001& &        &        &        &        \\
        O.ED    &$<$0.001&$<$0.001&$<$0.001&$<$0.001&$<$0.001&$<$0.001&$<$0.001&$<$0.001&&        &        &        \\
        O.FP    &$<$0.001&$<$0.001&$<$0.001&$<$0.001&$<$0.001&$<$0.001&$<$0.001&$<$0.001&$<$0.001 &&    &        \\
        O.TiR   & {0.28} & {0.79} &$<$0.001& {0.225} &$<$0.001& {0.414} & {0.088} & {0.028} & {0.052} & {0.12} & &        \\
        O.MeC   &$<$0.001&$<$0.001&$<$0.001&$<$0.001&{0.04} &$<$0.001&$<$0.001&$<$0.001&$<$0.001&$<$0.001& {0.157} & \\
        \bottomrule
    \end{tabular}
    \label{tab:metric_corr_pvalue}

\end{table*}

\begin{figure*}[!t]
    \centering
    \centering
    \includegraphics[width=0.6\textwidth]{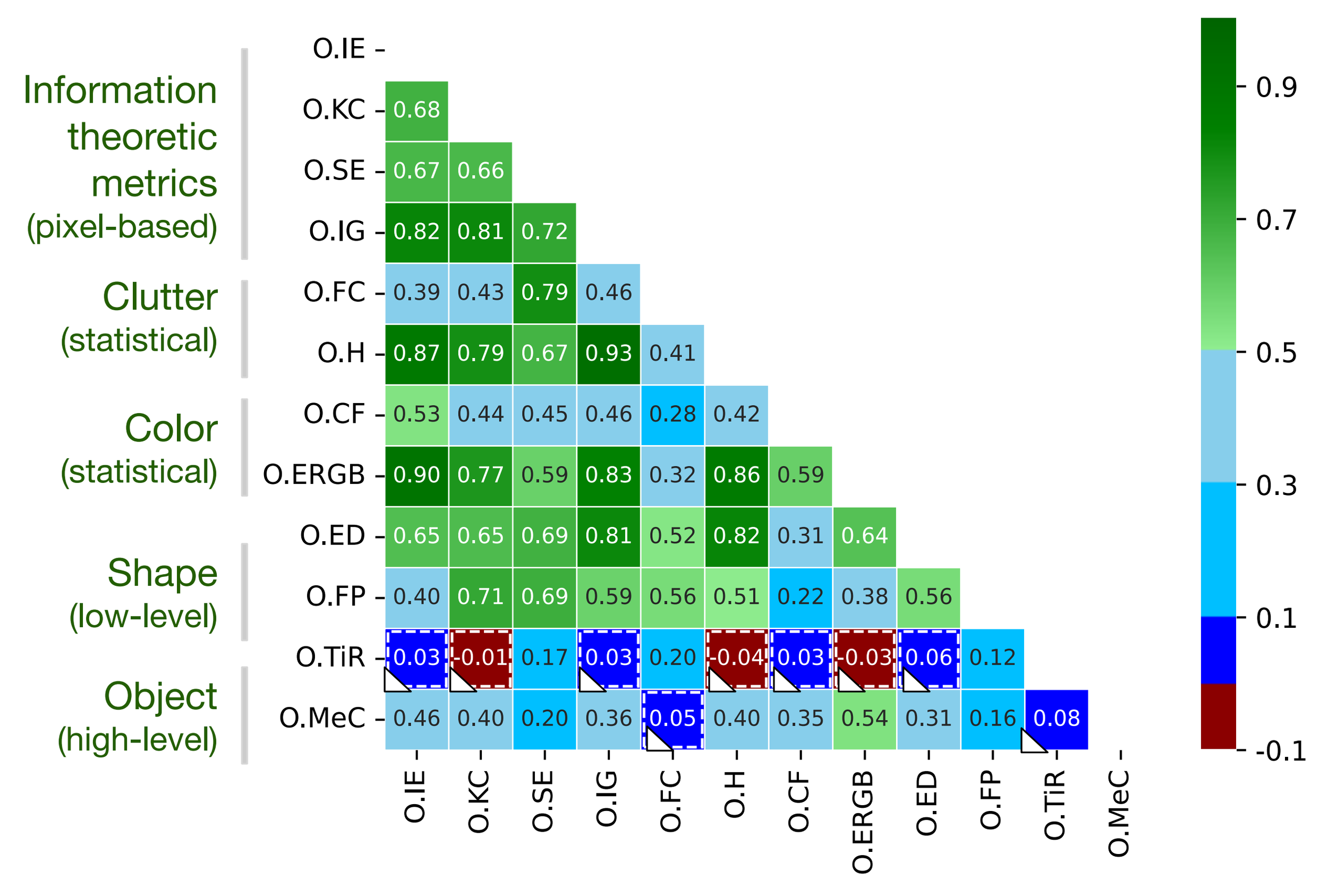}
    \vspace{-10pt}
    \caption{\textbf{Metric correlation analyses.} Pearson correlation coefficients ($r$) between the image quality metrics~(\autoref{tab:ovc}), computed across all images. Cells with a white triangle indicate non-significant correlations at the 0.01 level; all other correlations are statistically significant with $p<0.001$ (see \autoref{tab:metric_corr_pvalue}).
    \textbf{Observations.}
    \textbf{(1) Strong within-factor correlations} were observed among the four mathematical \textit{information-theoretic metrics} (O.IE, O.KC, O.SE, O.IG) and between the two shape-based metrics (O.ED, O.FP). In the case of information-theoretic metrics, complexity is mathematically defined by the length of the system description needed to encode an image. When the same input is processed through different computational models, the resulting compressions tend to be similar. Similarly, for shape-based metrics, the relationship is intuitive: more edges (O.ED) inherently lead to more feature points (O.FP).
    \textbf{(2) Metrics O.IG, O.FC and O.H are sensitive to texture and show strong mutual correlations.} (3)\textbf{Cross-factor correlations were also present}. Shape-based metrics (O.ED, O.FP), along with O.H and O.ERGB, exhibited strong correlations with information-theoretic metrics (O.IE, O.KC, O.SE, O.IG), with $r\in [0.58, 0.93]$ except for $r=0.41$ between O.FP and O.IE. These findings suggest that different metric categories may capture overlapping aspects of visual complexity. (4) In contrast, \textbf{object-based metric O.TiR showed little to no correlation with other metrics}, suggesting a fundamental distinction between text-based and graphical features. O.MeC also had relatively weak correlations with other metrics, supporting its role in capturing semantic or structural complexity rather than low-level image statistics. Finally, there was no meaningful correlation between O.FC and O.MeC, indicating they capture distinct dimensions of visual information. The p-values for between-metric correlations are reported in~\autoref{tab:metric_corr_pvalue} (see the main text~\autoref{sec:12metrics}). }
    \label{fig:metricsCorrelationfig}
\end{figure*}

\clearpage

\begin{table*}[!t]
    \centering
    \caption{\textbf{Effect sizes from partial least squares (PLS) analysis, measured by $f^2$.} These values indicate the contribution of each metric to the overall explained variance in the perceived VC across four experiments. Bold values represent small ($f^2 \geq 0.02$) or medium ($f^2 \geq 0.15$) effect sizes. A value of $f^2\geq0.35$ is considered a large effect size.
    \textbf{Observation.} O.ED emerged as the strongest metric, showing a significant effect in three out of four PLS models. It was followed by O.MeC, O.FP, and O.FC, each significant in two of the four models, and O.IE, which was significant in one model. Notably, O.ED mainly contributed to node-link diagrams; O.FC is the only factor that has a relatively large effect-size to represent the continuous color and texture pattern representations (see the main text ~\autoref{sec:FactorizingPerceivedVC} and~\autoref{sec:compareresults}). }
    \resizebox{\textwidth}{!}{\begin{tabular}{lrrrrrrrrrrrr}
        \toprule
        \textbf{Stimulus} & \textbf{O.IE} & \textbf{O.KC} & \textbf{O.SE} & \textbf{O.IG} & \textbf{O.FC} & \textbf{O.H} & \textbf{O.CF} & \textbf{O.ERGB} & \textbf{O.ED} & \textbf{O.FP} & \textbf{O.TiR} & \textbf{O.MeC} \\
        \midrule
        Overall (\autoref{fig:factorization}) & -0.000 & 0.000 & 0.002 & 0.001 & 0.003 & 0.001 & 0.001 & 0.003 & \textbf{0.050} & \textbf{0.031} & 0.007 & \textbf{0.068} \\
        HP-Node-link (\autoref{fig:networks}) & 0.001 & 0.006 & -0.003 & 0.001 & 0.007 & -0.001 & 0.005 & \textbf{0.026} & \textbf{0.288} & 0.012 & 0.001 & 0.019 \\
        Dsct.-Grid \& Matrix (\autoref{fig:gridmatrix}) & \textbf{0.026} & -0.000 & 0.014 & 0.003 & \textbf{0.021} & 0.001 & -0.001 & 0.007 & \textbf{0.059} & \textbf{0.058} & 0.003 & \textbf{0.148} \\
        RR-Cont.-Grid \& Matrix (ColorPatn) (\autoref{fig:gridmatrix}) & 0.000 & 0.009 & 0.015 & 0.016 & \textbf{0.152} & -0.001 & 0.007 & 0.011 & 0.014 & 0.001 & 0.001 & 0.001 \\
        \bottomrule
    \end{tabular} }
    \label{tab:effectsize}
\end{table*}

\begin{figure*}[!t]
    \centering
    \setlength{\picturewidth}{0.33\textwidth}
    \setlength{\pictureheight}{5cm}

    \begin{subfigure}{\picturewidth}
        \centering
        \includegraphics[height=\pictureheight]{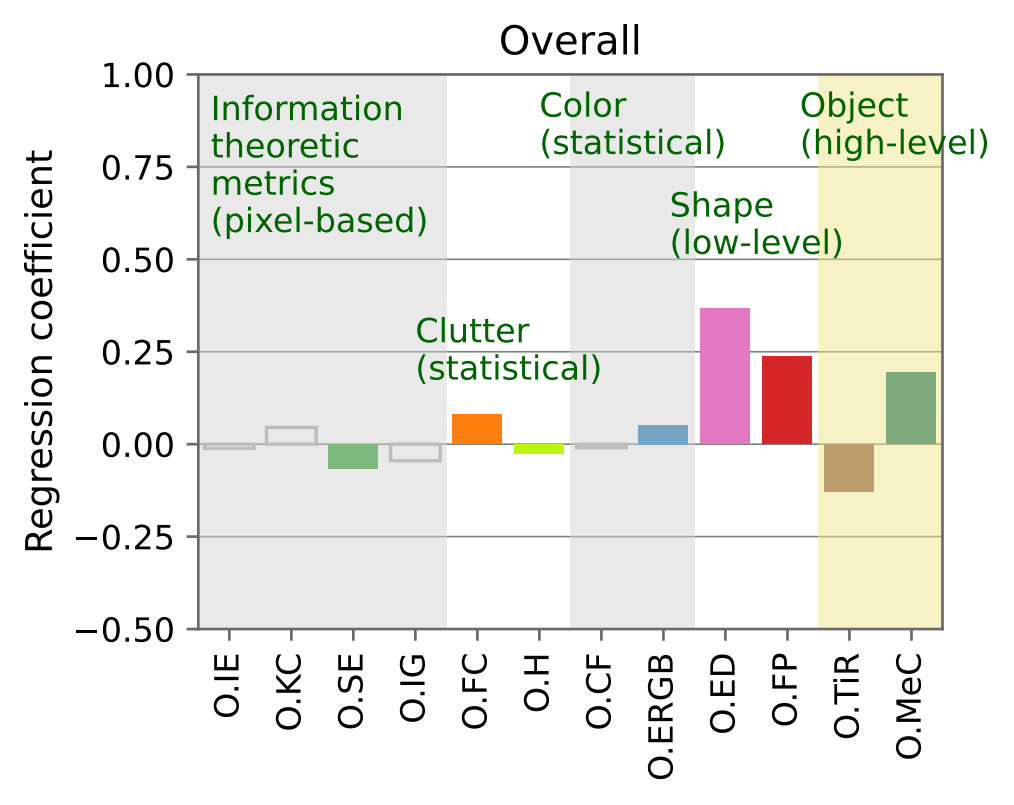}
        \caption{$\Delta E = 10$, $R^2 = 0.407$}
    \end{subfigure}
    \begin{subfigure}{\picturewidth}
        \centering \includegraphics[height=\pictureheight]{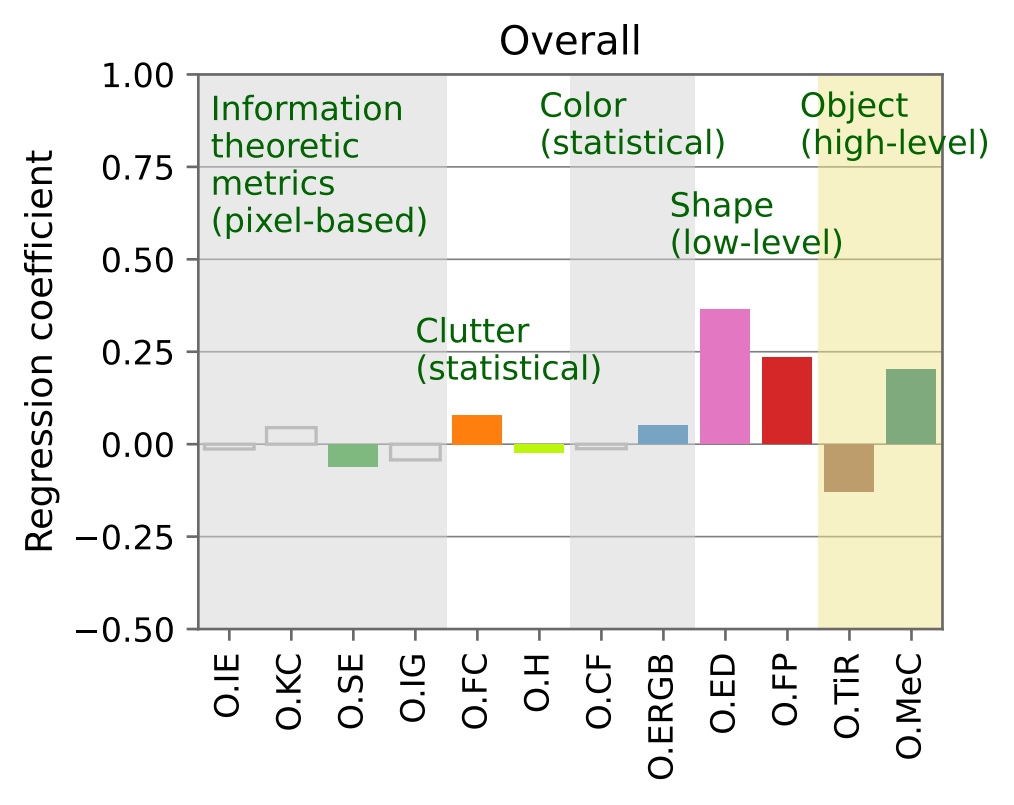}
        \caption{$\Delta E = 12$, $R^2 = 0.408$}
    \end{subfigure}
    \begin{subfigure}{\picturewidth}
        \centering
        \includegraphics[height=\pictureheight]{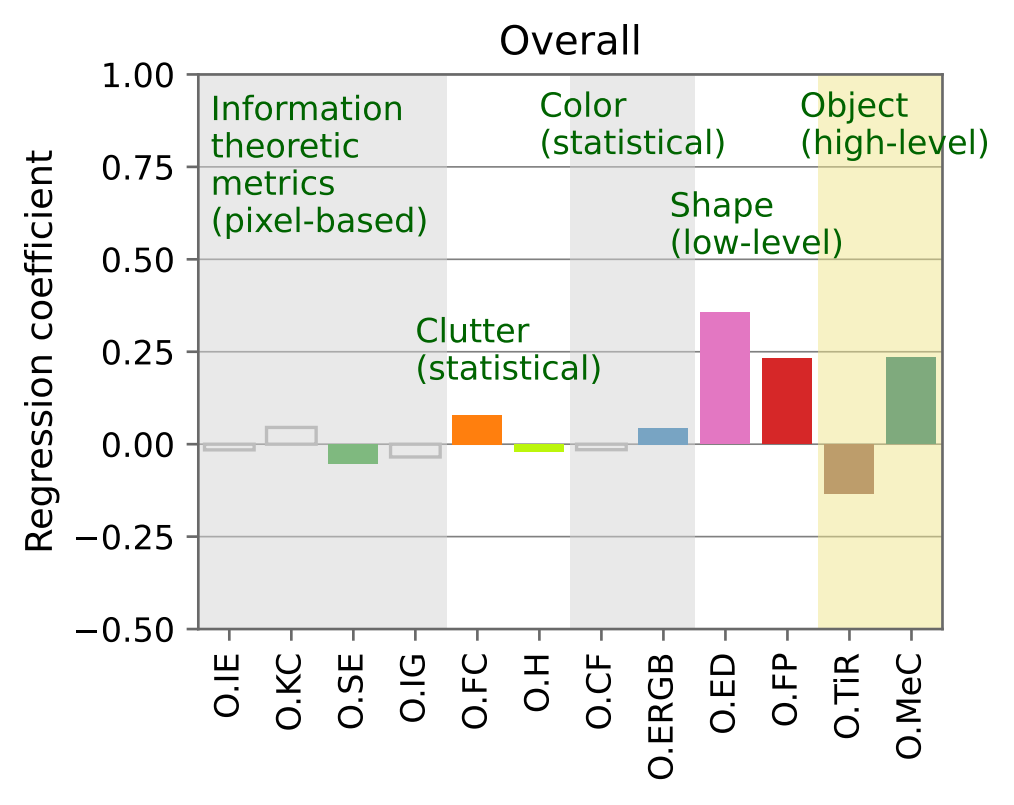}
        \caption{$\delta E = 16$, $R^2 = 0.411$}
    \end{subfigure}

    \begin{subfigure}{\picturewidth}
        \centering \includegraphics[height=\pictureheight]{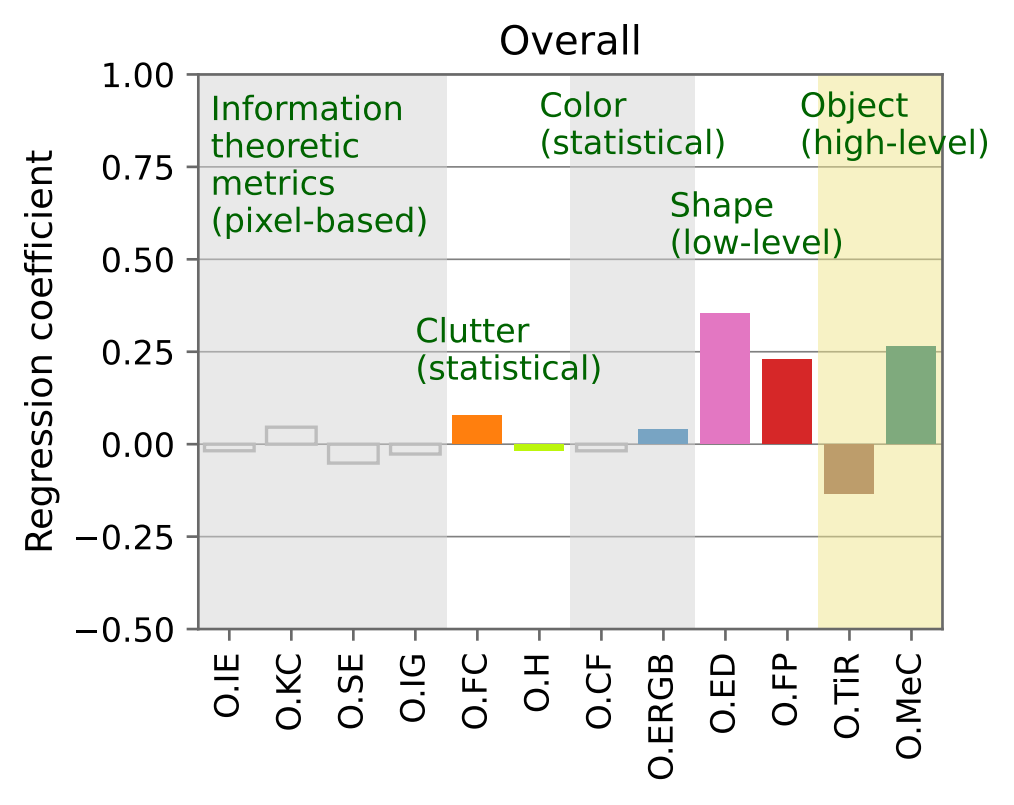}
        \caption{$\delta E = 18$, $R^2 = 0.415$}
    \end{subfigure}
    \begin{subfigure}{\picturewidth}
        \centering
        \includegraphics[height=\pictureheight]{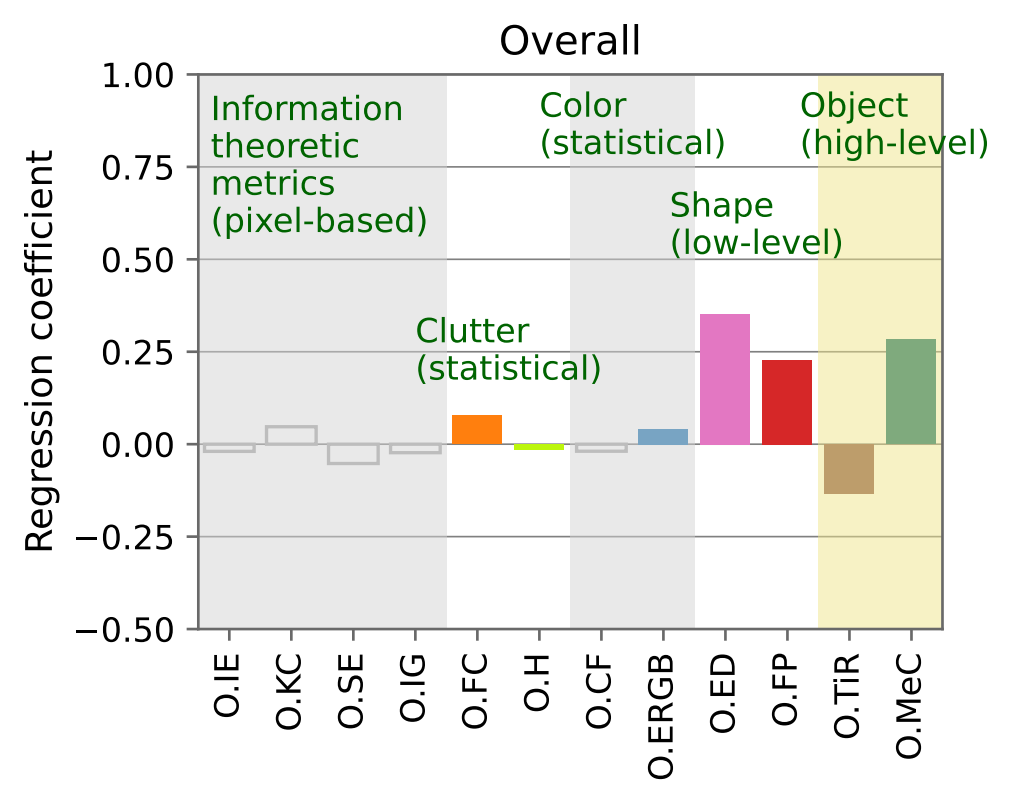}
        \caption{$\delta E = 20$, $R^2 = 0.415$}
    \end{subfigure}
    \begin{subfigure}{\picturewidth}
        \centering
        \includegraphics[height=\pictureheight]{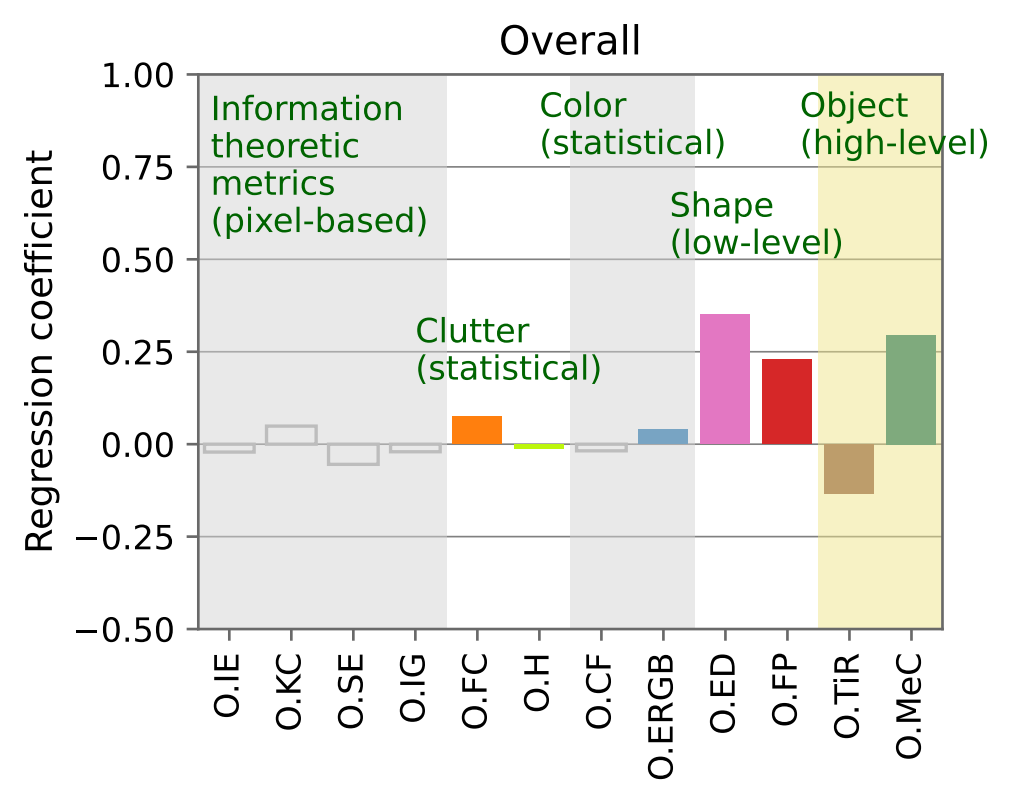}
        \caption{$\delta E = 22$, $R^2 = 0.416$}
    \end{subfigure}
    \caption{\rvision{\textbf{PLS modeling results of VC}, when O.MeC is computed using different threshold values ranging from $\Delta E$ = {10, 12, 14, 16, 18, 20, and 22}. Each subplot corresponds to a specific threshold $\Delta E$ and the corresponding PLS regression coefficient.
    \textbf{Observations.} Despite variations in threshold values, the modeling results remain largely consistent, indicating minimal impact of the thresholding on the overall model structure (see the main text~\autoref{sec:FactorizingPerceivedVC}).}}
    \label{fig:threshold}
\end{figure*}

\begin{figure*}[t!]
    \centering

    {\includegraphics[width=0.9\textwidth]{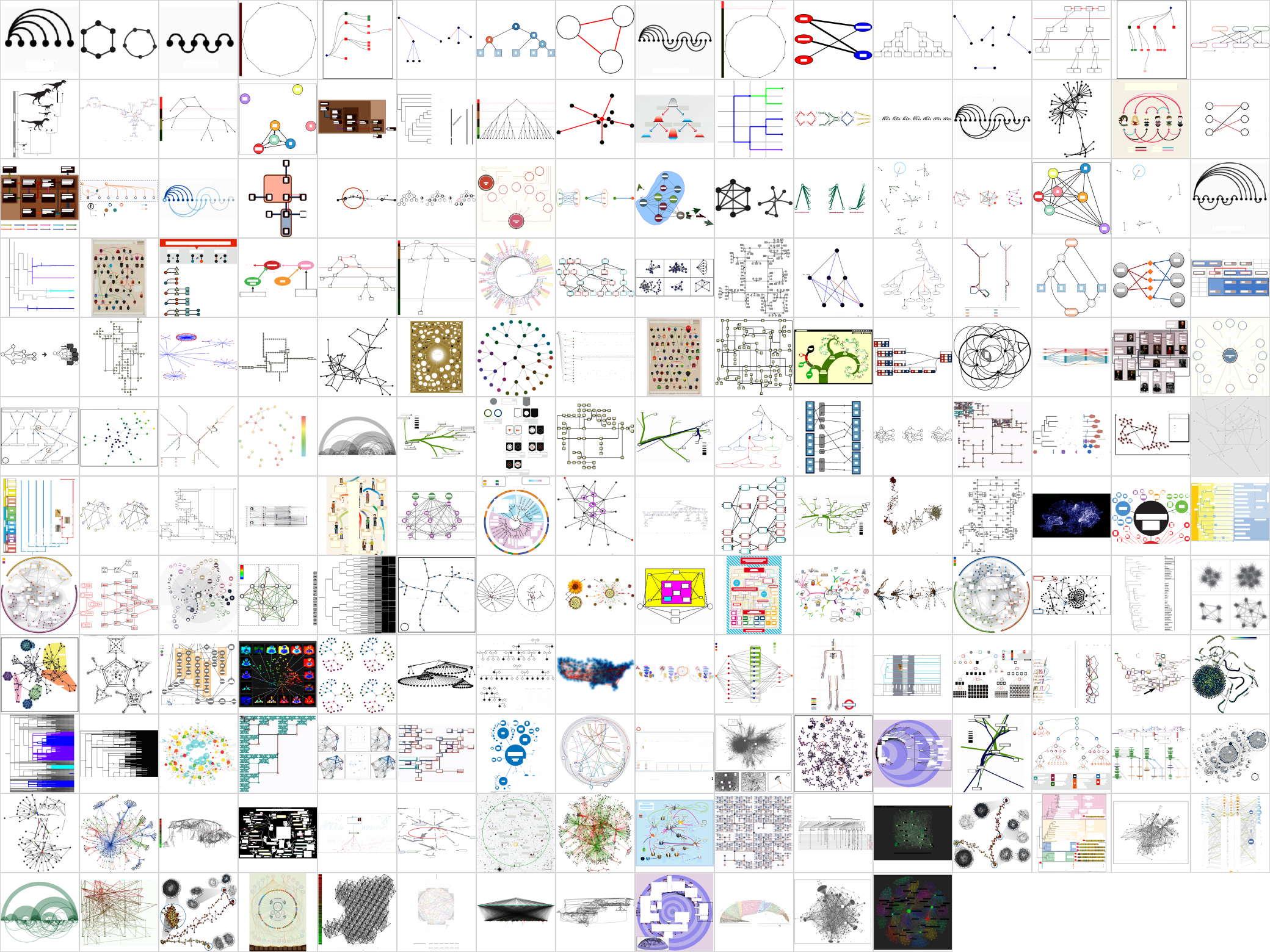}}
    \caption{\textbf{Direct node-link diagrams} in the style of Purchase~\cite{purchase2012exploration} (see the main text~\autoref{sec:purchase}).}
    \label{fig:TreesNetworks}
\end{figure*}

\begin{figure*}[!t]
    \centering

    \subfloat[Pixel location is given (183 images).] {\includegraphics[width=0.9\textwidth]{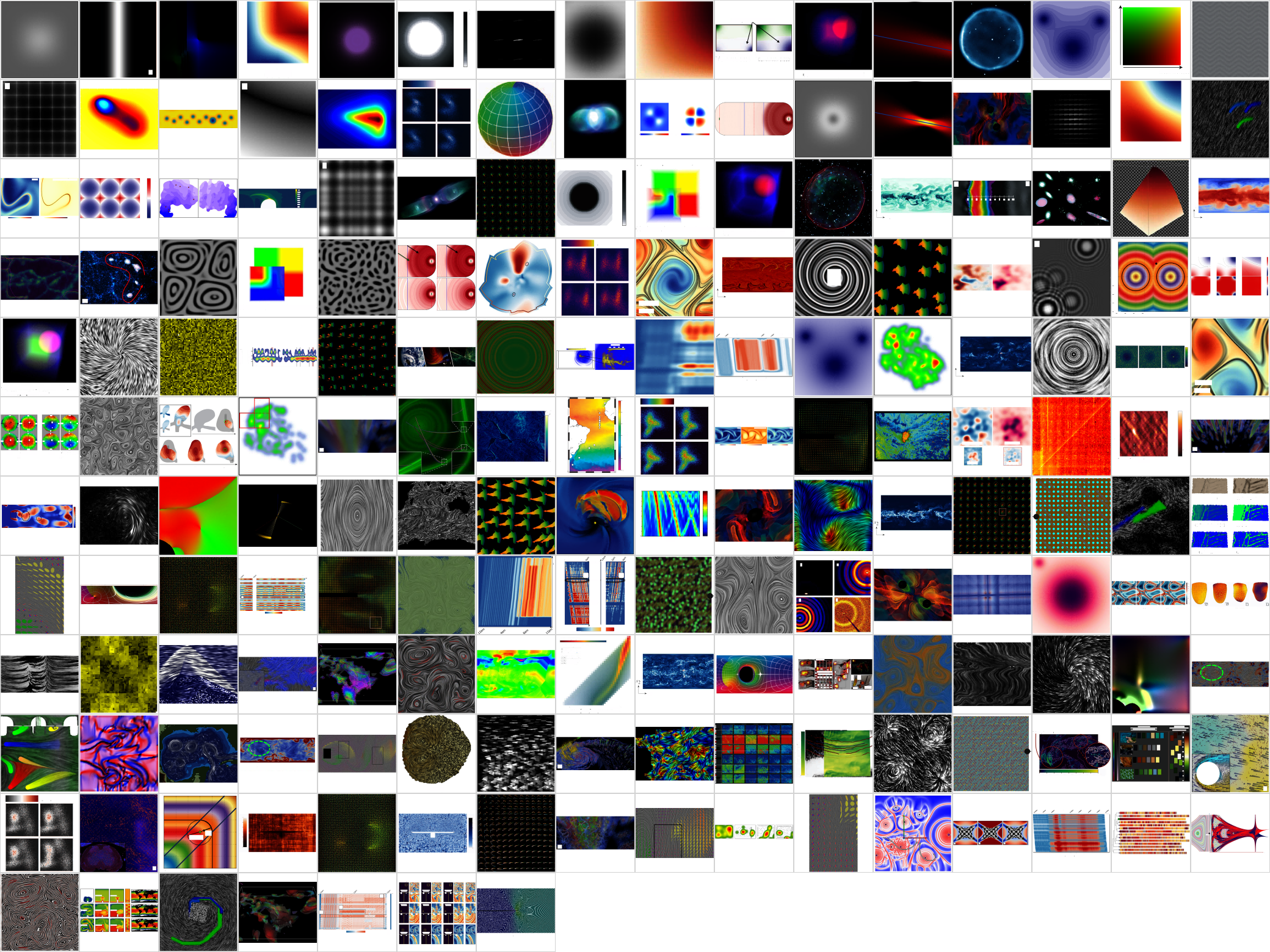}}

    \vspace{1em}
    \subfloat[Pixel location is not given (128 images).]{\includegraphics[width=0.9\textwidth]{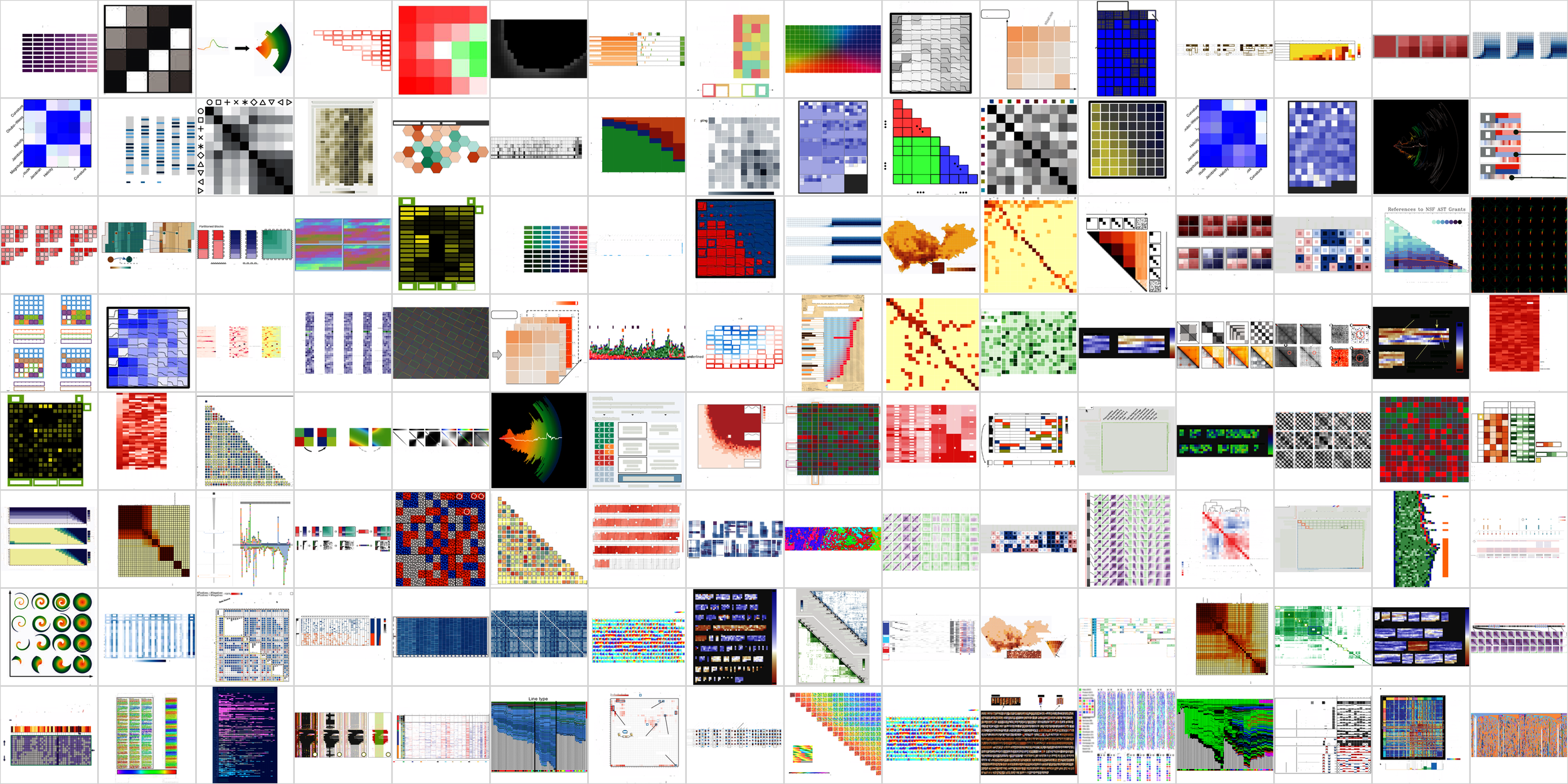}}

    \caption{\textit{Top (a):} {Images in the style of Rosenholtz et al.~\cite{rosenholtz2007measuring}, where colors and textures dominate the visual representation.} \textit{Bottom (b):} Examples represent discrete grid and matrix images (see the main text~\autoref{sec:rosenholtz}).}
    \label{fig:SpatialGridMatrix}
\end{figure*}

\clearpage
\begin{table*}[t!]
    \caption{\textbf{Top-10 keywords generated from the post-questionnaire comments.} Keywords in italics are unique to their category. The numbers in brackets and the width of the bar denote the number of participant comments mentioning that keyword or topic (see \sm\autoref{sec:verbal}).}
    \label{tab:keywordCountTop10}
    \resizebox{\textwidth}{!}{\begin{tabular}{@{}llllllll@{}}
        \toprule
        \databar[A1C9F4]{317}{698}{\textbf{Clarity and Readability (317)}}\hspace{10pt} & \databar[A1C9F4]{271}{698}{\textbf{Color Usage (271)}} & \databar[A1C9F4]{226}{698}{\textbf{Number of Elements (226)}} & \databar[A1C9F4]{221}{698}{\textbf{Information Density (221)}} & \databar[A1C9F4]{173}{698}{\textbf{Abstractness and Familiarity (173)}}\hspace{30pt} & \databar[A1C9F4]{113}{698}{\textbf{Visual Clutter (113)}} & \databar[A1C9F4]{49}{698}{\textbf{Interconnectedness (49)}} & \databar[A1C9F4]{16}{698}{\textbf{Beautifulness (16)}} \\ \midrule
        \databar[A1C9F4]{55}{698}{(55) clear}                              & \databar[A1C9F4]{249}{698}{(249) color}                & \databar[A1C9F4]{40}{698}{(40) line}                          & \databar[A1C9F4]{95}{698}{(95) information}                    & \databar[A1C9F4]{29}{698}{(29) complex}                                 & \databar[A1C9F4]{39}{698}{(39) detail}                    & \databar[A1C9F4]{17}{698}{(17) line}                        & \databar[A1C9F4]{3}{698}{(3) appealing}                \\
        \databar[A1C9F4]{39}{698}{(39) easy}                               & \databar[A1C9F4]{21}{698}{(21) \textit{contrast}}      & \databar[A1C9F4]{32}{698}{(32) number}                        & \databar[A1C9F4]{30}{698}{(30) detail}                         & \databar[A1C9F4]{19}{698}{(19) \textit{familiar}}                       & \databar[A1C9F4]{23}{698}{(23) element}                   & \databar[A1C9F4]{8}{698}{(8) \textit{connection}}           & \databar[A1C9F4]{3}{698}{(3) \textit{like}}            \\
        \databar[A1C9F4]{39}{698}{(39) hard}                               & \databar[A1C9F4]{18}{698}{(18) variety}                & \databar[A1C9F4]{31}{698}{(31) element}                       & \databar[A1C9F4]{20}{698}{(20) legend}                         & \databar[A1C9F4]{17}{698}{(17) pattern}                                 & \databar[A1C9F4]{21}{698}{(21) clutter}                   & \databar[A1C9F4]{7}{698}{(7) overlap}                       & \databar[A1C9F4]{3}{698}{(3) pattern}                  \\
        \databar[A1C9F4]{39}{698}{(39) label}                              & \databar[A1C9F4]{16}{698}{(16) different}              & \databar[A1C9F4]{30}{698}{(30) text}                          & \databar[A1C9F4]{19}{698}{(19) complex}                        & \databar[A1C9F4]{17}{698}{(17) \textit{structure}}                      & \databar[A1C9F4]{15}{698}{(15) \textit{excessive}}        & \databar[A1C9F4]{6}{698}{(6) \textit{intersect}}            & \databar[A1C9F4]{3}{698}{(3) \textit{pleasing}}        \\
        \databar[A1C9F4]{29}{698}{(29) difficult}                          & \databar[A1C9F4]{10}{698}{(10) \textit{shade}}         & \databar[A1C9F4]{22}{698}{(22) lot}                           & \databar[A1C9F4]{17}{698}{(17) point}                          & \databar[A1C9F4]{13}{698}{(13) \textit{unfamiliar}}                     & \databar[A1C9F4]{15}{698}{(15) overlap}                   & \databar[A1C9F4]{5}{698}{(5) \textit{interconnect}}         & \databar[A1C9F4]{2}{698}{(2) color}                    \\
        \databar[A1C9F4]{28}{698}{(28) font}                               & \databar[A1C9F4]{9}{698}{(9) lot}                      & \databar[A1C9F4]{21}{698}{(21) detail}                        & \databar[A1C9F4]{16}{698}{(16) lot}                            & \databar[A1C9F4]{11}{698}{(11) difficult}                               & \databar[A1C9F4]{8}{698}{(8) layout}                      & \databar[A1C9F4]{5}{698}{(5) point}                         & \databar[A1C9F4]{2}{698}{(2) \textit{impact}}          \\
        \databar[A1C9F4]{27}{698}{(27) layout}                             & \databar[A1C9F4]{7}{698}{(7) \textit{dark}}            & \databar[A1C9F4]{16}{698}{(16) variety}                       & \databar[A1C9F4]{13}{698}{(13) \textit{concentration}}         & \databar[A1C9F4]{11}{698}{(11) \textit{subject}}                        & \databar[A1C9F4]{7}{698}{(7) complex}                     & \databar[A1C9F4]{4}{698}{(4) \textit{crossing}}             &                                                        \\
        \databar[A1C9F4]{27}{698}{(27) text}                               & \databar[A1C9F4]{7}{698}{(7) \textit{range}}           & \databar[A1C9F4]{14}{698}{(14) many}                          & \databar[A1C9F4]{13}{698}{(13) layer}                          & \databar[A1C9F4]{10}{698}{(10) graph}                                   & \databar[A1C9F4]{5}{698}{(5) dot}                         & \databar[A1C9F4]{4}{698}{(4) element}                       &                                                        \\
        \databar[A1C9F4]{21}{698}{(21) information}                        & \databar[A1C9F4]{7}{698}{(7) \textit{saturation}}      & \databar[A1C9F4]{13}{698}{(13) point}                         & \databar[A1C9F4]{12}{698}{(12) amount}                         & \databar[A1C9F4]{9}{698}{(9) \textit{idea}}                             & \databar[A1C9F4]{5}{698}{(5) information}                 & \databar[A1C9F4]{4}{698}{(4) lot}                           &                                                        \\
        \databar[A1C9F4]{21}{698}{(21) small}                              & \databar[A1C9F4]{6}{698}{(6) number}                   & \databar[A1C9F4]{12}{698}{(12) amount}                        & \databar[A1C9F4]{11}{698}{(11) graph}                          & \databar[A1C9F4]{8}{698}{(8) clear}                                     & \databar[A1C9F4]{5}{698}{(5) tidy}                        & \databar[A1C9F4]{3}{698}{(3) arrow}                         &                                                        \\ \bottomrule
    \end{tabular}}
\end{table*}

\begin{table*}[!t]
    \centering
    \caption{\textbf{Categorization of subject VC terms} from the user-studies (see \sm\autoref{sec:verbal}). }
    \small \begin{tabular}{p{3.2cm}p{14cm}}
        \toprule
        Terms & Reasoning \\
        \midrule
        \textbf{Clarity and Readability} &
        Poor readability due to small font sizes, cluttered layouts, or lack of clear labels can increase perceived complexity. Clear and concise labeling, along with a logical flow of information, tends to reduce complexity.
        \\
        \textbf{Color Usage} &
        The use of a wide range of colors, especially when they are vibrant or poorly contrasting, can add to the complexity. Images with more color variations and saturation are often seen as more visually complex.
        \\

        \textbf{Information Density} &
        High information density, where a lot of data is presented in a small space, can make an image appear more complex. This includes the presence of detailed legends, annotations, and multiple layers of information.
        \\
        \textbf{Number of Elements} &
        Images with more elements, such as shapes, lines, and text, are often perceived as more complex.
        The presence of multiple data points, overlapping elements, and intricate patterns can increase the perceived complexity.
       % \glyphabbr, \nodelinkabbr, and \pointabbr were most concerning.
        \\

        \textbf{Abstractness and Familiarity} &
        Images that lack a clear structure or recognizable patterns are often perceived as more complex. Familiarity with the content also plays a role; images depicting unfamiliar subjects or using unfamiliar formats are seen as more complex.
        \\
        \textbf{Visual Clutter} &	Overlapping elements, excessive details, and lack of negative space can contribute to a sense of clutter, making an image appear more complex.
        \\
        \textbf{Beautifulness} &
        Visual appeal, good-looking, personal preference, emotional feeling.
        \\
        \bottomrule
    \end{tabular}
    \label{tab:vcDimensionNames}
\end{table*}

\clearpage

\end{document}